\documentclass[aps, prd, amsmath, floats, floatfix, twocolumn,
superscriptaddress, nofootinbib, showpacs]{revtex4-1}

\usepackage[colorlinks, citecolor=blue, pdfborder={0 0 0}, plainpages=false]{hyperref}
\usepackage{graphicx}
\usepackage{xspace}
\usepackage[usenames,dvipsnames]{color}
\usepackage{amssymb}
\usepackage{mathrsfs}

\allowdisplaybreaks

\usepackage{ulem}
\newcommand{\Caltech}{\affiliation{
    TAPIR,
    Walter Burke Institute for Theoretical Physics,
    California Institute of Technology, Pasadena, California 91125, USA}}

\begin{document}

\title{Gravitational waveforms of binary neutron star inspirals
using post-Newtonian tidal splicing}

\author{Kevin Barkett} \Caltech
\author{Yanbei Chen} \Caltech
\author{Mark A. Scheel} \Caltech
\author{Vijay Varma} \Caltech

\date{\today}

\begin{abstract}

The tidal deformations of neutron stars within an inspiraling compact binary
alter the orbital dynamics, imprinting a signature on the gravitational wave
signal. Modeling this signal could be done with numerical-relativity
simulations, but these are too computationally expensive for many applications.
Analytic post-Newtonian treatments are limited by unknown higher-order nontidal
terms. This paper further builds upon the ``tidal splicing'' model in which
post-Newtonian tidal terms are ``spliced'' onto numerical relativity simulations
of black-hole binaries. We improve on previous treatments of tidal splicing by
including spherical harmonic modes beyond the (2,2) mode, expanding the
post-Newtonian expressions for tidal effects to 2.5 order, including dynamical
tide corrections, and adding a partial treatment of the spin-tidal dynamics.
Furthermore, instead of numerical relativity simulations, we use the
spin-aligned binary black hole (BBH) surrogate model ``NRHybSur3dq8'' to provide the BBH waveforms
that are input into the tidal slicing procedure. This allows us to construct
spin-aligned, inspiraling TaylorT2 and TaylorT4 splicing waveform models that
can be evaluated quickly. These models are tested against existing binary
neutron star and black hole--neutron star simulations. We implement the TaylorT2
splicing model as an extension to ``NRHybSur3dq8,'' creating a model that we
call ``NRHybSur3dq8Tidal.''

\end{abstract}


\maketitle


\section{Introduction}\label{sec:Intro}

In August 2017, the network of interferometers consisting of Advanced LIGO~\cite{aLIGO2}
and VIRGO~\cite{aVirgo2} first observed the gravitational radiation from the
inspiral and merger of a binary neutron star (BNS)~\cite{TheLIGOScientific:2017qsa}, opening
the door to exploring extremely compact objects other than binary black holes
(BBH). Coincident detection with the electromagnetic observational counterpart
GRB 170817A~\cite{2017ApJ...848L..13A}, showed that these systems are the progenitors
of short gamma-ray bursts, and herald the start of multimessenger astronomy.
Additionally, this detection served as a probe of the neutron star
equation of state (EOS)~\cite{TheLIGOScientific:2017qsa,2017ApJL2041}, and
provided constraints on the gravitational wave speed~\cite{2017ApJ...848L..13A}. As
the detectors' sensitivities improve, observing more such systems will
further constrain the governing
physics~\cite{DelPozzo:13,Kumar:2016zlj}. However, capturing all of the
information encoded in the measured signals requires detailed, accurate
templates that precisely describe waveforms from BNS or black hole--neutron star
(BHNS) systems.

A common approach to generate BHNS and BNS waveforms is to create analytic and
phenomenological models that capture the behavior of BHNS and BNS systems, often
in the form of additional corrections to BBH waveform models. Within the
post-Newtonian (PN) formalism, Ref~\cite{Vines2011} first computed the leading-order
tidal effects on the orbital evolution, characterized by the static
quadrupolar tidal deformability, $\bar\lambda_2$. However, recent work suggests
that the choice of the BBH model used as the background for tidal corrections
can impact parameter estimation~\cite{Chakravarti:2018uyi,Samajdar:2018sd}, suggesting that
errors in PN waveforms for BNS and BHNS systems might be dominated by unknown
higher-order vacuum terms rather than tidal terms. Within the frequency
domain, work has gone into mitigating biases caused by this problem through
partially expanding the nonspinning point-mass equations to higher-orders~\cite{Messina:2019qta}.

An alternative to the PN formalism is to implement tidal corrections as an
extension to the effective one-body (EOB) formalism~\cite{Damour:2009wj}.
Current tidal EOB models include the time domain model
SEOBNRv4T~\cite{Hinderer:2016eia,Steinhoff:2016rfi}, whereby the calibrated EOB
model SEOBNRv4~\cite{Bohe:2016gbl} is extended by including additional static
higher-order effects~\cite{Bini:2012gu, Baiotti2011, Steinhoff:2016rfi,
Dietrich:2017dh}. Another distinct EOB model, TEOBResumS~\cite{Nagar:2018zoe,
Nagar:2018plt}, also includes spin aligned as well as self-spin effects through
next-to-next-to leading order, and a postmerger description informed by
black-hole perturbation theory and NR BBH waveforms.

Additional models have been constructed by calibrating
  to numerical simulations of BNS and BHNS binaries.
The LEA+ model~\cite{Kumar:2016zlj} is a frequency domain waveform model
calibrated to a series of BHNS simulations with mass ratios of
$q>2$~\cite{Lackey:2013axa} as an enhancement to the SEOBNRv2
model~\cite{Taracchini:2013rva}. While LEA+ is a full waveform model, it is
valid only over the limited parameter space of the simulations used to calibrate it,
i.e., $q>2$.

The frequency domain models SEOBNRv4\_ROM\_NRTidal, PhenomD\_NRTidal, and
PhenomPv2\_NRTidal~\cite{Dietrich:2018nrt} are built by combining the
BBH models SEOBNRv4\_ROM~\cite{Bohe:2016gbl}, PhenomD~\cite{Husa:2015iqa,Khan:2015jqa}, and
PhenomPv2~\cite{Schmidt:2012rh,Schmidt:2014iyl} with the NRTidal
model~\cite{Dietrich:2017aum}. NRTidal is a phenomenological fit of BNS/BHNS
simulation data to PN-like coefficients. While these models cover a wide range
of BNS parameter space, the fits are calibrated to a limited number of waveforms
and seem to overestimate the tidal effects during the
inspiral~\cite{Dietrich:2018nrt}, though
improvements to these models are expected to eliminate this
problem~\cite{Dietrich:2019nrt}.

The most accurate means of obtaining BNS and BHNS waveform templates would be to
run full numerical simulations. Running simulations that incorporate the
relevant matter physics for BHNS/BNS is a field of active
development~\cite{Foucart:2013psa, Lackey:2013axa, Lovelace:2013vma,
Kawaguchi:2015, Haas:2016, Hinderer:2016a, Dietrich:2017aum, 2018arXiv180601625D,
Foucart:2018inp, Kiuchi:2019kzt}.
However, the range of possible systems spans not just the masses and spins of
the components, but also includes all
allowable EOSs.
This would require a large number of simulations to populate such
a high dimensional parameter space.
Furthermore, the large computational cost of such simulations makes them
impractical for parameter estimation purposes.

On the other hand, numerical simulations of BBH systems have made great strides
over recent years, and there are now public repositories of hundreds of
simulations for binaries with a variety of different initial masses and
spins~\cite{Aylott:2009tn, Ajith:2012az, Hinder:2013oqa, Mroue:2013PRL,
  Jani:2016wkt, Healy:2017bsc, Healy:2019bsc, Huerta:2019hrh, Boyle:2019kee}.
Furthermore, surrogate models now allow interpolation of numerical-relativity waveforms
to desired values of initial masses and spins~\cite{Field:2013cfa, Purrer:2014fza, Blackman:2015pia, Blackman:2017dfb,
Blackman:2017pcm, Varma:2018mmi, Varma:2019csw}. Reference~\cite{Varma:2018mmi}
showed that surrogate models can robustly generate faithful representations of
binary black hole systems with spin magnitudes $\chi < 0.8$ and masses
low enough to be valid for BNS systems ($M\geq 2.25 M_{\odot}$).

We build on a hybrid method called ``tidal splicing''~\cite{Barkett2015}, which
computes inspiral waveforms for BNS and BHNS by combining the accuracy of
numerical BBH simulations with the efficiency of PN models for tidally
deformable systems. This method does so by decomposing the numerical BBH waveform
in a manner akin to the PN formalism and using this decomposition to replace all
orders of the vacuum terms in the analytic PN expansion with their numerical
equivalents. We combine these vacuum terms with the analytic tidal PN terms to
build up a waveform that models the inspiral of a BNS/BHNS.

In this paper, we continue the development of tidal splicing
beyond Ref.~\cite{Barkett2015} by extending the
method to spinning systems and spherical harmonic modes beyond the (2,2) mode.
Using results from
newly available EOB models that incorporate higher PN effects~\cite{Bini:2012gu,
Baiotti2011, Steinhoff:2016rfi, Dietrich:2017dh}, we also
extend the known higher-order tidal effects from EOB to the time domain PN
approximants. Previous explorations of tidal splicing~\cite{Barkett2015} were
based on particular individual numerical relativity BBH simulations, and
therefore could be tested only for the masses and spins of those simulations.
Here, instead of using numerical relativity simulations directly, we will use
the hybridized surrogate model `NRHybSur3dq8'~\cite{Varma:2018mmi} as our BBH
base.

We organize this paper as follows: in Sec~\ref{sec:PNTheory} we summarize the
current existing work on time domain tidal waveforms in the PN framework; in
Sec~\ref{sec:ExtendingPN} we discuss how we partially extend the PN tidal
approximants to 2.5PN order and how we correct the dynamical tide effects for
spinning NS; in Sec~\ref{sec:TidalSplicing} we explain our method of tidal
splicing; and in Sec~\ref{sec:Results} we compare tidal splicing with
some recent BHNS and BNS simulations.

Except where otherwise noted, we shall use the subscripts $A,B$ to refer to the
individual NS or BH objects, the subscripts $\ell = 2,3,\ldots$ refer usually to
the specific polar mode of the tidal effect in consideration (i.e.,
2=quadrupolar, 3=octopolar,\ldots), while the $\ell,m$ superscripts will
typically correspond to the spin-weighted spherical harmonic modes
${}_{-2}Y^{\ell m}$ of the waveform.
We chose units of $G=c=1$.

\section{Post-Newtonian Theory}\label{sec:PNTheory}

The PN approximation describes the binary's orbital behavior as series
expansions that are valid in the slow-moving, weak-field regime. The expansion
parameter is the characteristic velocity of the inspiraling objects, $v$
(another common parameter is $x=v^2$). We denote an expansion term of order
$\mathcal{O}(v^n)$ by the label $\frac{n}{2}$PN (e.g. 2.5PN corresponds to $v^5$
beyond leading order). A more detailed summary of PN theory as it pertains to
point-particle systems can be found in~\cite{Blanchet:2013haa}.

We start with a quasicircular binary system of a pair of compact objects with
component masses $m_A$ and $m_B$, with total mass $M$, and spins of
dimensionless magnitude $\chi_A$ and $\chi_B$ aligned with the orbital angular
momentum. Here $\chi_A = S_A/m_A^2$, where $S_A$ is the spin angular momentum.
We define the mass ratio $q$ as the larger mass over the smaller mass,
$m_A/m_B$, so that $q\geq 1$. For convenience, we also define the mass fraction
$X_A=m_A/M$, and symmetric mass ratio $\nu=X_AX_B$.

When the objects are not simply point particles, but extended objects
like neutron
stars, each object responds to the changing tidal
fields. The leading tidal effects are the result of the deformation of
the NS due to
the tidal field generated by the other object in the binary. This
effect is characterized by the dimensionless $\ell$-polar tidal deformability
parameter, $\bar\lambda_\ell$.
Other commonly used parameters are dimensionful
tidal parameter $\lambda_\ell$ or the tidal love number $k_\ell$, and are
related to $\bar\lambda_\ell$ by the NS radius $R_A$ or compactness
$C_A=m_A/R_A$ according to
\begin{align}
\bar\lambda_{\ell A} =& \frac{2}{(2\ell-1)!!}\frac{k_{\ell A}}{C_A^{2\ell+1}}
    = \frac{\lambda_{\ell A}}{m_A^{2\ell+1}}.
\label{eq:LambdaEllBar}
\end{align}
As each object in the binary can have its own deformability, we add the
subscript $A,B$ to specify the particular object.

In the following, we will often separate the PN expressions into two parts:
``BBH terms,'' the terms that describe a BBH inspiral, and ``tidal terms,'' the
terms that describe corrections due to one or more of the objects being
something other than a BH. This will be important for tidal splicing, for which
we replace the BBH terms with numerical relativity (or a surrogate model
thereof) but we use the PN expressions for the tidal terms. The tidal
deformability $\bar\lambda_{\text{BH}}$ of black holes is generally treated as
vanishing but is somewhat difficult to define~\cite{FangLovelace:2005}; here we
will set $\bar\lambda_{\text{BH}}=0$, so all terms that depend on
$\bar\lambda_{\ell}$ are tidal terms. The BBH terms are identical to
point-particle terms up to 4PN for nonspinning BHs and 2.5PN (with the exception
of 2PN quadrupole moment terms) for spinning
BHs~\cite{Alvi:2001mx}. We include the 2PN correction for spinning BHs in
Sec.~\ref{sec:partial-2.5pn-tidal}.

\subsection{Orbital evolution}
\label{sec:orbital-evolution}

Two equations govern the evolution of the quasicircular binary system in PN
theory. The first relates the orbital phase $\phi$ to $v$ by a correspondence
with the orbital frequency, $\omega$,
\begin{align}
\omega = \frac{d\phi}{dt} = \frac{v^3}{M}.
\label{eq:PNvDef}
\end{align}

The other equation is the energy balance equation as the emission of
gravitational radiation drives the adiabatic evolution by bleeding away the
orbital energy, $E(v)$. If the energy flux is given by $F(v)$, then this energy
balance equation is
\begin{align}
\frac{dE(v)}{dt} = -F(v).
\label{eq:PNEnergyBalance}
\end{align}

The quadrupolar tidal deformability $\bar\lambda_{2}$ enters $E(v)$ and $F(v)$
first at 0PN as a $\mathcal{O}(v^{10})$ term , and through 1PN corrections at
$\mathcal{O}(v^{12})$~\cite{Flanagan2008, Vines2011}.
While normally such high-order effects would be neglected, the relatively large
size of $\bar\lambda_{2} \sim \mathcal{O}(1000)$ suggest the tidal deformations
impact the waveform earlier in the inspiral than expected by their formal PN
order.

Reference~\cite{Vines2011} provides the energy and flux expansions to 1PN,
\begin{align}
E(v) =& -\frac{\nu v^2}{2}\bigg[1 + \left(-\frac{3}{4}
    - \frac{\nu}{12}\right)v^2 + \mathcal{O}(v^3) \nonumber\\
    +& \bar\lambda_{2A}v^{10}X_A^4\bigg(9(-1 + X_A)
    + \frac{11}{2}\left(-3 + X_A \right.\nonumber\\
    -&\left. X_A^2 + 3X_A^3\right)v^2 + \mathcal{O}(v^3)
    \bigg)
    \label{eq:Energy1PN}
    + (A\rightarrow B)\bigg], \\
F(v) =& \frac{32\nu^2v^{10}}{5}\bigg[1 + \left(-\frac{1247}{336}
    - \frac{35}{12}\nu\right)v^2 + \mathcal{O}(v^3) \nonumber\\
    +& \bar\lambda_{2A}v^{10}X_A^4\bigg(6(3 - 2X_A)
    + \frac{1}{28}\left(-704 - 1803X_A \right.\nonumber\\
    +&\left. 4501X_A^2 -2170X_A^3\right)v^2
    + \mathcal{O}(v^3) \bigg)
    + (A\rightarrow B)\bigg].
\label{eq:EnergyFlux1PN}
\end{align}

\subsection{TaylorT approximants}

We can now insert the energy and flux expressions into the orbital evolution,
Eq.~(\ref{eq:PNvDef}), and energy balance, Eq.~(\ref{eq:PNEnergyBalance}), equations
and solve these equations to describe the binary's evolution. These equations
are expanded in powers of $v$, and then truncated at a particular order. There
are many choices of how to do this truncation, and these choices give rise to
different families of PN approximants. All these families agree to the same
formal PN order but have different higher-order terms in $v$. The two
approximants that we will examine here are usually referred to as TaylorT4 and
TaylorT2.

Because the tidal terms are formally proportional to at least $v^{10}$,
naively truncating an expansion at a given power of $v$ would eliminate these
terms until we reached $v^{10}$, where the point-particle terms are unknown. To
ensure that tidal terms are included, and in light of the
fact that $\bar\lambda_2$ is large, we handle the leading tidal terms
as if they were the same order as the leading PN terms:
$\mathcal{O}(1) \sim \mathcal{O}(\bar\lambda_2v^{10})$~\cite{Vines2011}. (See
Appendix~\ref{app:LambdaLambda} for further discussion regarding this
correspondence in PN orders.) In the expansions below we then keep all terms
through 1PN beyond leading-order effects.

\subsubsection{TaylorT4}
\label{sec:taylort4}

The TaylorT4~\cite{Boyle2007}
method generates the orbital evolution by rewriting the energy balance equation as
\begin{align}
  \frac{dv}{dt} = -\frac{F(v)}{M\frac{dE(v)}{dv}},
  \label{eq:EnergyBalanceT4}
\end{align}
then expanding the ratio on the right-hand side as a power series in $v$
and truncating at the appropriate order, so that
\begin{align}
\frac{dv}{dt} = \mathcal F_{\text{BBH}}(v) + \mathcal F_{\text{Tid}}(v).
\label{eq:T4TidalEvo}
\end{align}
Here, we have broken the series into two parts: the terms corresponding to a BBH
system in $\mathcal F_{\text{BBH}}$ (i.e., $\bar\lambda_2=0$) and the terms
corresponding to the tidal correction in $\mathcal F_{\text{Tid}}$. We do not
reproduce $\mathcal F_{\text{BBH}}$ here (as it is not needed for our
methods), but $\mathcal F_{\text{Tid}}$ to 1PN order is~\cite{Vines2011},
\begin{align}
\mathcal F_{\text{Tid}}(v) =& \frac{32\nu v^9}{5M}\bigg[\bar\lambda_{2A}X_A^4v^{10}
    \bigg(72 - 66X_A + \bigg(\frac{4421}{56} \nonumber\\
    -& \frac{12263 X_A}{56}
    + \frac{1893 X_A^2}{4}-\frac{661 X_A^3}{2}\bigg)v^2\bigg) \nonumber\\
    +& (A\rightarrow B) \bigg].
\label{eq:T4Tidal1PN}
\end{align}

The TaylorT4 method computes the quantity $v(t)$ by integrating
Eq.~(\ref{eq:T4TidalEvo}), and then computes the orbital phase by integrating
\begin{align}
  \frac{d\phi}{dt} = v^3/M.
  \label{eq:dphidt}
\end{align}
The two constants arising from integrating both equations correspond to the
inherent freedom to choose the initial time and phase of the waveform.

\subsubsection{TaylorT2}
\label{sec:taylort2}

The TaylorT2~\cite{Damour:2000zb} expansion begins at the same point as TaylorT4
with the PN energy equation and definition of $v$, except the equations are
rearranged to get a pair of integral expressions parametric in $v$,
\begin{align}
t(v) =& t_0 + M\int\frac{\frac{dE(v)}{dv}}{F(v)}dv,
    \\
\phi(v) =& \phi_0 + \int v^3\frac{\frac{dE(v)}{dv}}{F(v)}dv.
\end{align}
The integration constants $t_0$ and $\phi_0$ are both freely specifiable, and
can be used to set the initial time and phase of the resulting waveform.

The above integrands are expanded as a power series, truncated to the
appropriate order, and then integrated to get
series expressions for both the time and the phase, which we break into a part
corresponding to a BBH system and a part comprised of all the additional tidal
effects,
\begin{align}
t(v) =& t_0 + \mathcal T_{\text{BBH}}(v) + \mathcal T_{\text{Tid}}(v), \\
\phi(v) =& \phi_0 + \mathcal P_{\text{BBH}}(v) + \mathcal P_{\text{Tid}}(v).
\label{eq:T2TidalEvo}
\end{align}
As before, we do not reproduce expressions for
$\mathcal T_{\text{BBH}}$ or $\mathcal
P_{\text{BBH}}$, but the 1PN tidal terms are
\begin{align}
\mathcal T_{\text{Tid}}(v) =& -\frac{5M}{256\nu v^8}\bigg[
    \bar\lambda_{2A}X_A^4v^{10}\bigg(288-264 X_A \nonumber\\
    +& \left(\frac{3179}{4}-\frac{919 X_A}{4}-\frac{1143 X_A^2}{2}
    + 65 X_A^3\right)v^2\bigg) \nonumber\\
    +& (A\rightarrow B)\bigg],
    \label{eq:T2TTidal1PN} \\
\mathcal P_{\text{Tid}}(v) =& -\frac{1}{32\nu v^5}\bigg[
    \bar\lambda_{2A}X_A^4v^{10}\bigg(72-66 X_A \nonumber\\
    +& \left(\frac{15895}{56}-\frac{4595 X_A}{56}
    - \frac{5715 X_A^2}{28}+\frac{325 X_A^3}{14}\right)v^2
    \bigg) \nonumber\\
    +& (A\rightarrow B)\bigg].
\label{eq:T2PTidal1PN}
\end{align}

\subsection{BBH strain modes}
\label{subsec:BBHModes}

The gravitational radiation emission pattern for distant observers can be
represented via a decomposition into spin-weighted spherical harmonics.
Following the PN formalism used in~\cite{BFIS}, we express the strain from
compact objects inspiraling in quasicircular orbits as
\begin{align}
h^{\ell m}_{\text{BBH}}(v) = \frac{2\nu v^2M}{r}\sqrt{\frac{16\pi}{5}}
    H^{\ell m}(v)e^{-im\Psi(v)},
\label{eq:BBHModesPN}
\end{align}
where $r$ is the distance from the source to the detector.
The various terms of $H^{\ell m}(v)$ are complex series expansions of the
individual modes and are distinct from the series expansions for the energy and
flux from above~\cite{BFIS}. $\Psi(v)$ is the tail-distorted orbital phase
variable~\cite{Blanchet93,Arun:2004}
\begin{align}
  \Psi(v) = \phi(v) - 2M\omega\ln\left(\frac{\omega}{\omega_0}\right).
  \label{eq:Psiversusphi}
\end{align}
The constant $\omega_0$ is the reference frequency, often chosen to
be the frequency the waveform enters the detector's frequency band.

We will now rewrite Eq.~(\ref{eq:BBHModesPN}) in a simpler and more
  convenient form.  The first simplification arises because $\phi(v)$
  is proportional to $v^{-5}$ to leading order, and because $\omega$
  is proportional to $v^3$. This means that the correction in
  Eq.~(\ref{eq:Psiversusphi}) is of 4PN order (i.e., $v^8$ beyond
  leading order), higher order than we consider in this paper, so we
  neglect it and set $\Psi(v) = \phi(v)$.

The second simplification is to rewrite $H^{\ell m}(v)$ in
  Eq.~(\ref{eq:BBHModesPN}), which is complex, in terms of real
  quantities. We do this by treating the imaginary part of $H^{\ell
    m}(v)$ as a phase correction as is done in~\cite{Kidder2008}. Thus we
  write
\begin{align}
h^{\ell m}_{\text{BBH}}(v) = A^{\ell m}_{\text{BBH}}(v)
    e^{i\left(\psi^{\ell m}_{\text{BBH}}(v)-m\phi(v)\right)},
\label{eq:PNBBHellm}
\end{align}
where $A^{\ell m}_{\text{BBH}}(v)$ and $\psi^{\ell m}_{\text{BBH}}(v)$
  are real. This is the expression we will use for the waveform amplitudes
in subsequent sections.

We make one further simplification for the special case of the
  $(2,2)$ mode: for that mode, $\psi^{2,2}_{\text{BBH}}$ can be
  neglected. To see why, we examine the expression for
  $H^{22}(v)$~\cite{BFIS} and find that the first imaginary terms
  enter at 2.5PN order. We then can write
\begin{align}
H^{22}(v)e^{-2i\phi(v)} =& A^{22}_{\text{BBH}}(v)(1+iv^5\delta+\mathcal{O}(v^6))
    e^{-2i\phi(v)}\nonumber\\
    \approx& A^{22}_{\text{BBH}}(v)e^{iv^5\delta}e^{-2i\phi(v)},
\end{align}
where $\delta$ is the imaginary 2.5PN coefficient of $H^{22}(v)$.
Because $\phi(v)$ is proportional to $v^{-5}$ to leading order,
$\psi^{2,2}_{\text{BBH}} = v^5 \delta$ is a 5PN phase correction
(i.e., a correction to $\phi(v)$ that is $v^{10}$ beyond leading order),
so we set  $\psi^{2,2}_{\text{BBH}}=0$. Therefore
\begin{align}
h^{22}_{\text{BBH}}(v) = A^{22}_{\text{BBH}}(v)e^{-2i\phi(v)}.
\label{eq:PNBBH22}
\end{align}
We will see later that in the tidal splicing procedure that replaces BBH terms
with numerical relativity, Eq.~(\ref{eq:PNBBH22}) allows us to extract $\phi(v)$
as the phase of the (2,2) mode.
  
\subsection{Tidal correction to strain}

Reference~\cite{Baiotti2011} computed the leading-order PN tidal corrections to the
strain modes [these are explicitly written out in the form we use here in
Eqs.~(A14)-(A17) of~\cite{damour:12}]. There are no corrections
to the phase of the individual modes at leading order, [i.e., $\psi^{\ell
m}_{\text{Tid}}(v) = 0$], so the
strain modes for systems with tidally deformed objects are then
\begin{align}
h^{\ell m}_{\text{Tid}}(t) = \left(A^{\ell m}_{\text{BBH}}(v)
    + A^{\ell m}_{\text{Tid}}(v)\right)
    e^{i\left(\psi^{\ell m}_{\text{BBH}}(v)-m\phi(v)\right)}.
\end{align}
The additive corrections to the strain amplitudes are then given by
\begin{align}
A^{22}_{\text{Tid}}(v) =& \Bigg|24\sqrt{\frac{\pi}{5}}v^{12}\bar\lambda_{2A}X_A^5
    \left(3 - 5X_A + 2X_A^2\right) \nonumber\\
    \times& \left(1 + \alpha^{22}_{2A}v^2
    + \alpha^{22}_{4A}v^4\right) + (A\rightarrow B)\Bigg|, \nonumber\\
A^{21}_{\text{Tid}}(v) =& \Bigg|8\sqrt{\frac{\pi}{5}}v^{13}\bar\lambda_{2A}X_A^5
    \left(\frac{9}{2} - 15X_A + \frac{33X_A^2}{2} - 6X_A^3\right)\nonumber\\
    \times& \left(1 + \alpha^{21}_{2A}v^2\right) - (A\rightarrow B)\Bigg|, \nonumber\\
A^{33}_{\text{Tid}}(v) =& \Bigg|108\sqrt{\frac{3\pi}{14}}v^{13}\bar\lambda_{2A}X_A^5
    \left(1 - 2X_A + X_A^2\right)\nonumber\\
    \times& \left(1 + \alpha^{33}_{2A}v^2\right)
    - (A\rightarrow B)\Bigg|, \nonumber\\
A^{31}_{\text{Tid}}(v) =& \Bigg|12\sqrt{\frac{\pi}{70}}v^{13}\bar\lambda_{2A}X_A^5
    \left(1 - 2X_A + X_A^2\right)\nonumber\\
    \times& \left(1 + \alpha^{31}_{2A}v^2\right)
    - (A\rightarrow B)\Bigg|,
\label{eq:TidalStrainAmplification}
\end{align}
where $\alpha^{\ell m}_i$ are coefficients that depend on the masses.
The term $\alpha^{22}_2$ is
\begin{align}
\alpha^{22}_{2A} = \frac{-202 + 560X_A - 340X_A^2 + 45X_A^3}{42(3-2X_A)},
\end{align}
while the rest of the $\alpha^{\ell m}_i$ are currently not known. The
corrections arising from $A^{\ell m}_{\text{Tid}}$ not listed in
Eq.~(\ref{eq:TidalStrainAmplification}) enter at higher PN orders and so we
ignore them here.

For $m=$ odd modes, note that the $(A\rightarrow B)$ terms in
Eqs.~(\ref{eq:TidalStrainAmplification}) appear with an overall minus sign. This
can be understood by considering that for identical objects $A$ and $B$, the
$m=$odd modes must vanish because of symmetry under an azimuthal rotation by
$\pi$.

\subsection{Dynamical tides}
\label{sec:dynamical-tides}

The tidal corrections considered so far are based on a tidal
  deformability parameter, which describes the deformation of a
  stationary object in the presence of a stationary tidal field.  For
  dynamical objects in a binary system, this amounts to treating the
  tidal deformation as proportional to the instantaneous tidal field
  of the companion.  However, the objects also have internal
  $f$-modes described by the resonant frequencies $\omega_{f\ell}$.
  In late inspiral, as the orbital frequency approaches $\omega_{f\ell}$,
  it is no longer
  appropriate to treat tides as stationary, and dynamical tidal
  effects must be considered.
Reference~\cite{Steinhoff:2016rfi,Hinderer:2016eia} analyzed
how these dynamical tides affect the orbital motion for nonspinning systems. The
approximate solution they derive treats the
dynamical tidal deformabilities as frequency-dependent scalings of their static
values. We summarize their results here.

In effect, the dynamical tides serve primarily to amplify the static
deformability during the evolution, peaking when the orbital frequency is on
resonance with one of the object's internal $f$-modes, with frequency
$\omega_{f\ell A}$. We
shall also make use of the dimensionless $f$-mode resonance frequency,
\begin{align}
\bar \omega_{f\ell A} =& M\omega_{f\ell A}.
\label{eq:OmegaFBar}
\end{align}
Note that our choice of defining $\bar \omega_{f\ell A}$ by scaling it as the
binary's total mass $M$, rather than the object mass $m_A$ as other works often
use, is for our convenience when comparing with the dimensionless
orbital frequency, $\bar\omega=M\omega$.

In the nonspinning case, we denote the characteristic parameter governing the
resonance as
\begin{align}
\gamma_{\ell A} = \frac{\ell \omega}{\omega_{f\ell A}}
    = \frac{\ell v^3}{\bar\omega_{f\ell A}}.
\label{eq:GammaNonspinning}
\end{align}
This parameter characterizes how close the system is to resonance.

Dynamical tidal effects result in a multiplicative correction factor for the
deformability. There are two such correction factors: one that appears in the
orbital evolution equations and one that appears in the strain amplitudes.

For the orbital phase evolution,
Ref~\cite{Steinhoff:2016rfi} expressed the effective enhancement factor
$\kappa_{\ell A}(v)$ on the static deformability of object $A$ by
\begin{align}
\kappa_{\ell A}(v) =& a_\ell + b_\ell\left[\frac{1}{1-\gamma_{\ell A}^2}
    + \frac{4}{3\sqrt{\epsilon_\ell}\hat t_\ell\gamma_{\ell A}^2}
    + \sqrt{\frac{\pi}{3\epsilon_\ell}}\frac{\mathcal Q_\ell}{\gamma_{\ell A}^2}
    \right],
\label{eq:DynamicalTides}
\end{align}
where
\begin{align}
\epsilon_\ell =& \frac{256\nu\omega_{f\ell A}^{5/3}}{5\ell^{5/3}}, \nonumber\\
\hat t_{\ell} =& \frac{8}{5\sqrt{\epsilon_\ell}}\left(1
    - \gamma_{\ell A}^{-5/3}\right), \nonumber\\
\mathcal Q_\ell =& \cos\left(\frac{3\hat t_\ell}{8}\right)\left[1
    + 2 \text F_\text S\left(\frac{\sqrt3}{2\sqrt\pi}\hat t_\ell\right)\right]
    \nonumber\\
    -& \sin\left(\frac{3\hat t_\ell}{8}\right)\left[1
    + 2 \text F_\text C\left(\frac{\sqrt3}{2\sqrt\pi}\hat t_\ell\right)\right],
\end{align}
with $\text F_\text S$ and $\text F_\text C$ as the Fresnel sine and cosine
integrals. The coefficients $(a_\ell, b_\ell)$ are given by $(a_2, b_2) = (1/4,
3/4)$ and $(a_3, b_3) = (3/8, 5/8)$. The PN orbital evolution equations,
Eqs.~(\ref{eq:T4Tidal1PN}), (\ref{eq:T2TTidal1PN}), and
(\ref{eq:T2PTidal1PN}),
can be modified to incorporate the dynamical
tides by taking $\bar\lambda_{\ell A} \rightarrow \bar\lambda_{\ell
A}\kappa_{\ell A}(v)$.

The correction to the strain amplitudes is computed in
Ref.~\cite{Dietrich:2017dh}:
\begin{align}
\hat\kappa_{2A}(v) = \frac{(\kappa_{2A}(v)-1)\omega_{f2A}^2
    + 6(1-X_A)\kappa_{2A}\omega^2}{(9-6X_A)\omega^2}.
\label{eq:DynamicalTidesStrain}
\end{align}
As in the case of the deformability amplification for the orbital evolution, the
deformability amplification for the dynamical strain can be incorporated into
Eq.~(\ref{eq:TidalStrainAmplification}) by the substitution $\bar\lambda_{2A}
\rightarrow \bar\lambda_{2A}\hat\kappa_{2A}(v)$.

Note that these dynamical tide corrections do not produce proper power series
expansions of $v$, because the Fresnel integrals
do not have a well-defined power series expansion
about $v=0$ (even though those terms vanish as $v\rightarrow0$, they also
oscillate infinitely fast). So including $\kappa_{\ell A}(v)$ and
$\hat\kappa_{2A}(v)$ into the evolution and strain formula means the tidal effects
cannot be represented to a formal PN order. While this is not a problem
for the actual generation of the waveforms, it does mean that the differences
between the different families of PN approximants no
longer diverge at a well-defined PN order.

\section{Additional tidal corrections}\label{sec:ExtendingPN}

\subsection{Partial 2.5PN Tidal Terms}
\label{sec:partial-2.5pn-tidal}

The static tidal corrections to the orbital evolution discussed so far have been
those derived from Eqs.~(\ref{eq:Energy1PN}) and~(\ref{eq:EnergyFlux1PN}), which
are 1PN order expressions computed by Ref.~\cite{Vines2011}. Since then,
higher-order corrections have been computed, but only within the EOB formalism
and not for standard PN approximants. These higher-order terms include not only
nonspinning
$\bar\lambda_2$ tidal corrections to the energy (through the EOB
Hamiltonian) up through 2.5PN order, but also corrections according to the static
octopolar deformability $\bar\lambda_3$~\cite{Bini:2012gu} through 2.5PN order.
These 2.5PN tidal effects are already included both for
SEOBNRv4T~\cite{Hinderer:2016eia, Steinhoff:2016rfi}, and for the frequency
domain approximant TaylorF2~\cite{damour:12}.
We use the EOB results to obtain the time domain Taylor approximants here.

The details of how to convert the $\bar\lambda_2$ and $\bar\lambda_3$ 2.5PN
tidal terms from the EOB Hamiltonian to corrections to Eq.~(\ref{eq:Energy1PN})
are given in Appendix~\ref{app:EOB2PN}. That computation yields
\begin{align}
E_{\bar\lambda_2}(v) =& -\frac{\nu v^2}{2}\bigg[
    \bar\lambda_{2A}v^{10}X_A^4\bigg(9(-1 + X_A) \nonumber\\
    +& \frac{11}{2}\left(-3 + X_A - X_A^2 + 3X_A^3\right)v^2
    + \frac{13}{2}\bigg(-\frac{51}{4} \nonumber\\
    -& \frac{15}{4}X_A + \frac{361}{42}X_A^2
    + \frac{47}{21}X_A^3 + \frac{47}{12}X_A^4 + \frac{7}{4}X_A^5\bigg)v^4
    \nonumber\\
    +& \mathcal O(v^6)\bigg) + (A\rightarrow B)\bigg], \\
E_{\bar\lambda_3}(v) =& -\frac{\nu v^2}{2}\bigg[
    \bar\lambda_{3A}v^{14}X_A^6\bigg((-65+65X_A) \nonumber\\
    +& \left(\frac{75}{2} - \frac{875}{2}X_A + \frac{475}{2}X_A^2
    + \frac{325}{2}X_A^3\right)v^2 \nonumber\\
    +& \bigg(-\frac{6205}{24}
    - \frac{4165}{8}X_A + \frac{425}{36}X_A^2
    - \frac{1955}{4}X_A^3  \nonumber\\
    +& \frac{26095}{24}X_A^4 + \frac{12155}{72}X_A^5\bigg)v^4
    + \mathcal O(v^6)\bigg) \nonumber\\
    +& (A\rightarrow B)\bigg].
\label{eq:PNTidalEnergy}
\end{align}
From these expressions
we can see that the leading-order octopolar deformability terms enter
at $\mathcal{O}(v^{14})$.
Numerically, $\bar\lambda_3$ is typically larger than $\bar\lambda_2$, with
$\bar\lambda_3\sim\mathcal{O}(1000-10000)$. Thus, as we did for $\bar\lambda_2$,
we treat the leading-order $\bar\lambda_3$ terms
as being the same formal order as the
leading-order PN terms, i.e., $\mathcal{O}(1) \sim
\mathcal{O}(\bar\lambda_2v^{10}) \sim \mathcal{O}(\bar\lambda_3v^{14})$. Terms
for both deformabilities are computed up to 2.5PN order.

Unfortunately, unlike the energy, the fluxes are not known to 2.5PN order.
We shall introduce undefined coefficients for the missing terms, $\alpha_4,
\beta_0, \beta_2, \beta_4$ so that we may expand the PN expressions to a
consistent order. We write the flux terms as
\begin{align}
F_{\bar\lambda_2}(v) =& \frac{32\nu^2v^{10}}{5}\bigg[
    \bar\lambda_{2A}v^{10}X_A^4\bigg(6(3 - 2X_A) \nonumber\\
    +& \frac{1}{28}\left(-704 - 1803X_A + 4501X_A^2 -2170X_A^3\right)v^2
    \nonumber\\
    +& \alpha_{4}v^4 + \mathcal O(v^6)\bigg)
    + (A\rightarrow B)\bigg], \\
F_{\bar\lambda_3}(v) =& \frac{32\nu^2v^{10}}{5}\bigg[
    \bar\lambda_{3A}v^{14}X_A^6\bigg(\beta_{0}
    + \beta_{2}v^2 + \beta_{4}v^4 \nonumber\\
    +& \mathcal O(v^6)\bigg) + (A\rightarrow B)\bigg].
\label{eq:PNTidalFlux}
\end{align}

We keep track of
the undefined coefficients $\alpha_4, \beta_0, \beta_2,
\beta_4$ for the purposes of completeness;
when the value of those coefficients are computed in some future work, those
values can simply be substituted into the flux expression and the Taylor expansions
of the orbital evolution below. For our results in Sec.~\ref{sec:Results}, we
set these coefficients to zero.

For aligned spin systems, there will also be terms corresponding to
spin-tidal terms, interactions in the Hamiltonian between object spins and the
tidal deformations of the NS. The spin-tidal connection terms are not included
in our expansions because of discrepancies in the
literature regarding the calculation of the leading-order 1.5PN coefficients; a
discussion of those coefficients, including how to
add them when those discrepancies are resolved, can be found in
Appendix~\ref{app:SpinTidal}.

Because we have added tidal terms up to 2.5PN order, for a consistent expansion
we also introduce nontidal 2.5PN spin-orbit, spin-spin, and rotationally
induced quadrupolar moment effects~\cite{Jaranowski99a, Andrade01,
Blanchet:2000ub, Damour01, Blanchet02a, Blanchet04a,
Blanchet-Buonanno-Faye:2006, Kidder:1995zr, Will96, Poisson:1997ha} to both the
orbital energy and the flux expressions. The nontidal parts of these
expressions are:

\begin{widetext}
\begin{align}
\label{eq:PNEnergyBBH}
E_{\text{BBH}}(v) =& -\frac{\nu v^2}{2}\bigg[1 + \left(-\frac{3}{4}
    - \frac{\nu}{12}\right)v^2
    + \left(2\chi_AX_A\left(1+\frac{X_A}{3}\right)
    - 2\chi_BX_B\left(1+\frac{X_B}{3}\right)\right)v^3 \nonumber\\
    +& \left(\frac{1}{8}\left(-27+19\nu-\frac{\nu^2}{3}\right)
    + 2\chi_A\chi_B\nu - (\bar Q_A+1)\chi_A^2X_A^2 - (\bar Q_B+1)\chi_B^2X_B^2
    \right)v^4 \nonumber\\
    +& \left(\chi_AX_A\left(3+\frac{5X_A}{3}
    + \frac{29X_A^2}{9}+\frac{X_A^3}{9}\right)
    + \chi_BX_B\left(3+\frac{5X_B}{3}+\frac{29X_B^2}{9}+\frac{X_B^3}{9}\right)
    \right)v^5 + \mathcal O(v^6) \bigg],\\
F_{\text{BBH}}(v) =& \frac{32\nu^2v^{10}}{5}\bigg[1 + \left(-\frac{1247}{336}
    - \frac{35}{12}\nu\right)v^2
    + \left(4\pi+ \chi_AX_A\left(-\frac{5}{4} - \frac{3X_A}{2}\right)
    + \chi_BX_B\left(-\frac{5}{4} - \frac{3X_B}{2}\right)\right)v^3 \nonumber\\
    +& \left(\left(-\frac{44711}{9072} + \frac{9271\nu}{504}
    + \frac{65\nu^2}{18}\right) + \frac{31}{8}\chi_A\chi_B\nu
    + \left(\frac{33}{16} + 2\bar Q_A\right)\chi_A^2X_A^2
    + \left(\frac{33}{16} + 2\bar Q_B\right)\chi_B^2X_B^2\right)v^4 \nonumber\\
    +& \left(\left(\frac{-8191}{672}-\frac{583\nu}{25}\right)\pi
    + \chi_AX_A\left(-\frac{13}{16} + \frac{63X_A}{8}
    - \frac{73X_A^2}{36} - \frac{157X_A^3}{18}\right) \right.\nonumber\\
    +& \left.\chi_BX_B\left(-\frac{13}{16} + \frac{63X_B}{8}
    - \frac{73X_B^2}{36} - \frac{157X_B^3}{18}\right)
    \right)v^5 + \mathcal O(v^6)\bigg].
\label{eq:PNFluxBBH}
\end{align}
\end{widetext}

In the above expressions, $\bar Q_A$ is the dimensionless quadrupole moment,
which is unity for black holes, $\bar Q_{\text{BH}}=1$, and is related to the
dimensionful quadrupole moment $Q_A$ by
\begin{align}
\bar Q_A =& -\frac{Q_A}{m_A^3\chi_A^2}.
\label{eq:Qbar}
\end{align}
  
Therefore the full 2.5PN energy and flux expressions (replacing the 1PN versions
given by Eqs.~(\ref{eq:Energy1PN}) and~(\ref{eq:EnergyFlux1PN})) are simply
\begin{align}
E(v) =& E_{\text{BBH}}(v) + E_{\bar\lambda_2}(v) + E_{\bar\lambda_3}(v), \\
F(v) =& F_{\text{BBH}}(v) + F_{\bar\lambda_2}(v) + F_{\bar\lambda_3}(v).
\end{align}

At this point, we can repeat the expansion procedure from
Secs.~\ref{sec:taylort4} and~\ref{sec:taylort2} to generate the TaylorT4
and TaylorT2 approximants. Again, we treat terms such as $\mathcal{O}(1) \sim
\mathcal{O}(\bar\lambda_2v^{10}) \sim \mathcal{O}(\bar\lambda_3v^{14})$ as
leading order and expand through 2.5PN order.

\subsubsection{TaylorT4}
\label{sec:taylort4-1}

With the 2.5PN expressions for the energy and flux, we can now recompute the
TaylorT4 approximant from Eq.~(\ref{eq:T4TidalEvo}) up to 2.5PN order. As before,
the terms in $dv/dt$ corresponding to a BBH system ($\bar\lambda_2 =
\bar\lambda_3 = 0, \bar Q=1$) are denoted $\mathcal F_{\text{BBH}}(v)$ (which we
do not reproduce here), and the terms describing tidal corrections are denoted
$\mathcal F_{\text{Tid}}(v)$. Here
\begin{align}
\mathcal F_{\text{Tid}}(v) =& \frac{32\nu v^9}{5M}\Bigg[
    \left(5(\bar Q_A-1)\chi_A^2X_A^2\right)v^4 \nonumber\\
    +& \bar\lambda_{2A}X_A^4v^{10}\left(\sum^5_{i=0}\mathcal F_{2A,i}v^i\right)
    \nonumber\\
    +& \bar\lambda_{3A}X_A^6v^{14}\left(\sum^5_{i=0}\mathcal F_{3A,i}v^i\right)
    + \left(A \leftrightarrow B\right)\Bigg].
\label{eq:T4TidalPartial}
\end{align}
The coefficients $\mathcal F_{2A,i}$ and $\mathcal F_{3A,i}$ are given in
Appendix~\ref{sec:2.5pn-tidal-expr}, Eq.~(\ref{eq:T4TidalFull}).

The $\bar\lambda_2\times\chi_{A,B}$ and $\bar\lambda_3\times\chi_{A,B}$ cross
terms appearing in Eqs.~(\ref{eq:T4TidalPartial}) and (\ref{eq:T4TidalFull}) are
not due to spin-tidal interaction terms in the Hamiltonian, but
instead are a consequence of the series expansion power counting.
To properly account for how $\bar Q_A$ appears in these equations, recall that
the part of the quadrupole moment $v^4$ term corresponding to the BH ($\bar
Q_{BH}=1$) is already included in $\mathcal F_{\text{BBH}}(v)$. Then the
only part of $\bar Q_A$ that will appear as a $v^4$ tidal term in the expansion
is the part not already accounted for by $\mathcal F_{\text{BBH}}(v)$, i.e., $\bar
Q_{\text{Tid}}=\bar Q_A-\bar Q_{\text{BH}}=\bar Q_A-1$. However, $\mathcal
F_{\text{BBH}}(v)$ does not have any terms corresponding to the
$\bar\lambda_{2,3}\times\bar Q_A$ cross terms that appear in
Eq.~(\ref{eq:T4TidalFull}), so the entire quadrupole moment must
be included in those terms, i.e., $\bar Q_A$, not $(\bar Q_A-1)$.

\subsubsection{TaylorT2}
\label{sec:taylort2-1}

Similarly for TaylorT2, the updated time and phase expressions corresponding to
Eqs.~(\ref{eq:T2TTidal1PN}) and (\ref{eq:T2PTidal1PN}) are
\begin{align}
\mathcal T_{\text{Tid}}(v) =& -\frac{5M}{256\nu v^8}\Bigg[
    -\left(10(\bar Q_A-1)\chi_A^2X_A^2\right)v^4 \nonumber\\
    +& \bar\lambda_{2A}X_A^4v^{10}\left(\sum^5_{i=0}\mathcal T_{2A,i}v^i\right)
    \nonumber\\
    +& \bar\lambda_{3A}X_A^6v^{14}\left(\sum^5_{i=0}\mathcal T_{3A,i}v^i\right)
    + \left(A \leftrightarrow B\right)\Bigg],
\label{eq:T2TTidalPartial}\\
\mathcal P_{\text{Tid}}(v) =& -\frac{1}{32\nu v^5}\Bigg[
    -\left(25(\bar Q_A-1)\chi_A^2X_A^2\right)v^4 \nonumber\\
    +& \bar\lambda_{2A}X_A^4v^{10}\left(\sum^5_{i=0}\mathcal P_{2A,i}v^i\right)
    \nonumber\\
    +& \bar\lambda_{3A}X_A^6v^{14}\left(\sum^5_{i=0}\mathcal P_{3A,i}v^i\right)
    + \left(A \leftrightarrow B\right)\Bigg].
\label{eq:T2PTidalPartial}
\end{align}
The individual coefficients $\mathcal T_{2A,i}, \mathcal T_{3A,i}, \mathcal
P_{2A,i}, \mathcal P_{3A,i}$ are given in Appendix~\ref{sec:2.5pn-tidal-expr},
Eqs.~(\ref{eq:T2TTidalFull}) and (\ref{eq:T2PTidalFull}).
The discussion regarding the $\bar\lambda_{2,3}\times\chi_{A,B}$ and
$\bar Q_A$ terms given above for TaylorT4 hold true here as well.

\subsection{Spinning dynamical tides}

\begin{figure*}
\includegraphics[width=.85\textwidth]{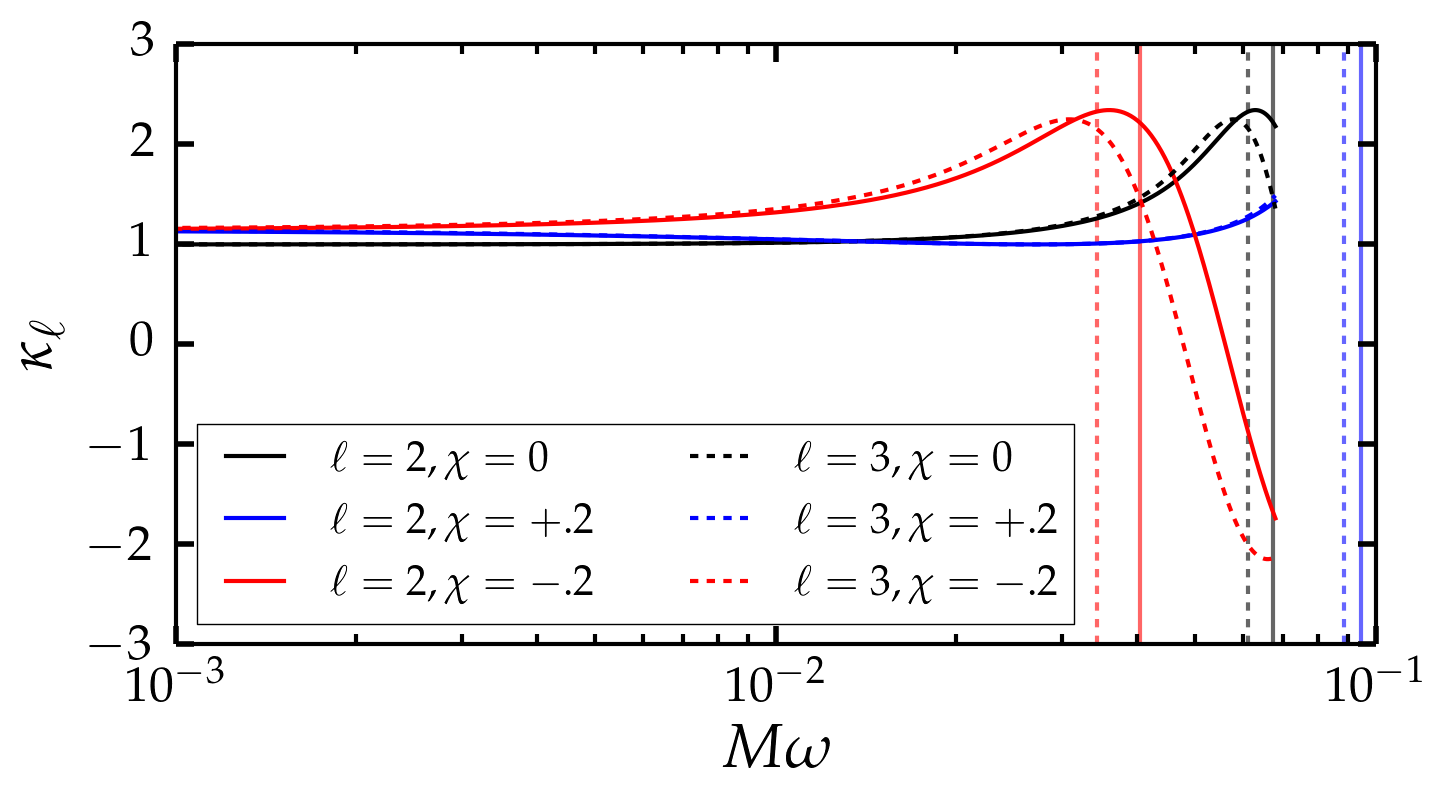}
\caption{
The dynamical tide amplification $\kappa_\ell$ from Eq.~(\ref{eq:DynamicalTides})
modified by Eq.~(\ref{eq:GammaSpinning}), as a function of orbital
frequency, for the nonspinning case (black) and the spin aligned/antialigned
(blue/red) cases for both $\ell=2$ (solid curves) and $\ell=3$ (dashed curves) tidal
deformabilities. The vertical lines represent the resonance frequencies of both
modes in every case. The parameters correspond to an NS with $\bar\lambda_2
\approx 800$ in an equal mass system with total mass $2.8M_{\odot}$. For this
NS, a spin magnitude of $|\chi_{\text{NS}}| = 0.2$ corresponds to a rotational
frequency of $f_{\text{NS}}\approx312$Hz.
\label{fig:DynTidesResPeak}
}
\end{figure*}

The dynamical tidal corrections discussed earlier in
Sec.~\ref{sec:dynamical-tides} do not take into account the spin of the NS.
The dynamical tides are caused by the
changing tidal field due to the orbital motion interacting with the internal
$f$-modes of the deformable object. So when the object is also spinning, its
internal modes will effectively experience a driving frequency equal to the
orbital frequency shifted by the object's rotational frequency.
We characterize this frequency shift by making a
slight correction to the characteristic parameter $\gamma_{\ell A}$ in
Eq.~(\ref{eq:GammaNonspinning}).

Given an aligned or antialigned
spin of $\chi_A$ and a moment of inertia $I_A$, then we can
compute the rotation frequency of the deformable object as
\begin{align}
\bar\omega_A = M\omega_A = \frac{\chi_A}{X_A\bar I_A},
\end{align}
where the dimensionless moment of inertia is
\begin{align}
\bar I_A =& \frac{I_A}{m_A^3}.
\label{eq:Ibar}
\end{align}

Thus, the effective orbital frequency the resonant modes will experience is
simply the difference between the orbital frequency and the rotational frequency
of the object. With that in mind, we rewrite $\gamma_{\ell A}$ as
\begin{align}
\gamma_{\ell A} = \left|\frac{\ell\left(v^3
    - \bar\omega_A\right)}{\bar\omega_{f\ell A}}\right|.
\label{eq:GammaSpinning}
\end{align}
We take the absolute value because Eq.~(\ref{eq:GammaNonspinning}) is undefined for
negative values of $\gamma_{\ell A}$, which can occur at low orbital frequencies
with aligned spin objects. This corresponds to saying that the resonance modes
of the object only care about the magnitude of the frequency of the changing
tidal field. Within Eq.~(\ref{eq:DynamicalTidesStrain}), we also need to
make the change $\omega\rightarrow\left|v^3 - \bar\omega_A\right|/M$.

To show how spin affects the profile of the dynamical tide correction, in
Fig~\ref{fig:DynTidesResPeak} we plot the profile of $\kappa_\ell$ from
Eq.~(\ref{eq:DynamicalTides}) as a function of orbital frequency. We assume an NS
with $\bar\lambda_2 \approx 800$ in an equal mass system and use the universal
relations (see Table~\ref{tab:UniversalRelations}) to compute the
other relevant tidal parameters. We compare the nonspinning NS against both
aligned and antialigned spinning NS with magnitude $|\chi_{\text{NS}}| = 0.2$,
which corresponds to a rotational frequency of $f_{\text{NS}}\approx312$Hz. The
upper frequency termination point is at an orbital frequency of
$M\omega_{\text{ISCO}}=6^{-3/2}$, which has been used as an approximate BNS
inspiral termination criterion~\cite{Bernuzzi:2014kca}.

From Fig~\ref{fig:DynTidesResPeak} we can see that aligned spins push the
resonance peak later in the inspiral while antialigned spins move the peak to
smaller frequencies. In fact, with large enough aligned spins it is possible
that the peak of the resonance is never reached before the system enters the
merger/ringdown phase.

At low frequencies, the nonspinning system behaves
the same as for static tides ($\kappa_\ell=1$). However, this is not true
for the spinning systems, both of which asymptote to a slightly different value.
Physically, these differences are due to the deformations of the object
experiencing a driving frequency not from the orbital frequency (which is
vanishingly small), but from its own rotation along its axis. We expect this
difference to result in a negligible contribution to the waveform as the relative
size of the tidal effects already vanishes [$\mathcal{O}(v^{10})$] at low
frequencies. For the aligned spin object, note that there is a point in the
evolution where the rotational and orbital frequencies match and the effective
dynamical tidal field vanishes.

One concern is that for antialigned spins, $\kappa_\ell$ becomes negative before
$M\omega_{\text{ISCO}}$. We recognize that Eq.~(\ref{eq:DynamicalTides}) is
derived assuming that the driving frequency is not much larger than
  the resonance peak~\cite{Steinhoff:2016rfi}.
While for nonspinning systems and spin aligned systems the resonance peak occurs
near the end of the inspiral or after merger/ringdown and is thus within the range
of validity, that condition does not necessarily apply in the antialigned case.
If the antialigned spin is large enough, as seems to be the case in
Fig~\ref{fig:DynTidesResPeak}, the resonance frequency occurs early enough in
the evolution that this approximate formalism potentially breaks down while
still in the inspiral, necessitating a more delicate handling of the dynamical
tides.

Until such a formalism is developed for antialigned NS spins, we instead assume
for antialigned spin
that the object is nonspinning (i.e., we set $\omega_A=0$) for the
purposes of Eqs.~(\ref{eq:DynamicalTides}) and~(\ref{eq:DynamicalTidesStrain}); the
aligned spin case will use Eq.~(\ref{eq:GammaSpinning}) as expected.

\section{Tidal Splicing}\label{sec:TidalSplicing}

Putting together the results of the previous sections, we can write the PN
equations of motion in the form
\begin{align}
\frac{d\phi}{dt} =& \frac{v^3}{M}, 
\label{eq:T4dphidt}\\
\frac{dv}{dt} =& \mathcal{F}_{\text{BBH}}(v)
    + \mathcal{F}_{\text{Tid}}(v),
\label{eq:T4dvdt}\\
h^{\ell m}(t) =& \left(A^{\ell m}_{\text{BBH}}(v)+A^{\ell m}_{\text{Tid}}(v)\right)
    e^{i(\psi^{\ell m}_{\text{BBH}}(v) - m\phi)},
\label{eq:T4EoM}
\end{align}
where we have expressed the equations within the TaylorT4 framework. The
expression for $\mathcal{F}_{\text{Tid}}(v)$ is given by
Eq.~(\ref{eq:T4TidalPartial}), while $A^{\ell
m}_{\text{BBH}}(v)$ and $\psi^{\ell m}_{\text{BBH}}(v)$ can be extracted from
the expansions of Eq.~(\ref{eq:BBHModesPN}) given in~\cite{BFIS} using the
procedure we describe in Sec~\ref{subsec:BBHModes}. However,
$\mathcal{F}_{\text{BBH}}(v)$, $A^{\ell m}_{\text{BBH}}(v)$ and $\psi^{\ell
m}_{\text{BBH}}(v)$ are all unimportant for our purposes and so we do not
present those formulas here. 

Equivalently, the PN equations of motion
can be written within the TaylorT2 framework as
\begin{align}
t(v) =&t_0 + \mathcal{T}_{\text{BBH}}(v)
+ \mathcal{T}_{\text{Tid}}(v),
\label{eq:T2time} \\
\phi(v) =& \phi_0 + \mathcal{P}_{\text{BBH}}(v)
+ \mathcal{P}_{\text{Tid}}(v),
\label{eq:T2phi} \\
h^{\ell m}(t) =& \left(A^{\ell m}_{\text{BBH}}(v)+A^{\ell m}_{\text{Tid}}(v)\right)
    e^{i(\psi^{\ell m}_{\text{BBH}}(v) - m\phi)}.
\label{eq:T2EoM}
\end{align}
The expression for $\mathcal{T}_{\text{Tid}}(v)$ and $\mathcal{P}_{\text{Tid}}(v)$ are given by
Eqs.~(\ref{eq:T2TTidalPartial}) and (\ref{eq:T2PTidalPartial}).

Consider the case of a BBH system. In
principle, if we knew the PN expansions
for the expressions of $\mathcal{F}_{\text{BBH}}$ (or $\mathcal{P}_{\text{BBH}}$
and $\mathcal{T}_{\text{BBH}}$), $A^{\ell m}_{\text{BBH}}$ and $\psi^{\ell
m}_{\text{BBH}}$ up through arbitrarily large order, then we could perfectly
reproduce the gravitational waveforms of those inspiraling systems.
But unfortunately these terms are known only to a limited order.

However, numerical simulations of BBH systems are able to accurately solve the
full Einstein equations. Thus, if we can represent these numerical waveforms in
a form akin to the systems of equations in either
Eqs.~(\ref{eq:T4dphidt})-(\ref{eq:T4EoM}) or
Eqs.~(\ref{eq:T2phi}) and (\ref{eq:T2EoM})
(with vanishing tidal terms), they would provide perfect
representations of the PN expressions up to the numerical resolution error.
This is the main idea of tidal splicing:  We use numerical relativity BBH
  waveforms to effectively obtain the functions
  $\mathcal{F}_{\text{BBH}}(v)$,$\mathcal{P}_{\text{BBH}}(v)$,
  $\mathcal{T}_{\text{BBH}}(v)$, $A^{\ell m}_{\text{BBH}}(v)$, and $\psi^{\ell
  m}_{\text{BBH}}(v)$ to all orders in $v$. Then we add the analytic expressions
for the tidal terms described above, and we integrate the PN equations of motion
to generate waveforms corresponding to the inspirals of
BHNS and BNS systems.

\subsection{Decomposition of NR waveforms}

We start with a numerical BBH waveform corresponding to a system with a
particular mass ratio $q$ and spins $\chi_A$ and $\chi_B$. The spins are assumed
here to be aligned or antialigned with the orbital angular momentum.
We decompose the $(2,2)$ mode of the BBH waveform in the form
\begin{align}
h^{22}_{\text{NR}}(t) = A^{22}_{\text{NR}}(t)
    e^{-2i\phi_{\text{NR}}(t)},
\end{align}
where $A^{22}_{\text{NR}}(t)$ is real. This is the same decomposition as
Eq.~(\ref{eq:PNBBH22}), so we interpret $\phi_{\text{NR}}(t)$ as the NR orbital
phase. We then compute the effective PN expansion parameter $v_{\text{NR}}$
using Eq.~(\ref{eq:dphidt}):
\begin{align}
v_{\text{NR}}(t) = \sqrt[3]{M\frac{d\phi_{\text{NR}}(t)}{dt}}.
\end{align}
Because the numerical waveform is known at a finite set of time samples, we use
a sixth order finite difference scheme to compute the $d/dt$ derivatives
numerically.

With the orbital phase and effective PN parameters in hand, we decompose each
mode from the waveform into an amplitude and phase,
\begin{align}
h^{\ell m}_{\text{NR}}(t) = A^{\ell m}_{\text{NR}}(t)
    e^{i\Phi^{\ell m}_{\text{NR}}(t)}.
\end{align}
Comparing with the expression for strain from Eq.~(\ref{eq:PNBBHellm}), we break up
the phase as
\begin{align}
\Phi^{\ell m}_{\text{NR}}(t) =
    \psi^{\ell m}_{\text{NR}}(t) - m\phi_{\text{NR}}(t).
\label{eq:NRPhi}
\end{align}
Since we know $\phi_{\text{NR}}$ we can compute $\psi^{\ell m}_{\text{NR}}$ from
the total phase by rearranging Eq.~(\ref{eq:NRPhi}),
\begin{align}
\psi^{\ell m}_{\text{NR}}(t) = 
    \Phi^{\ell m}_{\text{NR}}(t) + m\phi_{\text{NR}}(t).
\end{align}

Up to this point, we have been treating $t_{\text{NR}}$ as the
independent variable for the purposes of decomposition. Since the PN formalism
considers the frequency expansion parameter $v$ as the independent variable, we
invert $v_{\text{NR}}(t)$ to get $t_{\text{NR}}(v)$ as a function of $v$. This
inversion is straightforward numerically, since we have a finite number of time
samples $t_{\text{NR}}$. Thus we can represent all of the individual parts of
our waveform as functions of $v$, e.g.,
$\phi_{\text{NR}}(t_{\text{NR}}(v))=\phi_{\text{NR}}(v)$. Then we can write the
strain for each mode in the form corresponding to Eq.~(\ref{eq:PNBBHellm}):
\begin{align}
h^{\ell m}_{\text{NR}}(v) =& A^{\ell m}_{\text{NR}}(v)
e^{i(\psi^{\ell m}_{\text{NR}}(v)-m\phi_{\text{NR}}(v))}.
\label{eq:NrWaveformDecomposition}
\end{align}
We now have numerical equivalents for the various PN expansions for BBH systems;
these are correct up to an arbitrary PN order and limited only by the errors
from the simulations themselves.

\subsection{Tidal parameters}

With the numerical decomposition in hand, we will need
the tidal parameters for the particular BHNS or BNS system under consideration.
A review of the different tidal effects explored in the previous section show
there are six different parameters that
characterize the tidal behavior of each
object: the dimensionless quadrupolar and octopolar tidal deformabilities
$\bar\lambda_{2}$ and $\bar\lambda_{3}$,
their corresponding $f$-mode resonance frequencies
 $\omega_{f2}$ and $\omega_{f3}$, the dimensionless rotationally induced
quadrupole moment $\bar Q$, and the dimensionless moment of inertia $\bar I$.
For our model, we choose only $\bar\lambda_2$, and we compute the values of the
other parameters from $\bar\lambda_2$ using the universal relations, which are
approximate relations between $\bar\lambda_2$ and the other tidal parameters;
the details are given in Appendix~\ref{app:ModelParameters}. The choice of
$\bar\lambda_2$ depends on the physical properties of the deformable object in
consideration. Once we have chosen $\bar\lambda_2$ (and thus the other
parameters via the universal relations), the next step in tidal splicing is the
recomputation of the orbital evolution.

\subsection{Splicing of the orbital evolution equations}

The details of how to specifically splice the PN tidal information into the
orbital evolution depends on the specific Taylor expansion considered. We shall
discuss the details of tidal splicing with TaylorT4 and TaylorT2.
This method can also be performed for TaylorT3~\cite{Damour:2000zb}
which involves expanding about an intermediate dimensionless time variable.
However, TaylorT3 is known to do a poor job in general of reproducing the
results of BBH numerical simulations even in the equal mass, nonspinning case
\cite{Boyle2007, Buonanno:2009}, so we ignore that method here. We
also ignore TaylorT1~\cite{Damour:2000zb} because we do not know how to compute
the BBH contribution to the PN energy and flux separately using only the BBH
waveform. While Ref.~\cite{Barkett2015} discussed tidal splicing under a
TaylorF2~\cite{DIS00} framework, in this paper we only examine time
domain approximants.

\subsubsection{TaylorT4 splicing}

Splicing with TaylorT4, originally introduced within ~\cite{Barkett2015}, begins
by examining how the tidal terms manifest in the TaylorT4 framework. The
evolution of the PN parameter as seen in
Eq.~(\ref{eq:T4dvdt}) for a BBH system is
\begin{align}
\frac{dv}{dt} =& \mathcal F_{\text{BBH}}(v).
\label{eq:TaylorT4Basic}
\end{align}
With $t_{\text{NR}}(v)$ in hand from the simulation, we can compute a
numerically accurate version of $\mathcal F_{\text{BBH}}(v)$ which we shall call
$\mathcal F_{\text{NR}}(v)$,
\begin{align}
\mathcal F_{\text{NR}}(v) = \left(\frac{dt_{\text{NR}}(v)}{dv}\right)^{-1}
\end{align}

The tidal terms, $\mathcal F_{\text{Tid}}(v_{\text{NR}})$, represent the sum of
the additional tidal effects in the evolution and we compute them according
to Eq.~(\ref{eq:T4TidalFull}). We incorporate the dynamical tides by scaling the
deformabilities according to Eq.~(\ref{eq:DynamicalTides}), i.e.,
$\bar\lambda_{\ell} \rightarrow \bar\lambda_{\ell}\kappa_{\ell}(v)$.

We introduce a new spliced time coordinate, $t_{\text{Spl}}(v)$, which we
compute by integrating the differential equation
\begin{align}
\frac{dt_{\text{Spl}}}{dv} =&
    \frac{1}{\mathcal F_{\text{NR}}(v)
    + \mathcal F_{\text{Tid}}(v)},
\label{eq:TaylorT4Splicing}
\end{align}
which is the inverse of Eq.~(\ref{eq:T4dvdt}).
Once we have $t_{\text{Spl}}$, we find the orbital phase of this new waveform
by integrating Eq.~(\ref{eq:T4dphidt}):
\begin{align}
  \phi_{\text{Spl}}(v) =& \frac{1}{M}\int v(t_{\text{Spl}})^3dt_{\text{Spl}}.
  \label{eq:TaylorT4SplicingPhi}
\end{align}
We use Simpson's method for integration. We choose the integration constants in
Eqs.~(\ref{eq:TaylorT4Splicing}) and~(\ref{eq:TaylorT4SplicingPhi}) to align the
waveform to the numerical waveform at the initial time.

\subsubsection{TaylorT2 splicing}

For BBH systems Eqs.~(\ref{eq:T2time}) and (\ref{eq:T2phi}) take the form
\begin{align}
t(v) =& t_0 + \mathcal T_{\text{BBH}}(v), \nonumber\\
\phi(v) =& \phi_0 + \mathcal P_{\text{BBH}}(v).
\label{eq:TaylorT2BBH}
\end{align}
The constants $t_0$ and $\phi_0$ correspond simply to the starting time and
phase of the waveform.
Our numerically corrected versions of $\mathcal{T}_{\text{BBH}}$ and
$\mathcal{P}_{\text{BBH}}$ are simply
\begin{align}
\mathcal T_{\text{NR}}(v) =& t_{\text{NR}}(v), \nonumber\\
\mathcal P_{\text{NR}}(v) =& \phi_{\text{NR}}(v).
\end{align}

We compute $\mathcal{T}_{\text{Tid}}(v)$ according to Eq.~(\ref{eq:T2TTidalFull})
and $\mathcal{P}_{\text{Tid}}(v)$ according to Eq.~(\ref{eq:T2PTidalFull}),
incorporating the dynamical tides by making the frequency dependent adjustment
to $\bar\lambda_{\ell}$ from Eq.~(\ref{eq:DynamicalTides}).

The spliced waveform's time $t_{\text{Spl}}$ and phase $\phi_{\text{Spl}}$ are
then given by examining Eqs.~(\ref{eq:T2time}) and (\ref{eq:T2phi}) and making
the appropriate substitutions,
\begin{align}
t_{\text{Spl}}(v) =& t_0 + t_{\text{NR}}(v)
    + \mathcal T_{\text{Tid}}(v), \nonumber\\
\phi_{\text{Spl}}(v) =& \phi_0 + \phi_{\text{NR}}(v)
    + \mathcal P_{\text{Tid}}(v).
\label{eq:TaylorT2Splicing}
\end{align}
We use the freedom inherent in choosing $t_0$ and $\phi_0$ to align the spliced
waveform to the numerical waveform at the initial time

As the waveform nears the merger phase of the evolution, the effect from
$\mathcal T_{\text{Tid}}(v)$ might grow larger than that of $t_{\text{NR}}(v)$.
This may cause $t_{\text{Spl}}(v)$ to become nonmonotonic at some $v$; if this
happens, we end the waveform at that value of $v$.

\subsection{Waveform reconstruction}

Once $t_{\text{Spl}}(v)$ and $\phi_{\text{Spl}}(v)$ are computed, then the final
step is reconstructing the spliced waveform from Eqs.~(\ref{eq:T4EoM})
or~(\ref{eq:T2EoM}) (which are the same equation). For $A^{\ell
m}_{\text{BBH}}(v)$ and $\psi^{\ell m}_{\text{BBH}}(v)$ we use $A^{\ell
m}_{\text{NR}}(v)$ and $\psi^{\ell m}_{\text{NR}}(v)$ computed from
Eq.~(\ref{eq:NrWaveformDecomposition}). For $A^{\ell m}_{\text{Tid}}(v)$ we use
the expressions in Eq.~(\ref{eq:TidalStrainAmplification}), with the dynamical
tides accounted for by the replacement rule $\bar\lambda_{2} \rightarrow
\bar\lambda_{2A}\hat\kappa_{2A}(v)$ from Eq.~(\ref{eq:DynamicalTidesStrain}). We
thus arrive at the final formula for the spliced waveform modes:
\begin{align}
h^{\ell m}_{\text{Spl}}\left(v\right) =&
    \left(A^{\ell m}_{\text{NR}}(v)
    + A^{\ell m}_{\text{Tid}}(v)\right)
    e^{i(\psi^{\ell m}_{\text{NR}}(v)-m\phi_{\text{Spl}}(v))}.
\end{align}
To get a time-domain waveform, we invert the function $t_{\text{Spl}}(v)$.
Because $v$ is known only at discrete values, we interpolate the amplitudes and
phases of the waveforms onto a set of uniformly spaced values of
$t_{\text{Spl}}$ using a cubic spline.

\section{Results}\label{sec:Results}

\begin{figure*}
\includegraphics[width=.75\textwidth]{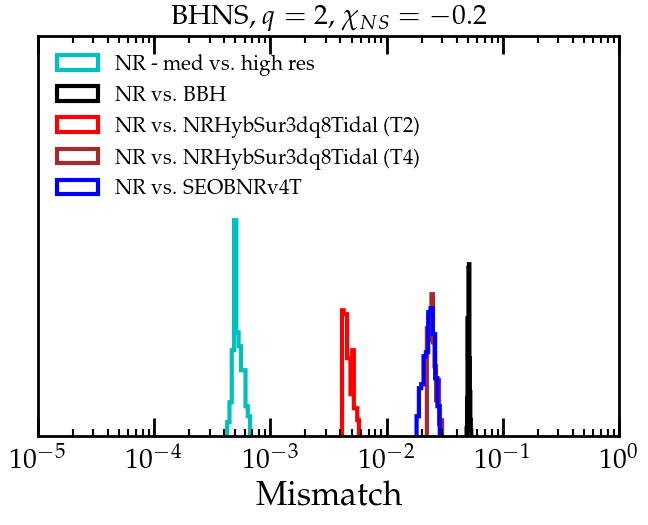}
\caption{
Distribution of mismatches across sky locations for the $q=2$,
$\chi_{NS}=-0.2$ BHNS simulation. The cyan histogram
show mismatches between waveforms from the highest two resolutions of the
simulation, whereas all other histograms shows mismatches between the
highest resolution waveform of the simulation versus the
labeled waveform model.  The BBH waveform model is the surrogate
NRHybSur3dq8.
\label{fig:mismatches_BHNSQ2Spin}
}
\end{figure*}

\begin{figure*}
\includegraphics[width=.95\textwidth]{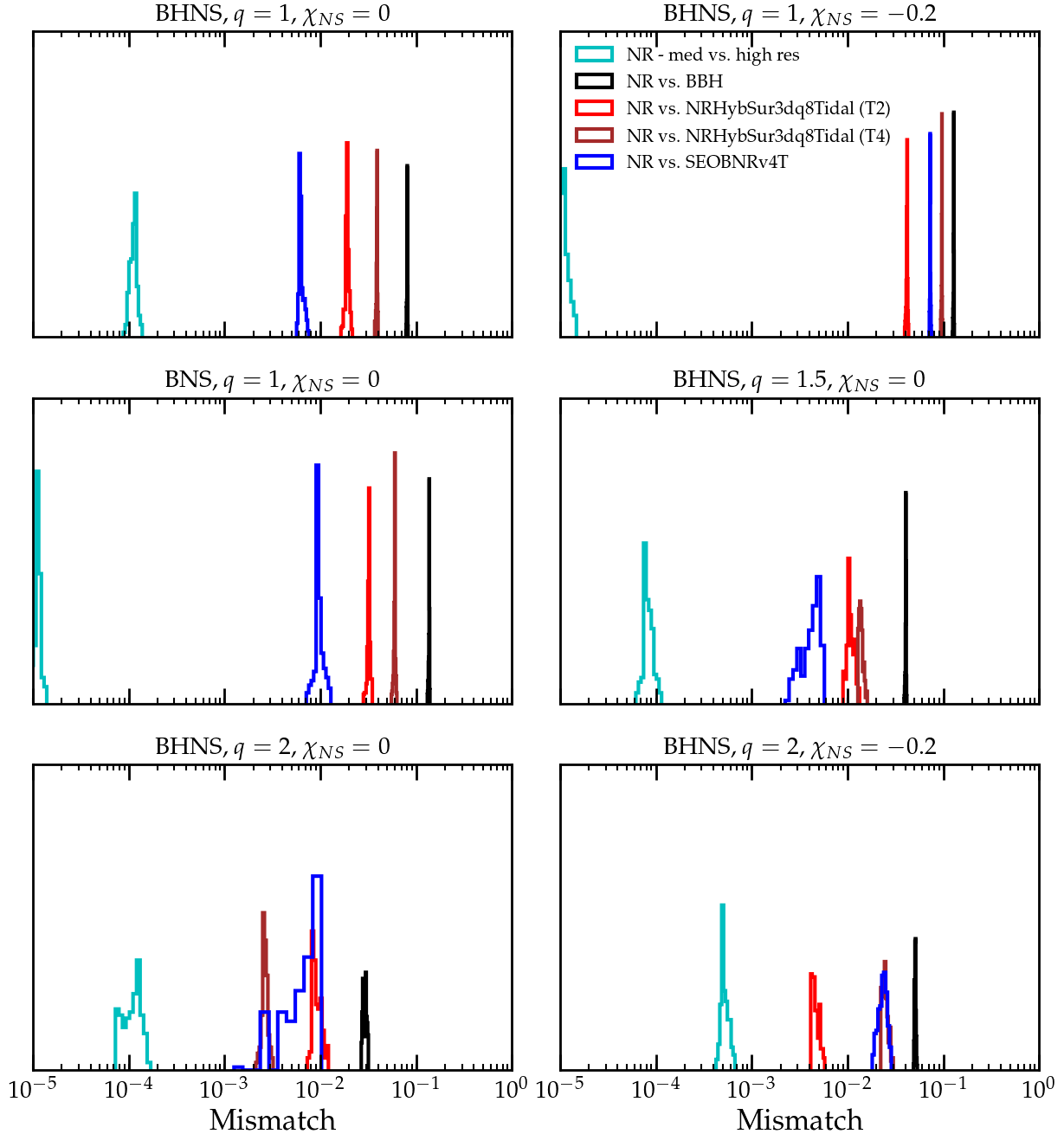}
\caption{
Distribution of mismatches across sky locations between the highest resolution waveform of the simulation versus the labeled waveform model for all six numerical simulations we consider.
\label{fig:mismatches_All}
}
\end{figure*}

\begin{figure*}
\includegraphics[width=.95\textwidth]{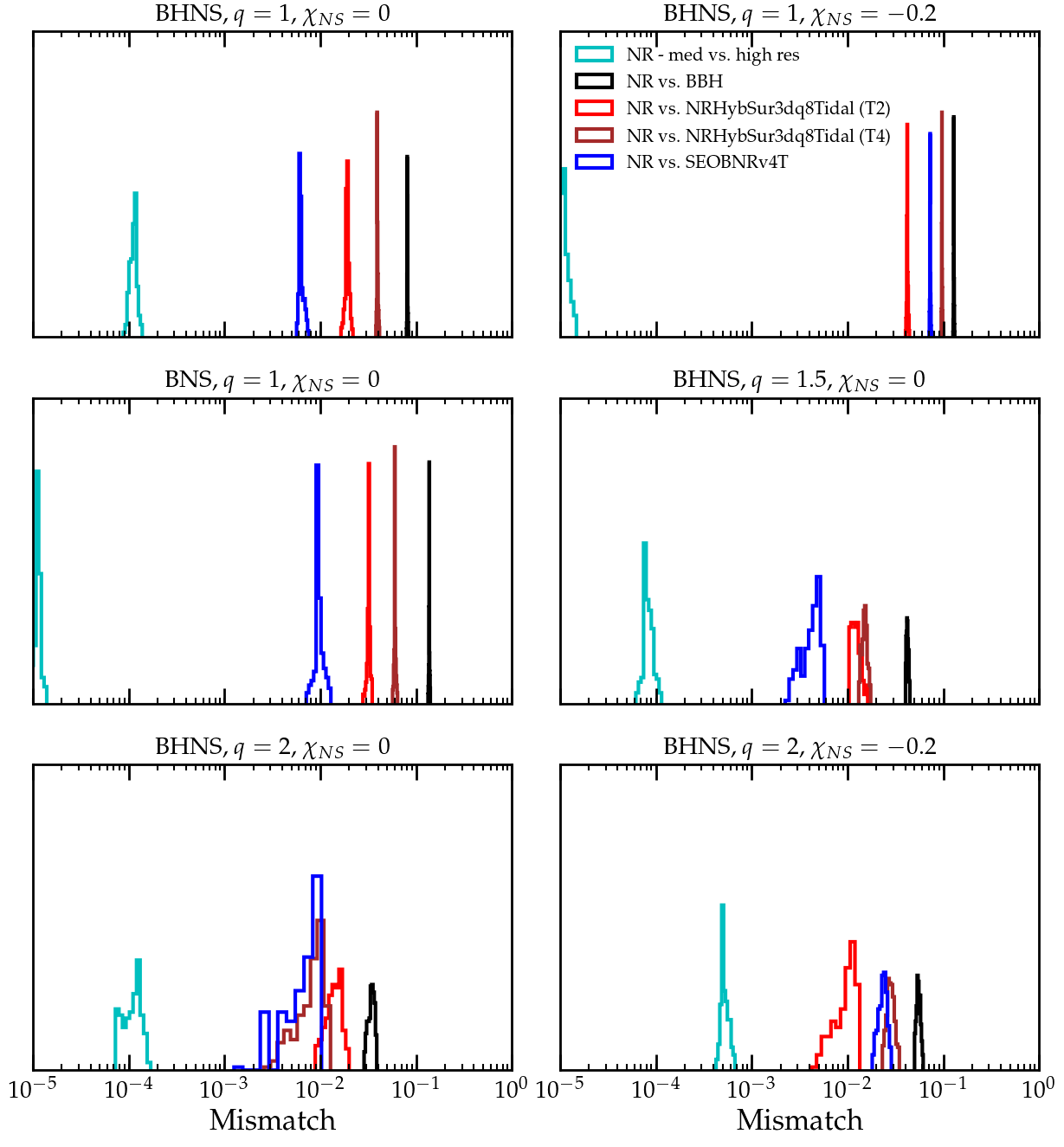}
\caption{
Distribution of mismatches across sky locations between the highest resolution
waveform of the simulation versus the labeled waveform model for all six
numerical simulations we consider, except that the BBH and tidally spliced
waveforms are generated with only the (2,2) mode. See Sec.~\ref{sec:mism-comp-2} for discussion of higher-order modes.
\label{fig:mismatches_All_22}
}
\end{figure*}

\subsection{Models for comparison}

To measure the accuracy of the tidal splicing method, we compare our spliced
waveforms against numerical simulations of BHNS/BNS inspirals. In particular, we
use some of the recent numerical simulations from~\cite{Foucart:2018inp}.
In all simulations we compare against, the NSs were generated according to an
EOS of $\Gamma=2$ polytrope with a mass $M_{\text{ADM}}=1.4M_{\odot}$ and
compactness of $C_{\text{NS}}=0.1444$ so that the quadrupolar tidal
deformability is $\bar\lambda_2 \sim 800$. Comparing against one specific EOS at
a single NS mass is a small slice of the full possible BHNS/BNS parameter space,
but should give some idea for how well tidal splicing can perform. See
Table~\ref{tab:NRParams} for the full list of the simulations we consider here.
In particular, there are 1 BNS and 5 BHNS runs, and two of the BHNS runs have a
small antialigned spin on the NS while the rest of the runs have zero spins.

We generated our tidally spliced waveforms for each of these cases using the
hybridized surrogate model `NRHybSur3dq8'~\cite{Varma:2018mmi} to compute the
underlying BBH signal and making use of the universal relations to obtain the
other tidal parameters from $\bar\lambda_2$; the details are given in
Appendix~\ref{app:ModelParameters}. The TaylorT2 splicing model has also been
implemented as a tidal extension of ``NRHybSur3dq8'' under the name
``NRHybSur3dq8Tidal''.

\begin{table}
\centering
\def\arraystretch{1.5}
\setlength\tabcolsep{.7em}
\begin{tabular}{| c | c | c | c | c | c |}
\hline
Type & $q$ & $\chi_{\text{NS}}$ & $f_0$ [Hz] & $f_1$ [Hz] & $N^{22}_{\text{cyc}}$ \\
\hline\hline
BHNS & 1 & 0 & 218 & 578 & 19.9 \\\hline
BNS  & 1 & 0 & 211 & 629 & 20.8 \\\hline
BHNS & 1 & -0.2 & 217 & 505 & 17.0 \\\hline
BHNS & 1.5 & 0 & 154 & 537 & 28.9 \\\hline
BHNS & 2 & 0 & 156 & 505 & 21.0 \\\hline
BHNS & 2 & -0.2 & 156 & 485 & 19.8 \\\hline
\end{tabular}
\par
\bigskip
\caption{List of parameters for numerical simulations
from Ref.~\cite{Foucart:2018inp} considered in this paper, namely the mass
ratio $q$, the dimensionless
spin of the neutron star $\chi_{\text{NS}}$, the lower and upper
orbital frequency cutoffs of the waveforms $f_0$ and $f_1$, and the number of
cycles in the (2,2) mode of the waveform between the cutoff frequencies.
}
\label{tab:NRParams}
\end{table}

To provide an additional point of comparison, we also test another waveform
model, SEOBNRv4T, which is the time domain model SEOBNRv4~\cite{Bohe:2016gbl}
augmented with most of the same effects we have used here, including
higher-order corrections to the static tides in the EOB potential
~\cite{Bini:2012gu}, strain corrections~\cite{Baiotti2011}, and dynamical tides
(without the resonance frequency correction for spinning
NSs)~\cite{Hinderer:2016eia}. These effects correspond to our
Eq.~(\ref{eq:EOBRadialPotential}), Eq.~(\ref{eq:TidalStrainAmplification}), and
Eqs.~(\ref{eq:DynamicalTides}) and~(\ref{eq:DynamicalTidesStrain}).
This comparison is meant to serve as a proof of concept of tidal splicing
  against a similar model rather than a comprehensive survey of all models;
  we leave a more detailed comparison with other state-of-the-art models like
  TEOBResumS, LEA+, and the various extensions of the NRTidal model for
  future work.

\subsection{Waveforms}\label{subsec:waveforms}

The numerical waveforms we compare against include all modes through $\ell=5$,
while the surrogate and spliced waveforms model only the modes [(2, 0), (2, 1),
(2, 2), (3, 2), (3, 0), (3, 1), (3, 3), (4, 2), (4, 3), (4, 4), (5, 5)], and
SEOBNRv4T models only the (2,2) mode. Since all systems we consider have spins
parallel to the orbital angular momentum, we
need only $m\geq0$ modes; the
$m<0$ modes are obtained from symmetry. The strain measured along a particular
direction in the sky can then be written as
\begin{align}
h_{+}(\iota,\varphi_0) - ih_{\times}(\iota,\varphi_0)
    = \sum_{\ell m}{h_{\ell m}}~_{-2}Y_{\ell m}(\iota,\varphi_0)
\end{align}
Here $_{-2}Y_{\ell m}$ are spin-weighted spherical harmonics,
and the angles $(\iota,\varphi_0)$ are defined so that $\iota$ is the
inclination angle between the binary angular momentum and the line of sight to
the observer while $\varphi_0$ is the binary orbital phase when it enters the
detector sensitivity band.

We choose the beginning of the waveform to be at a time after the initial burst
of junk radiation, $t=200M$. That time also sets
the starting orbital frequency
of the waveform (chosen as half the time derivative of the phase of
the (2,2) mode). To prevent the starting frequency $f_{\text{Initial}}$
from being contaminated by residual junk radiation in
the imperfect BHNS/BNS initial data, we use a quadratic fit of the simulations'
frequency against time over the interval $t\in(200M,700M)$ to estimate the precise starting
frequency. We window the waveform with a Planck-Taper window~\cite{McKechan:2010kp}
over that early
$500M$ region of the waveform. We label the orbital frequency at the end of this
window as $f_0$; this frequency will serve as the initial frequency
considered in our mismatches below.

At late times, we set the upper frequency cutoff $\omega_{\text{Cutoff}}$ by the
frequency attained by the simulation at its peak power,
and window the waveform (again with Planck-Taper) over the times from $f_1
= 0.85f_{\text{Cutoff}}$ to $f_{\text{Cutoff}}$.
This gives us an
inspiraling waveform from orbital frequency $f_0$ to $f_1$.
The other waveforms we generated from
$f_{\text{Initial}}$ to $f_{\text{Cutoff}}$ and windowed in a similar
manner. In Table~\ref{tab:NRParams}, we also list the number of cycles
$N^{22}_{\text{cyc}}$ in the (2,2) mode of 
the numerical simulation within the listed
orbital frequency bounds. While this is not necessarily a large frequency range
to be making our comparisons, it is during the late inspiral where we expect the
tidal effects to make the strongest contributions to the binary's evolution.

After we transform all of the waveforms into the frequency domain, we calculate
the mismatches with the numerical waveforms.
To evaluate the mismatch, we optimize over time, orbital phase,
  and polarization angle shifts between the waveforms, following the
  procedure of Appendix D of~\cite{Blackman:2017dfb}.  We assume a
  flat noise spectrum.
The starting and ending frequencies of the windowed waveforms, as given in
Table~\ref{tab:NRParams}, bound the frequency range for the mismatch
computations.
The mismatch between each of the models and the
numerical simulation is then computed across 363 uniformly distributed sky locations.

We also attempted to evaluate a pair of frequency domain
tidal waveform models, SEOBNRv4ROM\_NRTidal and IMRPhenomD\_NRTidal, but found
mismatches between those models and the numerical simulations
that we believed to be artificially large, near that of the BBH
waveform. This is likely a consequence of having short numerical waveforms, which require windowing over a relatively short time interval, whereas the
  frequency-domain waveforms are not windowed. Note that all time-domain waveforms here are windowed in the same way, so systematic errors introduced by windowing should be the same for all of them.
We have excluded the frequency domain waveforms from the results below.

\subsection{Mismatch comparison: All modes}

In Fig~\ref{fig:mismatches_BHNSQ2Spin}, we plot a histogram
of mismatches against the numerical waveform across sky locations for a
$q=2$, $\chi_{\text{NS}}=-0.2$ BHNS system. We do not normalize the vertical axis
since the exact heights of the histograms are dependent on the binning choice;
the locations of the histogram peaks correspond to how well the model does while
the spread measures how dependent the model is on the sky location.

We estimate the numerical simulation error during the inspiral by the mismatch
between simulations at the two highest numerical resolutions (cyan). Although
this estimate is only a rough measure of the simulation error, and is possibly
optimistic, the size of this mismatch suggests that the numerical error is
smaller than the effects we are examining. Recall that we are not including
merger and ringdown; the numerical error for the merger and postmerger
evolution is expected to be much larger. The mismatch between the BHNS system
and the surrogate BBH waveform (black) measures the strength of the tidal
effects in the system, showing how poorly the waveforms will perform if tidal
effects are neglected entirely.

As expected, both of the tidal splicing methods we try here, TaylorT4 (magenta)
and TaylorT2 (red), improve upon the BBH waveform, as does
the SEOBNRv4T model (blue).
In this particular case, both the TaylorT4
and the SEOBNRv4T models
show moderate improvements compared with the BBH waveform, accounting for some of
the NSBH tidal effects, while the TaylorT2 model has mismatches about an order
of magnitude smaller.

In Fig~\ref{fig:mismatches_All}, we display the mismatch histograms across all
simulations we consider. In all cases, the
estimated numerical error of the inspiral is smaller than
any of the mismatches from the waveforms considered here, and the
size of the tidal effects behaves qualitatively as expected (i.e., more
extreme mass ratios have smaller tidal effects, a spinning NS has larger tidal
effects, \ldots). The
hierarchy between the TaylorT2 splicing, TaylorT4 splicing, and SEOBNRv4T
changes across the parameter space, with each model performing the best in at
least one of the cases.

In the case of low mass ratio ($q=1,1.5$), nonspinning BHNS and BNS
(top left, middle left, and middle right), SEOBNRv4T has the lowest mismatches,
followed by TaylorT2 splicing, and then TaylorT4 splicing. Across this region of the
parameter space, there is very little change in behavior of the mismatches, with
the BNS mismatches slightly larger, presumably because the tidal effects are at
least twice as large (since both objects in the binary are
being deformed rather than one). For nonspinning neutron stars, increasing
the BHNS mass ratio from $q=1$ (top left) to $q=1.5$ (middle right)
to $q=2$ (bottom left) shows relative improvement in the spliced waveforms,
until in the $q=2$ case, the TaylorT4 splicing model has the smallest
mismatches. The distribution of mismatches for SEOBNRv4T widens significantly
with the increasing mass ratio, likely at least in part because of the growing
significance of modes beyond (2,2), as we will discuss below.

The most significant change to the mismatches arises in the case of the spinning
NS ($q=1$ top right; $q=2$ bottom right),
where TaylorT2 splicing performs
appreciably better than the other two models. In both cases, the mismatches of
the
waveforms worsen significantly compared to the corresponding nonspinning cases.
At best, the effects included here only account for some of the changes the
spinning NS has on the evolution of the system and on the gravitational
radiation. Remaining errors in the tidal-splicing
  waveform models may be
due to the missing spin-tidal terms (see
Appendix~\ref{app:SpinTidal}), the inaccurate handling of the dynamical tides in
the case of antialigned NS, or some other unaccounted tidal effect. Further
work will need to be done in order to properly capture the full behavior of
systems with spinning NSs.

\subsection{Mismatch comparison: (2,$\pm$2) Modes}
\label{sec:mism-comp-2}

The nonquadrupole modes are expected to become more important for systems with
large mass ratios $q$ and/or large inclination angles
$\iota$~\cite{Varma:2016dnf}. In
order to characterize how much of the disparity between the spliced and
SEOBNRv4T waveforms is due to the inclusion of higher-order modes in the spliced
models, we recompute the mismatches after restricting the BBH and spliced
waveforms to only the $(2,\pm2)$ modes (see Fig~\ref{fig:mismatches_All_22}).
The numerical simulations still utilize all the same modes as before.

For the nonspinning $q=1$ BHNS and BNS systems, there is very little change in
the mismatches of the spliced waveforms when excluding
the higher-order modes from the splicing model. This is
expected, as a majority of the power in those waveforms is
concentrated in the $(2,2)$ mode so leaving out the other modes
produces a negligible effect.
Thus the change in the histograms between Figs~\ref{fig:mismatches_All}
and~\ref{fig:mismatches_All_22} for these systems is smaller than the width of
the histograms. This also holds true for the equal mass
spinning system and the $q=1.5$ system.

For the $q=2$ waveforms, restricting the splicing models to the $(2,\pm2)$
modes leads to noticeably wider mismatch distributions. In the nonspinning case,
the TaylorT4 splicing profile is similar to that of
SEOBNRv4T, suggesting that much of the discrepancy between those
  two models in the bottom left panel of Fig.\ref{fig:mismatches_All}
arises from the inclusion of higher-order modes.
In the spinning case, while the TaylorT2 splicing waveform is wider than before,
it still performs better than the other waveforms.

\section{Conclusions}\label{sec:Conclusions}

We have demonstrated how tidal splicing combines the accuracy of numerical BBH
systems with the cheap computation of PN tidal formulas
to generate waveforms
corresponding to inspiraling BHNS and BNS systems. We expanded the tidal terms
of the TaylorT2 and TaylorT4 PN approximants up to 2.5PN order and
incorporated dynamical
corrections to the approximants in order to improve our model. We also included
a partial expansion of aligned spin-tidal corrections, though these additions
are not complete to the same order as the nonspinning tidal effects.
The tidal splicing method is now able to generate waveforms
not only for the (2,2) mode, but for all
modes supplied by the underlying BBH waveform. In
particular, we applied tidal splicing to a surrogate model
that spans the spin-aligned region of the
BBH parameter space of interest to BNS and BHNS systems.

We measure the accuracy of tidal splicing against a series of inspiraling BNS
and BHNS numerical simulations, and
against the results of the SEOBNRv4T
model. These simulations include systems with a mass ratio from $q=1$ to $q=2$ and
both nonspinning and antialigned spin NSs. Across all cases, the tidally
spliced waveforms capture an appreciable fraction of the tidal effects in the
system. In different regions of the parameter space, different models perform
best, with TaylorT4 splicing best in the $q=2$, nonspinning case, TaylorT2
splicing best in the spinning cases, and SEOBNRv4T best in the low mass ratio,
nonspinning cases. We implement a model we call ``NRHybSur3dq8Tidal'',
which uses TaylorT2 splicing to extend the ``NRHybSur3dq8'' surrogate model for
inspiraling tidal systems.

The accuracy of tidal splicing during the inspiral is limited
in principle by the analytic tidal information fed into it, so the most natural
extension of tidal splicing is through the inclusion of additional PN effects.
As higher-order terms are computed, they can be appended to the TaylorT2 and
TaylorT4 expressions. Adding more complete spin-tidal couplings and corrections
to the dynamical tides would be particularly useful as the mismatches between
the current spliced waveforms against the numerical simulations are worse for
a spinning NS than for a nonspinning NS.

The current splicing method generates waveforms for only
the inspiral portion of the signal, with no prescription for the
merger and ringdown portions. Completing the full waveform will require
developing a formalism for splicing BBH merger/ringdown signals, hybridizing the
end portion with numerical results, or some other method. This is complicated by
the fact that for BNS there are multiple possible final states (i.e.,
direct collapse to BH, long-lived hypermassive NS, \ldots), each of which can
imprint a unique signature on the merger/ringdown signal. We leave these possible
improvements to future work.

\begin{acknowledgments}
Research in this paper is supported by the the the Sherman Fairchild Foundation,
the Simons Foundation (Award Number 568762), and the National Science Foundation
(Grants PHY-1708212 and PHY-1708213).
\end{acknowledgments}
\appendix

\section{The 2.5PN Tidal Energy}\label{app:EOB2PN}

Within the EOB framework, the energy terms associated with the quadrupolar and
octopolar static deformations are known to 2PN~\cite{Bini:2012gu}. We can
convert these terms to an equivalent expression for the energy within the PN
framework.
For circular orbits, the full Hamiltonian is
\begin{align}
H_{\text{EOB}}(u,J) =& M\sqrt{1+2\nu\left(-1+\sqrt{\mathcal A(u)
    \left(1+\frac{J^2u^2}{m_1^2m_2^2}\right)}\right)},
\label{eq:EOBHamiltonianCirc}
\end{align}
where $u=M/r$ is the dimensionless inverse EOB radial
coordinate, $J$ is the orbital angular momentum, and $\mathcal A(u)$ is the
nonspinning radial PN potential, known through 2PN.

This $\mathcal A(u)$ is given in~\cite{Bini:2012gu}, which we reproduce here for
completeness,
\begin{widetext}
\begin{align}
\mathcal A(u) =& \mathcal A_{\text{BBH}}(u) + \mathcal A_{\bar\lambda_2}(u)
    + \mathcal A_{\bar\lambda_3}(u), \nonumber\\
\mathcal A_{\text{BBH}}(u) =& 1 - 2u + 2\nu u^3 + \mathcal O(u^4), \nonumber\\
\mathcal A_{\bar\lambda_2}(u) =& 3\bar\lambda_{2A}X_A^4(1-X_A)u^5\left(u +
    u^2\left(\frac{5}{2}X_A\right) + u^3\left(\frac{337}{28}X_A^2
    +\frac{1}{8}X_A+ 3\right) + \mathcal O(u^4)\right)
    + (A \rightarrow B), \nonumber\\
\mathcal A_{\bar\lambda_3}(u) =& 3\bar\lambda_{3A}x_A^6(1-X_A)u^7\left(u
    + u^2\left(\frac{15}{2}X_A - 2\right)
    + u^3\left(\frac{110}{2}X_A^2 - \frac{311}{24}X_A + \frac{8}{3}\right)
    + \mathcal O(u^4)\right) + (A \rightarrow B).
\label{eq:EOBRadialPotential}
\end{align}
\end{widetext}
The first unknown terms in $\mathcal A(u)$ are of order
$\mathcal O(u^3)$ beyond the leading order $1-2u$ Newtonian term,
which corresponds to $\mathcal O(v^{6})$ beyond leading-order,
or 3PN order. Since the tidal effects
only enter at full PN orders, this potential is valid until to the first
unknown terms at 3PN, i.e., up through 2.5PN.

To relate the EOB Hamiltonian to the PN energy expansions, we relate the EOB
radial variable $u$ to the PN expansion variable $v$ via the orbital phase
$\phi_{\text{orb}}$. Namely, $\phi_{\text{orb}}$ is both the conjugate variable
of $p_{\phi}=J$, and one of the PN evolution equations as defined in
Eq.~(\ref{eq:PNvDef}). Therefore we establish the relationship between the EOB and
PN energy equations with
\begin{align}
\frac{\partial\phi_{\text{orb}}}{\partial t} =&
  \frac{\partial H_{\text{EOB}}(u,J)}{\partial J} = \frac{v^3}{M}.
\label{eq:EOBPNConnection}
\end{align}

Because we are considering circular orbits, the radial conjugate variable
$p_{u}$ is constant over the orbit, or $p_{u} = -\partial
H_{\text{EOB}}(u,J)/\partial u = 0$, which reduces to
\begin{align}
0 = \frac{\partial}{\partial u}\left(\mathcal A(u)
    \left(1+\frac{J^2u^2}{m_1^2m_2^2}\right)\right),
\end{align}
providing a relation between $J$ and $u$~\cite{Bini:2012gu},
\begin{align}
J^2 = -\left(\frac{\partial \mathcal A(u)}{\partial u}\right)\bigg/
    \left(\frac{\partial\left(u^2\mathcal A(u)\right)}{\partial u}\right).
\end{align}
We can substitute this expression into Eq.~(\ref{eq:EOBPNConnection}), reducing
the expression to a formula connecting $u$ and $v$,
which we then expand to obtain $u$ as a power series of $v$. We then insert that expansion
into the EOB Hamiltonian Eq,~(\ref{eq:EOBHamiltonianCirc}). By expressing the EOB
Hamiltonian in powers of $v$, we obtain the PN energy
formula given in Eq.~(\ref{eq:PNTidalEnergy}), which is complete up
through 2.5PN order.

\section{The 2.5PN Tidal Expressions}
\label{sec:2.5pn-tidal-expr}

\subsection{TaylorT4}

Here we explicitly provide
the full PN coefficients for the tidal terms in the TaylorT4 expansion
in Eq.~(\ref{eq:T4TidalPartial}):
\begin{widetext}
\begin{align}
\mathcal F_{\text{Tid}}(v) =& \frac{32\nu v^9}{5M}\left[
    \left(5(\bar Q_A-1)\chi_A^2X_A^2\right)v^4
    + \bar\lambda_{2A}X_A^4v^{10}\left(\sum^5_{i=0}\mathcal F_{2A,i}v^i\right)
    + \bar\lambda_{3A}X_A^6v^{14}\left(\sum^5_{i=0}\mathcal F_{3A,i}v^i\right)
    + \left(A \leftrightarrow B\right)\right], \nonumber\\
\mathcal F_{2A,0} =& 72 - 66X_A, \nonumber\\
\mathcal F_{2A,2} =& \frac{4421}{56}-\frac{12263 X_A}{56}
    + \frac{1893 X_A^2}{4}-\frac{661 X_A^3}{2}, \nonumber\\
\mathcal F_{2A,3} =& 216 \pi -216 \pi X_A+\left(-\frac{1395 X_A}{2}
    + \frac{753 X_A^2}{2}+281 X_A^3\right) \chi _A+\left(-\frac{1977}{2}
    + 2228X_A-\frac{3041 X_A^2}{2}+281 X_A^3\right) \chi _B, \nonumber\\
\mathcal F_{2A,4} =& \frac{130225}{96}+\alpha_4-\frac{853193 X_A}{672}
    + \frac{664637 X_A^2}{672}-\frac{249689 X_A^3}{224}+\frac{5139 X_A^4}{8}
    - \frac{12931 X_A^5}{24} \nonumber\\
    +& \left(\frac{3915 X_A^2}{8}-\frac{3771 X_A^3}{8}
    + \bar Q_A \left(486 X_A^2-468 X_A^3\right)\right) \chi_A^2
    + \left(\frac{3861 X_A}{4}-\frac{3789 X_A^2}{2}
    + \frac{3717 X_A^3}{4}\right) \chi _A \chi _B \nonumber\\
    +& \left(\frac{3915}{8}-\frac{11601X_A}{8}+\frac{11457 X_A^2}{8}
    - \frac{3771 X_A^3}{8}+\bar Q_B \left(486-1440 X_A+1422 X_A^2
    - 468 X_A^3\right)\right)\chi _B^2, \nonumber\\
\mathcal F_{2A,5} =& \frac{50601 \pi }{112}-\frac{154957 \pi X_A}{112}
    + \frac{5267 \pi X_A^2}{2}-\frac{6807 \pi X_A^3}{4} \nonumber\\
    +& \left(-\frac{48145 X_A}{14}+\frac{236779X_A^2}{84}
    - \frac{650371 X_A^3}{168}+\frac{21111 X_A^4}{8}
    + \frac{19277 X_A^5}{12}\right)\chi _A \nonumber\\
    +& \left(-\frac{570401}{84}+\frac{849409X_A}{42}-\frac{4606397 X_A^2}{168}
    + \frac{319379 X_A^3}{14}-\frac{250409 X_A^4}{24}
    + \frac{19277 X_A^5}{12}\right) \chi_B,\nonumber\\
\mathcal F_{3A,0} =& 520-520 X_A+\beta_0, \nonumber\\
\mathcal F_{3A,2} =& -\frac{14855}{21}+\frac{3 \beta_0}{2}+\beta_2
    + X_A \left(\frac{62245}{21}+\frac{\beta_0}{6}\right)
    + \frac{X_A^2}{6}  (3295-\beta_0)-\frac{16835 X_A^3}{6}, \nonumber\\
\mathcal F_{3A,3} =& 2080 \pi -2080 \pi X_A
    + \left(- 5 X_A (1170+\beta_0)-\frac{5}{3} X_A^2 (-2002+\beta_0)
    + \frac{7540 X_A^3}{3}\right) \chi_A \nonumber\\
    +& \left(-\frac{10}{3} (2509+2 \beta_0)
    + X_A\left(19240+\frac{25 \beta_0}{3}\right)-\frac{5}{3} X_A^2(8034+\beta_0)
    + \frac{7540 X_A^3}{3}\right) \chi_B, \nonumber\\
\mathcal F_{3A,4} =& \frac{154460615}{18144}+\frac{99 \beta_0}{8}
    + \frac{3 \beta_2}{2}+\beta_4+X_A \left(-\frac{84699635}{18144}
    - \frac{53 \beta_0}{8}+\frac{\beta_2}{6}\right)
    + X_A^2 \left(-\frac{2405675}{672}+\frac{61 \beta_0}{9}
    - \frac{\beta_2}{6}\right) \nonumber\\
    +& X_A^3 \left(\frac{3298825}{224}-\frac{11 \beta_0}{36}\right)
    + X_A^4 \left(-7980+\frac{11\beta_0}{72}\right)-\frac{56095 X_A^5}{8} \nonumber\\
    +& \left(X_A^2 \left(\frac{8385}{2}+3 \beta_0\right)-\frac{8385 X_A^3}{2}
    + \bar{Q}_A \left(X_A^2 (4160+3\beta_0)-4160 X_A^3\right)\right)
    \chi_A^2 \nonumber\\
    +& \left(X_A (8255+6 \beta_0)-2 X_A^2 (8255+3 \beta_0)+8255 X_A^3\right)
    \chi_A \chi_B \nonumber\\
    +& \bigg[\frac{8385}{2}+3 \beta_0-\frac{3}{2} X_A (8385+4 \beta_0)
    + X_A^2\left(\frac{25155}{2}+3\beta_0\right)-\frac{8385 X_A^3}{2}\nonumber\\
    +& \bar{Q}_B \left(4160+3 \beta_0-6 X_A (2080+\beta_0)
    + 3 X_A^2 (4160+\beta_0)-4160 X_A^3\right)\bigg] \chi_B^2, \nonumber\\
\mathcal F_{3A,5} =& -\frac{121655 \pi }{84}+\frac{328435 \pi  X_A}{84}
    + \frac{45980 \pi  X_A^2}{3}-\frac{53365 \pi  X_A^3}{3} \nonumber\\
    +& \bigg[X_A \left(-\frac{2284265}{168}-\frac{51\beta_0}{2}-5 \beta_2\right)
    - \frac{5}{504} X_A^2 (1874947+1260 \beta_0+168 \beta_2)
    + X_A^3\left(-\frac{1280165}{56}-\frac{61 \beta_0}{6}\right) \nonumber\\
    +& X_A^4 \left(\frac{2816635}{72}+\frac{\beta_0}{6}\right)
    + \frac{573755 X_A^5}{36}\bigg] \chi_A \nonumber\\
    +& \bigg[-\frac{5}{504}(3565057+672 \beta_2)-48 \beta_0+X_A \left(\frac{9267085}{126}+\frac{241 \beta_0}{3}+\frac{25 \beta_2}{3}\right) \nonumber\\
    +& X_A^2 \left(-42 \beta_0-\frac{5}{252} (4784305+84 \beta_2)\right)
    + X_A^3 \left(\frac{31907395}{252}+\frac{19\beta_0}{2}\right)
    + X_A^4 \left(-\frac{6178195}{72}+\frac{\beta_0}{6}\right)
    + \frac{573755 X_A^5}{36}\bigg] \chi_B.
\label{eq:T4TidalFull}
\end{align}
\end{widetext}
Note that $\mathcal F_{2A,0}$ and $\mathcal F_{2A,2}$ reproduce the TaylorT4
coefficients from the quadrupolar deformability
first computed in Ref~\cite{Vines2011}.

\subsection{TaylorT2}

Here we explicitly provide
the full PN coefficients for the tidal terms in the TaylorT2 expansion
in Eqs.~(\ref{eq:T2TTidalPartial}) and (\ref{eq:T2PTidalPartial}).
For the correction to the time, we find
\begin{widetext}
\begin{align}
\mathcal T_{\text{Tid}}(v) =& -\frac{5M}{256\nu v^8}\left[
    -\left(10(\bar Q_A-1)\chi_A^2X_A^2\right)v^4
    + \bar\lambda_{2A}X_A^4v^{10}\left(\sum^5_{i=0}\mathcal T_{2A,i}v^i\right)
    + \bar\lambda_{3A}X_A^6v^{14}\left(\sum^5_{i=0}\mathcal T_{3A,i}v^i\right)
    + \left(A \leftrightarrow B\right)\right], \nonumber\\
\mathcal T_{2A,0} =& 288-264 X_A, \nonumber\\
\mathcal T_{2A,2} =& \frac{3179}{4}-\frac{919 X_A}{4}-\frac{1143 X_A^2}{2}
    + 65 X_A^3, \nonumber\\
\mathcal T_{2A,3} =& -576 \pi +\frac{2496 \pi X_A}{5}+\left(324 X_A+12 X_A^2
    - \frac{1096 X_A^3}{5}\right) \chi _A+\left(588-\frac{6616 X_A}{5}
    + \frac{4772X_A^2}{5}-\frac{1096 X_A^3}{5}\right) \chi _B, \nonumber\\
\mathcal T_{2A,4} =& \frac{70312133}{21168}+\frac{4 \alpha_4}{3}
    - \frac{147794303 X_A}{127008}-\frac{20905 X_A^2}{28}
    - \frac{432193 X_A^3}{504}-\frac{5848X_A^4}{9}+\frac{857 X_A^5}{3}
    \nonumber\\
    +& \left(-\frac{639 X_A^2}{2}+\frac{525 X_A^3}{2}
    + \bar Q_A \left(-312 X_A^2+256 X_A^3\right)\right)\chi _A^2
    + \left(-609 X_A+1108 X_A^2-499 X_A^3\right) \chi _A \chi _B \nonumber\\
    +& \left(-\frac{639}{2}+\frac{1803 X_A}{2}-\frac{1689 X_A^2}{2}
    + \frac{525X_A^3}{2}+\bar Q_B \left(-312+880 X_A-824 X_A^2+256 X_A^3\right)
    \right) \chi _B^2, \nonumber\\
\mathcal T_{2A,5} =& -\frac{241295 \pi }{98}+\frac{216921 \pi X_A}{98}
    - \frac{7528 \pi X_A^2}{7}+\frac{7142 \pi X_A^3}{7} \nonumber\\
    +& \left(\frac{101949 X_A}{49}+\frac{48875X_A^2}{294}
    - \frac{78373 X_A^3}{147}-\frac{3417 X_A^4}{7}
    - \frac{2574 X_A^5}{7}\right)\chi _A \nonumber\\
    +& \left(\frac{637447}{147}-\frac{1026647X_A}{98}+\frac{931999 X_A^2}{98}
    - \frac{713938 X_A^3}{147}+\frac{12977 X_A^4}{7}
    - \frac{2574 X_A^5}{7}\right) \chi _B, \nonumber\\
\mathcal T_{3A,0} =& \frac{4}{3} (520+\beta_0)-\frac{2080 X_A}{3},
    \nonumber\\
\mathcal T_{3A,2} =& \frac{33440}{21}+\frac{995 \beta_0}{168}+\beta_2
    + \left(\frac{24670}{7}+\frac{17 \beta_0}{3}\right) X_A
    - \frac{17}{6} (1825+2\beta_0) X_A^2+\frac{325 X_A^3}{6},
    \nonumber\\
\mathcal T_{3A,3} =& -\frac{64}{9} (260+\beta_0) \pi +\frac{16640 \pi X_A}{9}
    + \left(\frac{20}{9} (260+3 \beta_0) X_A+\frac{16}{27} (195+7 \beta_0)X_A^2
    - \frac{2080 X_A^3}{3}\right) \chi _A \nonumber\\
    +& \left(\frac{4}{27} (8580+73 \beta_0)-\frac{4}{27}\left(21840
    + 101\beta_0\right)X_A+\frac{16}{27} (4485+7 \beta_0) X_A^2
    - \frac{2080 X_A^3}{3}\right) \chi _B, \nonumber\\
\mathcal T_{3A,4} =& \frac{35111473}{3969}+\frac{6080015 \beta_0}{254016}
    + \frac{199 \beta_2}{42}+\frac{4 \beta_4}{5}+\left(\frac{68598877}{7938}
    + \frac{16999\beta_0}{840}+\frac{68 \beta_2}{15}\right) X_A \nonumber\\
    +& \frac{(7215185-16319 \beta_0-11424 \beta_2) X_A^2}{2520}
    + \left(-\frac{154907}{8}-\frac{2477\beta_0}{90}\right) X_A^3
    + \left(-\frac{3040}{3}+\frac{2477 \beta_0}{180}\right) X_A^4
    + \frac{455 X_A^5}{18} \nonumber\\
    +& \left(\left(-858-\frac{57\beta_0}{10}\right) X_A^2
    + 858 X_A^3+\bar Q_A \left(\left(-832-\frac{28}{5} \beta_0\right) X_A^2
    + 832 X_A^3\right)\right)\chi _A^2 \nonumber\\
    +& \left(\left(-1612-11 \beta_0\right) X_A+\left(3224
    + 11 \beta_0\right) X_A^2-1612 X_A^3\right)\chi _A \chi _B
    \nonumber\\
    +& \left[-858-\frac{57\beta_0}{10}+\left(2574+\frac{57}{5}\beta_0\right)X_A
    + \left(-2574-\frac{57 \beta_0}{10}\right)X_A^2+858 X_A^3
    \right.\nonumber\\
    &~~~+ \left.\bar Q_B \left(-832-\frac{28\beta_0}{5}
    + \left(2496+\frac{56\beta_0}{5}\right) X_A-\frac{4}{5}(3120+7 \beta_0) X_A^2
    + 832 X_A^3\right)\right] \chi _B^2, \nonumber\\
\mathcal T_{3A,5} =& -\frac{1}{462} (2604260+17705 \beta_0+2688 \beta_2) \pi
    + -\frac{2}{231}(516675+1687\beta_0) \pi X_A
    + \frac{2}{33}(90640+241 \beta_0) \pi X_A^2 + \frac{152360 \pi X_A^3}{33}
    \nonumber\\
    +& \left[\frac{1}{924} (3242620+49273 \beta_0+5040 \beta_2) X_A
    + \frac{(4691970+82637\beta_0+4704 \beta_2) X_A^2}{1386}
    + \frac{1}{693} (-221335+91 \beta_0) X_A^3 \right.\nonumber\\
    &~~~-\left.\frac{5}{99} (98807+254 \beta_0) X_A^4-\frac{156910X_A^5}{99}
    \right] \chi_A \nonumber\\
    +&\left[\frac{23393100+277897 \beta_0+24528 \beta_2}{2772}
    + \frac{(-44089360-337219 \beta_0-33936 \beta_2)X_A}{2772} \right.\nonumber\\
    &~~~\left.+\frac{(9394160-23497 \beta_0+4704 \beta_2) X_A^2}{1386}
    + \left(-\frac{272870}{231}+\frac{563 \beta_0}{11}\right)X_A^3
    - \frac{5}{99} (-68399+254 \beta_0) X_A^4-\frac{156910 X_A^5}{99}\right]
    \chi_B,
\label{eq:T2TTidalFull}
\end{align}
and for the correction to the phase we find
\begin{align}
\mathcal P_{\text{Tid}}(v) =& -\frac{1}{32\nu v^5}\left[
    -\left(25(\bar Q_A-1)\chi_A^2X_A^2\right)v^4
    + \bar\lambda_{2A}X_A^4v^{10}\left(\sum^5_{i=0}\mathcal P_{2A,i}v^i\right)
    + \bar\lambda_{3A}X_A^6v^{14}\left(\sum^5_{i=0}\mathcal P_{3A,i}v^i\right)
    + \left(A \leftrightarrow B\right)\right], \nonumber\\
\mathcal P_{2A,0} =& 72-66 X_A, \nonumber\\
\mathcal P_{2A,2} =& \frac{15895}{56}-\frac{4595 X_A}{56}
    - \frac{5715 X_A^2}{28}+\frac{325 X_A^3}{14}, \nonumber\\
\mathcal P_{2A,3} =& -225 \pi +195 \pi  X_A
    + \left(\frac{2025 X_A}{16}+\frac{75 X_A^2}{16}
    - \frac{685 X_A^3}{8}\right) \chi _A+\left(\frac{3675}{16}
    - \frac{4135X_A}{8}+\frac{5965 X_A^2}{16}
    - \frac{685 X_A^3}{8}\right) \chi _B, \nonumber\\
\mathcal P_{2A,4} =& \frac{351560665}{254016}+\frac{5\alpha_4}{9}
    - \frac{738971515 X_A}{1524096}-\frac{104525 X_A^2}{336}
    - \frac{2160965 X_A^3}{6048}-\frac{7310 X_A^4}{27}
    + \frac{4285 X_A^5}{36} \nonumber\\
    +& \left(-\frac{1065 X_A^2}{8}+\frac{875 X_A^3}{8}
    + \bar Q_A \left(-130 X_A^2+\frac{320X_A^3}{3}\right)\right) \chi _A^2
    + \left(-\frac{1015 X_A}{4}+\frac{1385 X_A^2}{3}
    - \frac{2495 X_A^3}{12}\right) \chi _A \chi_B \nonumber\\
    +& \left(-\frac{1065}{8}+\frac{3005 X_A}{8}-\frac{2815 X_A^2}{8}
    + \frac{875 X_A^3}{8}+\bar Q_B \left(-130+\frac{1100 X_A}{3}
    - \frac{1030X_A^2}{3}+\frac{320 X_A^3}{3}\right)\right)
    \chi _B^2, \nonumber\\
\mathcal P_{2A,5} =& -\frac{241295 \pi}{224}+\frac{216921 \pi X_A}{224}
    - \frac{941 \pi X_A^2}{2}+\frac{3571 \pi X_A^3}{8} \nonumber\\
    +& \left(\frac{101949 X_A}{112}+\frac{48875X_A^2}{672}
    - \frac{78373 X_A^3}{336}-\frac{3417 X_A^4}{16}
    - \frac{1287 X_A^5}{8}\right) \chi_A \nonumber\\
    +& \left(\frac{637447}{336}-\frac{1026647X_A}{224}+\frac{931999 X_A^2}{224}
    - \frac{356969 X_A^3}{168}+\frac{12977 X_A^4}{16}
    - \frac{1287 X_A^5}{8}\right) \chi_B, \nonumber\\
\mathcal P_{3A,0} =& \frac{5}{9} (520+\beta_0)-\frac{2600 X_A}{9},
    \nonumber\\
\mathcal P_{3A,2} =& \frac{5 (267520+995 \beta_0+168 \beta_2)}{1848}
    + \left(\frac{123350}{77}+\frac{85 \beta_0}{33}\right) X_A
    - \frac{85}{66} (1825+2\beta_0) X_A^2+\frac{1625 X_A^3}{66}, \nonumber\\
\mathcal P_{3A,3} =& -\frac{10}{3} (260+\beta_0) \pi +\frac{2600 \pi X_A}{3}
    + \left(\frac{25}{24} (260+3 \beta_0) X_A+\frac{5}{18} (195+7 \beta_0)X_A^2
    - 325 X_A^3\right) \chi _A \nonumber\\
    +& \left(\frac{5}{72} (8580+73 \beta_0)-\frac{5}{72}\left(21840
    + 101\beta_0\right) X_A+\frac{5}{18}(4485+7 \beta_0) X_A^2
    - 325 X_A^3\right) \chi _B, \nonumber\\
\mathcal P_{3A,4} =&
    \frac{25 (2247134272+6080015 \beta_0+1203552 \beta_2)}{13208832}
    + \frac{5 \beta_4}{13}+\left(\frac{1714971925}{412776}
    + \frac{84995\beta_0}{8736} + \frac{85 \beta_2}{39}\right) X_A \nonumber\\
    -& \frac{5 (-7215185+16319 \beta_0+11424 \beta_2) X_A^2}{26208}
    - \frac{5(6970815+9908 \beta_0) X_A^3}{3744}
    + \frac{5 (-182400+2477 \beta_0) X_A^4}{1872}+\frac{875 X_A^5}{72} \nonumber\\
    +& \left(\left(-\frac{825}{2}-\frac{285\beta_0}{104}\right) X_A^2
    + \frac{825 X_A^3}{2}+\bar Q_A \left(\left(-400
    - \frac{35 \beta_0}{13}\right) X_A^2+400 X_A^3\right)\right)\chi _A^2
    \nonumber\\
    +& \left(\left(-775-\frac{275 \beta_0}{52}\right) X_A+\left(1550
    + \frac{275 \beta_0}{52}\right) X_A^2-775 X_A^3\right)\chi _A \chi _B
    \nonumber\\
    +& \left[-\frac{825}{2}-\frac{285 \beta_0}{104}+\frac{15}{52} (4290
    + 19 \beta_0) X_A+\left(\frac{-2475}{2}-\frac{285 \beta_0}{104}\right)X_A^2
    + \frac{825 X_A^3}{2} \right.\nonumber\\
    &~~~+ \left.\bar Q_B \left(-400-\frac{35 \beta_0}{13}+\left(1200
    + \frac{70\beta_0}{13}\right) X_A+\left(-1200-\frac{35\beta_0}{13}\right)X_A^2
    + 400 X_A^3\right)\right] \chi _B^2, \nonumber\\
\mathcal P_{3A,5} =& -\frac{5 (2604260+17705 \beta_0+2688 \beta_2) \pi}{4704}
    - \frac{5 (516675+1687 \beta_0) \pi X_A}{1176}+\frac{5}{168} (90640
    + 241\beta_0) \pi X_A^2+\frac{95225 \pi X_A^3}{42} \nonumber\\
    +& \left[\frac{5 (3242620+49273 \beta_0+5040 \beta_2) X_A}{9408}
    + \frac{5(4691970+82637 \beta_0+4704 \beta_2) X_A^2}{14112}
    + \frac{5 (-221335+91 \beta_0) X_A^3}{7056} \right.\nonumber\\
    &~~~- \left.\frac{25 (98807+254 \beta_0)X_A^4}{1008}
    - \frac{392275 X_A^5}{504}\right] \chi _A \nonumber\\
    +& \left[\frac{5 (23393100+277897 \beta_0+24528 \beta_2)}{28224}
    - \frac{5 (44089360+337219\beta_0+33936 \beta_2) X_A}{28224}
    \right.\nonumber\\
    &~~~+ \left.\frac{5 (9394160-23497 \beta_0+4704 \beta_2)X_A^2}{14112}
    + \frac{5(-272870+11823 \beta_0) X_A^3}{2352}
    - \frac{25 (-68399+254 \beta_0) X_A^4}{1008}-\frac{392275 X_A^5}{504}\right]
    \chi_B.
\label{eq:T2PTidalFull}
\end{align}
\end{widetext}

\section{Model Parameters}\label{app:ModelParameters}

\begin{table*}
\centering
\def\arraystretch{1.5}
\setlength\tabcolsep{.7em}
\begin{tabular}{|c c | c c c c c | c|}
\hline
$x$ & $y$ & $a_0$ & $a_1$ & $a_2$ & $a_3$ & $a_4$ & Ref \\
\hline\hline
$\bar\lambda_2$ & $\ln\bar Q$ & 0.194 & 0.0936 & 0.0474 & $-4.21\times10^{-3}$ &
    $1.23\times10^{-4}$ & \cite{Yagi:2013ilq,Yagi:2013bca} \\\hline
$\bar\lambda_2$ & $\ln\bar I$ & 1.47 & 0.0817 & 0.0149 & $2.87\times10^{-4}$ &
    $-3.64\times10^{-5}$ & \cite{Yagi:2013ilq,Yagi:2013bca} \\\hline
$\bar\lambda_2$ & $\ln\bar\lambda_3$ & $-1.15$ & 1.18 & $-0.0251$ &
    $-1.31\times10^{-3}$ & $2.52\times10^{-5}$ & \cite{Yagi:2014mlr} \\\hline
$\bar\lambda_2$ & $X_A\bar\omega_{f2}$ & 0.1820 & $-6.836\times10^{-3}$ &
    $-4.196\times10^{-3}$ & $5.215\times10^{-4}$ & $-1.857\times10^{-5}$ &
    \cite{Chan:2014a} \\\hline
$\bar\lambda_3$ & $X_A\bar\omega_{f3}$ & 0.2245 & $-0.01500$ &
    $-1.412\times10^{-3}$ & $1.832\times10^{-4}$ & $-5.561\times10^{-6}$ &
    \cite{Chan:2014a} \\\hline
\end{tabular}
\par
\bigskip
\caption{Universal relations relating the static dimensionless deformability to
various other dimensionless tidal parameters using
Eq~(\ref{eq:UniversalRelationsFormula}).}
\label{tab:UniversalRelations}
\end{table*}

To serve as our underlying proxy for numerical BBH waveforms, we use the
hybridized surrogate model ``NRHybSur3dq8''~\cite{Varma:2018mmi}. This model
accurately captures the behavior of nonprecessing BBH systems for mass ratios up
to $q=8$ with component spins $\chi\leq0.8$ and including all of the following
modes: [(2, 0), (2, 1), (2, 2), (3, 2), (3, 0), (3, 1), (3, 3), (4, 2), (4, 3),
(4, 4), (5, 5)]. This region of parameter space spans the corresponding space of
BNS and BHNS systems because the expected breakup spin for a neutron star is
$\chi\sim0.7$, and the tidal effects rapidly diminish for mass ratios
significantly larger than unity (the leading-order tidal term in the Taylor
expansions goes as $X_A^4$).

The tidal effects of each object are dependent on the object's mass and the
choice of EOS. There are currently six tidal parameters that enter into our
model: the quadrupole and octopole static tidal deformabilities
$\bar\lambda_2$ and $\bar\lambda_3$, their corresponding $f$-mode resonant
frequencies $\bar\omega_{f2}$ and $\bar\omega_{f3}$, the dimensionless
rotationally induced quadrupole moment $\bar Q$, and the dimensionless moment of
inertia $\bar I$. (In Appendix~\ref{app:SpinTidal}, we also briefly discuss how
to include the four spin-tidal deformability parameters though they currently are
not a part of our model.)

While in general, all of these parameters depend on the specific details of the
object's mass and EOS, recent analysis of these numbers show that the various
parameters follow a series of universal relations that can accurately
approximate their values given just $\bar\lambda_2$. The universal relations we
use here all follow the same form~\cite{Yagi:2013ilq, Yagi:2013bca},
\begin{align}
y =& \sum_{i=0}^4a_i(\ln x)^i,
\label{eq:UniversalRelationsFormula}
\end{align}
where $x$ and $y$ are the two tidal parameters, and
$a_i$ are the numerically fitted coefficients relating
them (see Table~\ref{tab:UniversalRelations}).
Thus, the properties of the NS of different theoretical EOS could be represented
by how that EOS traces out the curve of allowable $\bar\lambda_2$ as a function
of the mass of the NS. The extra factor of $X_A$ that appears as part of the
dimensionless resonance frequencies $\bar\omega_{f\ell}$ in
Table~\ref{tab:UniversalRelations} arises because Ref.~\cite{Chan:2014a} defines
their dimensionless resonance frequency as $m_A\omega_{f\ell}$ whereas we use
$\bar\omega_{f\ell} = M\omega_{f\ell}$.

Utilizing these universal relations reduces the effective parameter space of the
tidal information from 12 (6 for each object) to just the static quadrupolar
tidal deformability for each object, since all tidal parameters
are derived simply from the
choice of $\bar\lambda_2$. All together our spliced waveforms effectively fill a
six-dimensional (6D) parameter space: $(q,M,\chi_A,\chi_B,\bar\lambda_{2A},\bar\lambda_{2B})$.

Equations~(\ref{eq:PNTidalFlux}) and~(\ref{eq:TidalStrainAmplification}) contain
various currently unknown higher-order coefficients ($\alpha_4$ from the
quadropolar tidal flux, $\beta_0, \beta_2, \beta_4$ from the octopolar tidal
flux, and $\alpha^{22}_{4}, \alpha^{21}_{2}, \alpha^{33}_{2}, \alpha^{31}_{2}$
from the tidal strain amplitude corrections) We set these coefficents to zero.
Further study is needed in order to characterize the error associated with such
a choice. In the case of the strain amplification corrections, we expect missing
coefficients to be subdominant contributions to the signal according to
Ref~\cite{damour:12}.

\section{Spin-Tidal Connection Terms}\label{app:SpinTidal}

In the expressions above, there are a number of terms that scale as
$\bar\lambda_{2A}\chi_{A}$, so one might be tempted to view them as connections
between the object's tidal deformation and spin. Tracing back to the original
energy and flux expressions, we know that instead these terms arise
naturally as a consequence of power counting in the series expansion, and are
merely cross terms between the tidal and spin-orbit or spin-spin effects.

Two different groups, Refs~\cite{Abdelsalhin:2018a,Landry:2018a},
derived the first spin-tidal connection terms in the PN expansion. However,
their two results are not consistent with each other, so we do not include
either within our current splicing model. We do look at each paper and summarize
how to implement either of these effects within our framework so that when this
discrepancy is resolved, tidal splicing can be updated to include these terms.

The leading-order spin-tidal terms all enter at the $v^{13}$ order. There are
four related effects that all enter at this order, each with its own dimensionless
tidal deformability coefficient: $\bar\lambda_{23}$, the mass quadrupole tidal
deformation arising due to the gravitomagnetic octopole tidal field;
$\bar\lambda_{32}$, the mass octopole tidal deformation arising due to the
gravitomagnetic quadrupole tidal field; $\bar\sigma_{23}$,
the current quadrupole
tidal deformation arising due to the gravitoelectric octopole tidal field; and
$\bar\sigma_{32}$, the current octopole tidal deformation arising due to the
gravitoelectric quadrupole tidal field.

To obtain the TaylorT terms, we examine leading-order terms in the PN energy and
flux, and find
\begin{align}
E_{\text{ST}} =& -\frac{\nu v^2}{2}\left(1 + v^{13}\chi_AX_A^6
    \sum_i\epsilon_{iA}\bar\Lambda_{iA} + (A\rightarrow B)\right),
\label{eq:EST}\\
F_{\text{ST}} =& \frac{32\nu^2v^{10}}{5}\left(1 + v^{13}\chi_AX_A^6
    \sum_i\rho_{iA}\bar\Lambda_{iA} + (A\rightarrow B)\right),
\label{eq:FST}
\end{align}
where the sums are over each of the ST parameters $\bar\Lambda_{iA} =
(\bar\lambda_{23A}, \bar\lambda_{32A}, \bar\sigma_{23A}, \bar\sigma_{32A})$,
with energy coefficients $\epsilon_{iA}$ and flux coefficients $\rho_{iA}$ that
are functions only of the mass fraction of the object $X_A$. In principle, we
would need to include these energy and flux corrections into the full energy and
flux equations, Eqs.~(\ref{eq:PNTidalEnergy}) and~(\ref{eq:PNTidalFlux}).

We can expand Eqs.~(\ref{eq:EST}) and~(\ref{eq:FST})
in both the TaylorT4 and TaylorT2 manners yielding
\begin{align}
\mathcal F_{\text{ST}}(v) =& \frac{32\nu v^9}{5M}\left[
    v^{13}\chi_AX_A^6\sum_i\left(-\frac{15}{2}\epsilon_{iA}
    + \rho_{iA}\right)\bar\Lambda_{iA}\right], \nonumber\\
\mathcal T_{\text{ST}}(v) =& -\frac{5M}{256\nu v^8}\left[
    v^{13}\chi_AX_A^6\sum_i\left(-12\epsilon_{iA}
    + \frac{8}{5}\rho_{iA}\right)\bar\Lambda_{iA}\right], \nonumber\\
\mathcal P_{\text{ST}}(v) =& -\frac{1}{32\nu v^5}\left[
    v^{13}\chi_AX_A^6\sum_i\left(-\frac{75}{16}\epsilon_{iA}
    + \frac{5}{8}\rho_{iA}\right)\bar\Lambda_{iA}\right].
\end{align}
We can linearly add these expressions to the full tidal PN expressions $\mathcal
F_{\text{Tid}}(v), \mathcal T_{\text{Tid}}(v)$, and $\mathcal
P_{\text{Tid}}(v)$, respectively.

We now compare the energy and flux coefficients $\epsilon_{iA}$
  and $\rho_{iA}$ as computed by Refs.~\cite{Abdelsalhin:2018a,
Landry:2018a}. In no particular order, we first examine~\cite{Landry:2018a},
  beginning with the correspondence between their (dimensionful) definitions
  of the
deformability parameters,
\begin{align}
\bar\lambda_{23} =& m_A^{-6}\hat\lambda_2, \nonumber\\
\bar\lambda_{32} =& m_A^{-6}\hat\lambda_3, \nonumber\\
\bar\sigma_{23} =& -m_A^{-6}\hat\sigma_2, \nonumber\\
\bar\sigma_{32} =& -m_A^{-6}\hat\sigma_3,
\end{align}
where the parameters of Ref.~\cite{Landry:2018a} are hatted.
From this we can write their energy and flux coefficients by reading off from
Eqs.~(28) and~(30) of~\cite{Landry:2018a},
\begin{align}
\epsilon_{iA} =& \frac{44(1-X_A)}{7X_A}\left(18,-2,-4,3\right), \nonumber\\
\rho_{iA} =& \left(144 - \frac{204}{X_A},-16 + \frac{16}{X_A},
    -38 + \frac{113}{3X_A}, 24 - \frac{24}{X_A} \right).
\end{align}

Repeating this same setup but with~\cite{Abdelsalhin:2018a}, we find the
correspondence between their definitions and ours as
\begin{align}
\bar\lambda_{23} =& \frac{M^2}{m_A^6}\lambda_{23}, \nonumber\\
\bar\lambda_{32} =& \frac{M^2}{m_A^6}\lambda_{32}, \nonumber\\
\bar\sigma_{23} =& \frac{M^2}{m_A^6}\sigma_{23}, \nonumber\\
\bar\sigma_{32} =& \frac{M^2}{m_A^6}\sigma_{32},
\end{align}
where the parameters of Ref.~\cite{Abdelsalhin:2018a} are unbarred. Using their
Eqs.~(90) and~(95),
\begin{align}
\epsilon_{iA} =& \frac{1-X_A}{X_A}\left(96,-32,-32,24\right), \nonumber\\
\rho_{iA} =& \left(96 - \frac{136}{X_A},-32 + \frac{32}{X_A},
    -38 + \frac{113}{3X_A},24 - \frac{24}{X_A} \right).
\end{align}

\section{$\bar\lambda_2\times\bar\lambda_2$ Self-Cross Terms}
\label{app:LambdaLambda}

In the tidal PN formulations in this paper, the leading-order tidal terms are
considered to be the same effective PN order as the Newtonian-order vacuum
terms. Or in other words, we view $\mathcal O(\bar\lambda_2v^{10})\sim\mathcal
O(1)${, so terms like $\bar\lambda_2v^{10}$ are treated as 0PN order, terms like
$\bar\lambda_2v^{12}$ are treated as 1PN order, and so on, for the purposes of
truncating the PN series. This means that terms like $\bar\lambda_2^2v^{20}$,
$\bar\lambda_2^3v^{30}$, and so on can also in principle be considered 0PN
order. So when multiplying or dividing PN series (for example, when dividing the
two series on the right-hand side of Eq.~(\ref{eq:EnergyBalanceT4})) and then
truncating the series to a consistent PN order, the question arises as to
whether one should eliminate terms of order $\mathcal O(\bar\lambda_2^2v^{20})$,
$O(\bar\lambda_2^3v^{30})$, and so on. In this paper we do eliminate such terms,
and we justify this choice in this appendix.

Note that the higher-order terms in question are not nonlinear static tidal
terms (where the deformed tidal field of one object perturbs the tidal
deformation of the other object) despite their appearance as such. These
$\bar\lambda_2\times\bar\lambda_2$ terms are simply linear cross terms, such as the
terms that go as $\bar\lambda_2\times\chi_{A,B}$ in Eqs.~(\ref{eq:T4TidalFull}),
(\ref{eq:T2TTidalFull}), and~(\ref{eq:T2PTidalFull}), which are cross terms
between spin-orbit and tidal contributions as opposed to spin-tidal effects.

Now while $v\rightarrow0$, the assumption that $\mathcal
O(\bar\lambda_2v^{10})\sim\mathcal O(1)$ is numerically not true, and instead we
have $\mathcal O(\bar\lambda_2v^{10})\ll\mathcal O(1)$, so one could argue to
neglect $\bar\lambda_2\times\bar\lambda_2$ terms. But as the binary approaches
merger, and $v$ is no longer small, then we cannot assume $\mathcal
O(\bar\lambda_2v^{10})\ll\mathcal O(1)$ holds true without further exploration.

To facilitate this discussion, we examine truncated versions of
the PN energy and flux equations keeping just the leading-order BBH and tidal
terms,
\begin{align}
E(v) =& -\frac{\nu v^2}{2}\left(1 + 9\bar\lambda_{2A}v^{10}X_A^4(-1 + X_A)
    \right.\nonumber\\
    +& \left.9\bar\lambda_{2B}v^{10}X_B^4(-1 + X_B) + \mathcal O(v)
    + \mathcal O(\bar\lambda_2v^{11})\right),\nonumber\\
F(v) =& \frac{32\nu^2v^{10}}{5}\left(1 + 6\bar\lambda_{2A}v^{10}X_A^4(3 - 2X_A)
    \right.\nonumber\\
    +& \left.6\bar\lambda_{2B}v^{10}X_B^4(3 - 2X_B) + \mathcal O(v)
    + \mathcal O(\bar\lambda_2v^{11})\right).\nonumber\\
\end{align}

We now perform the PN series expansion in both the TaylorT4 and the TaylorT2 manners
as in the rest of this paper, except we retain the next higher-order tidal cross
terms,
\begin{widetext}
\begin{align}
\mathcal F_{\bar\lambda_2\times\bar\lambda_2}(v) =& \frac{32\nu v^9}{5M}\left[
    1 + v^{10}\left(\bar\lambda_{2A}X_A^4(72 - 66X_A) + \bar\lambda_{2B}X_B^4(72
    - 66X_B)\right) + 324v^{20}\left(\bar\lambda_{2A}^2X_A^8\left(12 - 23X_A
    + 11X_A^2\right) \right.\right.\nonumber\\
    +& \left.\left.\bar\lambda_{2A}\bar\lambda_{2B}X_A^4X_B^4\left(24
    - 23(X_A+X_B) + 22X_AX_B\right) + \bar\lambda_{2B}^2X_B^8\left(12 - 23X_B
    + 11X_B^2\right)\right) \right.\nonumber\\
    +&\left. \mathcal O(v) + \mathcal O(\lambda_2v^{11})
    + \mathcal O(\lambda_2^3v^{30})\right], \nonumber\\
\mathcal T_{\bar\lambda_2\times\bar\lambda_2}(v) =&-\frac{5M}{256\nu v^8}\left[
    1 + v^{10}\left(\bar\lambda_{2A}X_A^4(288-264 X_A) + \bar\lambda_{2B}X_B^4(
    288 - 264 X_B)\right) + 24v^{20}\left(\bar\lambda_{2A}^2X_A^8\left(-36
    + 57X_A - 22X_A^2\right) \right.\right.\nonumber\\
    +&\left.\left. \bar\lambda_{2A}\bar\lambda_{2B}X_A^4X_B^4(-72 + 57(X_A+X_B)
    - 44X_AX_B) + \bar\lambda_{2B}^2X_B^8(-36 + 57X_B - 22X_B^2)\right)
    \right.\nonumber\\
    +&\left. \mathcal O(v) + \mathcal O(\lambda_2v^{11})
    + \mathcal O(\lambda_2^3v^{30})\right], \nonumber\\
\mathcal P_{\bar\lambda_2\times\bar\lambda_2}(v) =&-\frac{1}{32\nu v^5}\left[
    1 + v^{10}\left(\bar\lambda_{2A}X_A^4(72-66 X_A) + \bar\lambda_{2B}X_B^4(
    72 - 66 X_B)\right) + 12v^{20}\left(\bar\lambda_{2A}^2X_A^8\left(-36
    + 57X_A - 22X_A^2\right) \right.\right.\nonumber\\
    +&\left.\left. \bar\lambda_{2A}\bar\lambda_{2B}X_A^4X_B^4(-72 + 57(X_A+X_B)
    - 44X_AX_B) + \bar\lambda_{2B}^2X_B^8(-36 + 57X_B - 22X_B^2)\right)
    \right.\nonumber\\
    +&\left. \mathcal O(v) + \mathcal O(\lambda_2v^{11})
    + \mathcal O(\lambda_2^3v^{30})\right].
\label{eq:Lambda2Lambda2Cross}
\end{align}
\end{widetext}
Compare with Eqs.~(\ref{eq:T4Tidal1PN}), ~(\ref{eq:T2TTidal1PN}),
and~(\ref{eq:T2PTidal1PN}), or with the equations in
Appendix~\ref{sec:2.5pn-tidal-expr}.
  
\begin{table*}
\centering
\def\arraystretch{1.75}
\setlength\tabcolsep{.5em}
\begin{tabular}{| c | c c c c | c c c c |}
\hline
 & $\bar\omega_{\text{ISCO}}$ & $\bar\lambda_2v_{\text{ISCO}}^{10}$
    & $\bar\lambda_2v_{\text{ISCO}}^{15}$ & $\bar\lambda_2^2v_{\text{ISCO}}^{20}$
    & $\bar\omega_{\text{test}}$ & $\bar\lambda_2v_{\text{test}}^{10}$
    & $\bar\lambda_2v_{\text{test}}^{15}$ & $\bar\lambda_2^2v_{\text{test}}^{20}$ \\
\hline\hline
$\mathcal F_{\text{Tid}}(v)$ & $6^{-3/2}$ & 0.313465 &  0.0588958 & 0.0680262
    & $6^{-3/2}/5$ & $1.46652\times10^{-3}$ & $1.88467\times10^{-5}$
    & $1.48894\times10^{-6}$ \\
\hline
$\mathcal T_{\text{Tid}}(v)$ & $6^{-3/2}$ & 1.25386 & $-0.4286$ & $-0.0201559$
    & $6^{-3/2}/5$ & $5.86608\times10^{-3}$ & $-1.37152\times10^{-4}$
    & $-4.41166\times10^{-7}$ \\
\hline
$\mathcal P_{\text{Tid}}(v)$ & $6^{-3/2}$ & 0.313465 & $-0.187513$ & $-0.0100779$
    & $6^{-3/2}/5$ & $1.46652\times10^{-3}$ & $-6.0004\times10^{-5}$
    & $-2.20583\times10^{-7}$ \\
\hline
\end{tabular}
\par
\bigskip
\caption{Magnitudes of the PN tidal self-cross terms from
Eq.~(\ref{eq:Lambda2Lambda2Cross}), assuming a fiducial system where $q=1, \chi_A =
\chi_B = 0, \bar\lambda_{2A} = 1000, \bar\lambda_{2B} = 0$, evaluated at two
points in the inspiral, $\bar\omega_{\text{ISCO}}=6^{-3/2}$ and
$\bar\omega_{\text{test}}=\bar\omega_{\text{ISCO}}/5$.}
\label{tab:Lambda2Lambda2Cross}
\end{table*}

Because these cross terms will most strongly influence the waveform during the
final stages of the inspirals, we estimate the size of the
self-cross terms by comparing
the magnitude of the various terms here at two different frequencies
$\bar\omega_{\text{ISCO}}=6^{-3/2}$ and
$\bar\omega_{\text{test}}=\bar\omega_{\text{ISCO}}/5$. We assume a fiducial
binary system where $q=1, \chi_A = \chi_B = 0, \bar\lambda_{2A} = 1000,
\bar\lambda_{2B} = 0$. The results are displayed in
Table~\ref{tab:Lambda2Lambda2Cross}. The columns labeled
  $\bar\lambda_2v^{10}$ and $\bar\lambda_2^2v^{20}$ in
  Table~\ref{tab:Lambda2Lambda2Cross} correspond
  to the terms in Eqs~(\ref{eq:Lambda2Lambda2Cross}).
The columns labeled $\bar\lambda_2v^{15}$ correspond to the 2.5PN order
$\bar\lambda_2$ terms [$\mathcal F_{2A,5}$ from Eq.(\ref{eq:T4TidalFull}),
$\mathcal T_{2A,5}$ from Eq.(\ref{eq:T2TTidalFull}), and $\mathcal P_{2A,5}$
from Eq.(\ref{eq:T2PTidalFull})], which serve
as an error bound estimate arising from the unknown quadrupolar tidal terms. If
the self-cross terms are larger than these,
or near the size of the leading-order
tidal terms, then we should not neglect the self-cross terms.

At the end of the waveform (i.e., at $\bar\omega_{\text{ISCO}}$), in the TaylorT4
case, the size of the self-cross terms is an appreciable fraction of the
$\bar\lambda_2v^{10}$ term and the same size as the $\bar\lambda_2v^{15}$ term.
This is also true in TaylorT2, but to a lesser extent, as the self-cross terms
there are a factor of a few smaller. However, looking just a bit earlier in the
inspiral we find that the self-cross terms drop off until they are distinctly smaller
than even $\bar\lambda_2v^{15}$. Our general conclusion is that we are fine
neglecting the self-cross terms for the time being, but as more tidal terms are
introduced, we will eventually need to include these terms, especially to get the last
orbits of the inspiral before the start of merger/ringdown.

This entire argument generalizes to the octopolar static tides,
where the equivalent assumption is $\mathcal
O(1)\sim\mathcal O(\bar\lambda_3v^{14})\sim\mathcal O(\bar\lambda_3^2v^{28})$,
yet the octopolar effects are suppressed compared to the quadrupolar static
tides. All of the arguments in favor of neglecting the $\mathcal
O(\bar\lambda_2^2v^{20})$ terms should apply even more strongly to $\mathcal
O(\bar\lambda_3^2v^{28})$, and so we ignore the $\mathcal
O(\bar\lambda_3^2v^{28})$ terms as well.

\bibliography{References/References}

\begin{thebibliography}{89}%
\makeatletter
\providecommand \@ifxundefined [1]{%
 \@ifx{#1\undefined}
}%
\providecommand \@ifnum [1]{%
 \ifnum #1\expandafter \@firstoftwo
 \else \expandafter \@secondoftwo
 \fi
}%
\providecommand \@ifx [1]{%
 \ifx #1\expandafter \@firstoftwo
 \else \expandafter \@secondoftwo
 \fi
}%
\providecommand \natexlab [1]{#1}%
\providecommand \enquote  [1]{``#1''}%
\providecommand \bibnamefont  [1]{#1}%
\providecommand \bibfnamefont [1]{#1}%
\providecommand \citenamefont [1]{#1}%
\providecommand \href@noop [0]{\@secondoftwo}%
\providecommand \href [0]{\begingroup \@sanitize@url \@href}%
\providecommand \@href[1]{\@@startlink{#1}\@@href}%
\providecommand \@@href[1]{\endgroup#1\@@endlink}%
\providecommand \@sanitize@url [0]{\catcode `\\12\catcode `\$12\catcode
  `\&12\catcode `\#12\catcode `\^12\catcode `\_12\catcode `\%12\relax}%
\providecommand \@@startlink[1]{}%
\providecommand \@@endlink[0]{}%
\providecommand \url  [0]{\begingroup\@sanitize@url \@url }%
\providecommand \@url [1]{\endgroup\@href {#1}{\urlprefix }}%
\providecommand \urlprefix  [0]{URL }%
\providecommand \Eprint [0]{\href }%
\providecommand \doibase [0]{http://dx.doi.org/}%
\providecommand \selectlanguage [0]{\@gobble}%
\providecommand \bibinfo  [0]{\@secondoftwo}%
\providecommand \bibfield  [0]{\@secondoftwo}%
\providecommand \translation [1]{[#1]}%
\providecommand \BibitemOpen [0]{}%
\providecommand \bibitemStop [0]{}%
\providecommand \bibitemNoStop [0]{.\EOS\space}%
\providecommand \EOS [0]{\spacefactor3000\relax}%
\providecommand \BibitemShut  [1]{\csname bibitem#1\endcsname}%
\let\auto@bib@innerbib\@empty
\bibitem [{\citenamefont {Aasi}\ \emph {et~al.}(2015)\citenamefont {Aasi} \emph
  {et~al.}}]{aLIGO2}%
  \BibitemOpen
  \bibfield  {author} {\bibinfo {author} {\bibfnamefont {J.}~\bibnamefont
  {Aasi}} \emph {et~al.} (\bibinfo {collaboration} {LIGO Scientific
  Collaboration}),\ }\href {\doibase 10.1088/0264-9381/32/7/074001} {\bibfield
  {journal} {\bibinfo  {journal} {Class.\ Quantum Grav.}\ }\textbf {\bibinfo
  {volume} {32}},\ \bibinfo {pages} {074001} (\bibinfo {year} {2015})},\
  \Eprint {http://arxiv.org/abs/1411.4547} {arXiv:1411.4547 [gr-qc]}
  \BibitemShut {NoStop}%
\bibitem [{\citenamefont {Acernese}\ \emph {et~al.}(2015)\citenamefont
  {Acernese} \emph {et~al.}}]{aVirgo2}%
  \BibitemOpen
  \bibfield  {author} {\bibinfo {author} {\bibfnamefont {F.}~\bibnamefont
  {Acernese}} \emph {et~al.} (\bibinfo {collaboration} {Virgo Collaboration}),\
  }\href {\doibase 10.1088/0264-9381/32/2/024001} {\bibfield  {journal}
  {\bibinfo  {journal} {Class.\ Quantum Grav.}\ }\textbf {\bibinfo {volume}
  {32}},\ \bibinfo {pages} {024001} (\bibinfo {year} {2015})},\ \Eprint
  {http://arxiv.org/abs/1408.3978} {arXiv:1408.3978 [gr-qc]} \BibitemShut
  {NoStop}%
\bibitem [{\citenamefont {Abbott}\ \emph {et~al.}(2017)\citenamefont {Abbott}
  \emph {et~al.}}]{TheLIGOScientific:2017qsa}%
  \BibitemOpen
  \bibfield  {author} {\bibinfo {author} {\bibfnamefont {B.~P.}\ \bibnamefont
  {Abbott}} \emph {et~al.} (\bibinfo {collaboration} {Virgo, LIGO
  Scientific}),\ }\href {\doibase 10.1103/PhysRevLett.119.161101} {\bibfield
  {journal} {\bibinfo  {journal} {Phys. Rev. Lett.}\ }\textbf {\bibinfo
  {volume} {119}},\ \bibinfo {pages} {161101} (\bibinfo {year} {2017})},\
  \Eprint {http://arxiv.org/abs/1710.05832} {arXiv:1710.05832 [gr-qc]}
  \BibitemShut {NoStop}%
\bibitem [{\citenamefont {{Abbott}}\ \emph {et~al.}(2017)\citenamefont
  {{Abbott}}, \citenamefont {{Abbott}}, \citenamefont {{Abbott}}, \citenamefont
  {{Acernese}}, \citenamefont {{Ackley}}, \citenamefont {{Adams}},
  \citenamefont {{Adams}}, \citenamefont {{Addesso}}, \citenamefont
  {{Adhikari}}, \citenamefont {{Adya}},\ and\ \citenamefont
  {et~al.}}]{2017ApJ...848L..13A}%
  \BibitemOpen
  \bibfield  {author} {\bibinfo {author} {\bibfnamefont {B.~P.}\ \bibnamefont
  {{Abbott}}}, \bibinfo {author} {\bibfnamefont {R.}~\bibnamefont {{Abbott}}},
  \bibinfo {author} {\bibfnamefont {T.~D.}\ \bibnamefont {{Abbott}}}, \bibinfo
  {author} {\bibfnamefont {F.}~\bibnamefont {{Acernese}}}, \bibinfo {author}
  {\bibfnamefont {K.}~\bibnamefont {{Ackley}}}, \bibinfo {author}
  {\bibfnamefont {C.}~\bibnamefont {{Adams}}}, \bibinfo {author} {\bibfnamefont
  {T.}~\bibnamefont {{Adams}}}, \bibinfo {author} {\bibfnamefont
  {P.}~\bibnamefont {{Addesso}}}, \bibinfo {author} {\bibfnamefont {R.~X.}\
  \bibnamefont {{Adhikari}}}, \bibinfo {author} {\bibfnamefont {V.~B.}\
  \bibnamefont {{Adya}}}, \ and\ \bibinfo {author} {\bibnamefont {et~al.}},\
  }\href {\doibase 10.3847/2041-8213/aa920c} {\bibfield  {journal} {\bibinfo
  {journal} {"Astrophys.\ J. \ Lett."}\ }\textbf {\bibinfo {volume} {848}},\
  \bibinfo {eid} {L13} (\bibinfo {year} {2017})},\ \Eprint
  {http://arxiv.org/abs/1710.05834} {arXiv:1710.05834 [astro-ph.HE]}
  \BibitemShut {NoStop}%
\bibitem [{\citenamefont {Radice}\ \emph {et~al.}(2018)\citenamefont {Radice},
  \citenamefont {Perego}, \citenamefont {Zappa},\ and\ \citenamefont
  {Bernuzzi}}]{2017ApJL2041}%
  \BibitemOpen
  \bibfield  {author} {\bibinfo {author} {\bibfnamefont {D.}~\bibnamefont
  {Radice}}, \bibinfo {author} {\bibfnamefont {A.}~\bibnamefont {Perego}},
  \bibinfo {author} {\bibfnamefont {F.}~\bibnamefont {Zappa}}, \ and\ \bibinfo
  {author} {\bibfnamefont {S.}~\bibnamefont {Bernuzzi}},\ }\href
  {http://stacks.iop.org/2041-8205/852/i=2/a=L29} {\bibfield  {journal}
  {\bibinfo  {journal} {The Astrophysical Journal Letters}\ }\textbf {\bibinfo
  {volume} {852}},\ \bibinfo {pages} {L29} (\bibinfo {year} {2018})},\ \Eprint
  {http://arxiv.org/abs/1711.03647} {arXiv:1711.03647 [astro-ph.HE]}
  \BibitemShut {NoStop}%
\bibitem [{\citenamefont {{Del Pozzo}}\ \emph {et~al.}(2013)\citenamefont {{Del
  Pozzo}}, \citenamefont {{Li}}, \citenamefont {{Agathos}}, \citenamefont {{Van
  Den Broeck}},\ and\ \citenamefont {{Vitale}}}]{DelPozzo:13}%
  \BibitemOpen
  \bibfield  {author} {\bibinfo {author} {\bibfnamefont {W.}~\bibnamefont {{Del
  Pozzo}}}, \bibinfo {author} {\bibfnamefont {T.~G.~F.}\ \bibnamefont {{Li}}},
  \bibinfo {author} {\bibfnamefont {M.}~\bibnamefont {{Agathos}}}, \bibinfo
  {author} {\bibfnamefont {C.}~\bibnamefont {{Van Den Broeck}}}, \ and\
  \bibinfo {author} {\bibfnamefont {S.}~\bibnamefont {{Vitale}}},\ }\href
  {\doibase 10.1103/PhysRevLett.111.071101} {\bibfield  {journal} {\bibinfo
  {journal} {Phys.\ Rev.\ Lett.}\ }\textbf {\bibinfo {volume} {111}},\ \bibinfo
  {eid} {071101} (\bibinfo {year} {2013})},\ \Eprint
  {http://arxiv.org/abs/1307.8338} {arXiv:1307.8338 [gr-qc]} \BibitemShut
  {NoStop}%
\bibitem [{\citenamefont {Kumar}\ \emph {et~al.}(2017)\citenamefont {Kumar},
  \citenamefont {Pürrer},\ and\ \citenamefont {Pfeiffer}}]{Kumar:2016zlj}%
  \BibitemOpen
  \bibfield  {author} {\bibinfo {author} {\bibfnamefont {P.}~\bibnamefont
  {Kumar}}, \bibinfo {author} {\bibfnamefont {M.}~\bibnamefont {Pürrer}}, \
  and\ \bibinfo {author} {\bibfnamefont {H.~P.}\ \bibnamefont {Pfeiffer}},\
  }\href {\doibase 10.1103/PhysRevD.95.044039} {\bibfield  {journal} {\bibinfo
  {journal} {Phys. Rev.}\ }\textbf {\bibinfo {volume} {D95}},\ \bibinfo {pages}
  {044039} (\bibinfo {year} {2017})},\ \Eprint
  {http://arxiv.org/abs/1610.06155} {arXiv:1610.06155 [gr-qc]} \BibitemShut
  {NoStop}%
\bibitem [{\citenamefont {{Vines}}\ \emph {et~al.}(2011)\citenamefont
  {{Vines}}, \citenamefont {{Flanagan}},\ and\ \citenamefont
  {{Hinderer}}}]{Vines2011}%
  \BibitemOpen
  \bibfield  {author} {\bibinfo {author} {\bibfnamefont {J.}~\bibnamefont
  {{Vines}}}, \bibinfo {author} {\bibfnamefont {{\'E}.~{\'E}.}\ \bibnamefont
  {{Flanagan}}}, \ and\ \bibinfo {author} {\bibfnamefont {T.}~\bibnamefont
  {{Hinderer}}},\ }\href {\doibase 10.1103/PhysRevD.83.084051} {\bibfield
  {journal} {\bibinfo  {journal} {Phys.\ Rev.\ D}\ }\textbf {\bibinfo {volume}
  {83}},\ \bibinfo {eid} {084051} (\bibinfo {year} {2011})},\ \Eprint
  {http://arxiv.org/abs/1101.1673} {arXiv:1101.1673 [gr-qc]} \BibitemShut
  {NoStop}%
\bibitem [{\citenamefont {Chakravarti}\ \emph {et~al.}(2019)\citenamefont
  {Chakravarti} \emph {et~al.}}]{Chakravarti:2018uyi}%
  \BibitemOpen
  \bibfield  {author} {\bibinfo {author} {\bibfnamefont {K.}~\bibnamefont
  {Chakravarti}} \emph {et~al.},\ }\href {\doibase 10.1103/PhysRevD.99.024049}
  {\bibfield  {journal} {\bibinfo  {journal} {Phys. Rev.}\ }\textbf {\bibinfo
  {volume} {D99}},\ \bibinfo {pages} {024049} (\bibinfo {year} {2019})},\
  \Eprint {http://arxiv.org/abs/1809.04349} {arXiv:1809.04349 [gr-qc]}
  \BibitemShut {NoStop}%
\bibitem [{\citenamefont {Samajdar}\ and\ \citenamefont
  {Dietrich}(2018)}]{Samajdar:2018sd}%
  \BibitemOpen
  \bibfield  {author} {\bibinfo {author} {\bibfnamefont {A.}~\bibnamefont
  {Samajdar}}\ and\ \bibinfo {author} {\bibfnamefont {T.}~\bibnamefont
  {Dietrich}},\ }\href {\doibase 10.1103/PhysRevD.98.124030} {\bibfield
  {journal} {\bibinfo  {journal} {Phys. Rev. D}\ }\textbf {\bibinfo {volume}
  {98}},\ \bibinfo {pages} {124030} (\bibinfo {year} {2018})},\ \Eprint
  {http://arxiv.org/abs/1810.03936} {arXiv:1810.03936 [gr-qc]} \BibitemShut
  {NoStop}%
\bibitem [{\citenamefont {Messina}\ \emph {et~al.}(2019)\citenamefont
  {Messina}, \citenamefont {Dudi}, \citenamefont {Nagar},\ and\ \citenamefont
  {Bernuzzi}}]{Messina:2019qta}%
  \BibitemOpen
  \bibfield  {author} {\bibinfo {author} {\bibfnamefont {F.}~\bibnamefont
  {Messina}}, \bibinfo {author} {\bibfnamefont {R.}~\bibnamefont {Dudi}},
  \bibinfo {author} {\bibfnamefont {A.}~\bibnamefont {Nagar}}, \ and\ \bibinfo
  {author} {\bibfnamefont {S.}~\bibnamefont {Bernuzzi}},\ }\href {\doibase
  10.1103/PhysRevD.99.124051} {\bibfield  {journal} {\bibinfo  {journal} {Phys.
  Rev. D}\ }\textbf {\bibinfo {volume} {99}},\ \bibinfo {pages} {124051}
  (\bibinfo {year} {2019})},\ \Eprint {http://arxiv.org/abs/1904.09558}
  {arXiv:1904.09558 [gr-qc]} \BibitemShut {NoStop}%
\bibitem [{\citenamefont {Damour}\ and\ \citenamefont
  {Nagar}(2010)}]{Damour:2009wj}%
  \BibitemOpen
  \bibfield  {author} {\bibinfo {author} {\bibfnamefont {T.}~\bibnamefont
  {Damour}}\ and\ \bibinfo {author} {\bibfnamefont {A.}~\bibnamefont {Nagar}},\
  }\href {\doibase 10.1103/PhysRevD.81.084016} {\bibfield  {journal} {\bibinfo
  {journal} {Phys.\ Rev.\ D}\ }\textbf {\bibinfo {volume} {81}},\ \bibinfo
  {pages} {084016} (\bibinfo {year} {2010})},\ \Eprint
  {http://arxiv.org/abs/0911.5041} {arXiv:0911.5041 [gr-qc]} \BibitemShut
  {NoStop}%
\bibitem [{\citenamefont {Hinderer}\ \emph {et~al.}(2016)\citenamefont
  {Hinderer} \emph {et~al.}}]{Hinderer:2016eia}%
  \BibitemOpen
  \bibfield  {author} {\bibinfo {author} {\bibfnamefont {T.}~\bibnamefont
  {Hinderer}} \emph {et~al.},\ }\href {\doibase 10.1103/PhysRevLett.116.181101}
  {\bibfield  {journal} {\bibinfo  {journal} {Phys. Rev. Lett.}\ }\textbf
  {\bibinfo {volume} {116}},\ \bibinfo {pages} {181101} (\bibinfo {year}
  {2016})},\ \Eprint {http://arxiv.org/abs/1602.00599} {arXiv:1602.00599
  [gr-qc]} \BibitemShut {NoStop}%
\bibitem [{\citenamefont {Steinhoff}\ \emph {et~al.}(2016)\citenamefont
  {Steinhoff}, \citenamefont {Hinderer}, \citenamefont {Buonanno},\ and\
  \citenamefont {Taracchini}}]{Steinhoff:2016rfi}%
  \BibitemOpen
  \bibfield  {author} {\bibinfo {author} {\bibfnamefont {J.}~\bibnamefont
  {Steinhoff}}, \bibinfo {author} {\bibfnamefont {T.}~\bibnamefont {Hinderer}},
  \bibinfo {author} {\bibfnamefont {A.}~\bibnamefont {Buonanno}}, \ and\
  \bibinfo {author} {\bibfnamefont {A.}~\bibnamefont {Taracchini}},\ }\href
  {\doibase 10.1103/PhysRevD.94.104028} {\bibfield  {journal} {\bibinfo
  {journal} {Phys. Rev.}\ }\textbf {\bibinfo {volume} {D94}},\ \bibinfo {pages}
  {104028} (\bibinfo {year} {2016})},\ \Eprint
  {http://arxiv.org/abs/1608.01907} {arXiv:1608.01907 [gr-qc]} \BibitemShut
  {NoStop}%
\bibitem [{\citenamefont {Boh\'e}\ \emph {et~al.}(2017)\citenamefont {Boh\'e},
  \citenamefont {Shao}, \citenamefont {Taracchini}, \citenamefont {Buonanno},
  \citenamefont {Babak}, \citenamefont {Harry}, \citenamefont {Hinder},
  \citenamefont {Ossokine}, \citenamefont {P\"urrer}, \citenamefont {Raymond},
  \citenamefont {Chu}, \citenamefont {Fong}, \citenamefont {Kumar},
  \citenamefont {Pfeiffer}, \citenamefont {Boyle}, \citenamefont {Hemberger},
  \citenamefont {Kidder}, \citenamefont {Lovelace}, \citenamefont {Scheel},\
  and\ \citenamefont {Szil\'agyi}}]{Bohe:2016gbl}%
  \BibitemOpen
  \bibfield  {author} {\bibinfo {author} {\bibfnamefont {A.}~\bibnamefont
  {Boh\'e}}, \bibinfo {author} {\bibfnamefont {L.}~\bibnamefont {Shao}},
  \bibinfo {author} {\bibfnamefont {A.}~\bibnamefont {Taracchini}}, \bibinfo
  {author} {\bibfnamefont {A.}~\bibnamefont {Buonanno}}, \bibinfo {author}
  {\bibfnamefont {S.}~\bibnamefont {Babak}}, \bibinfo {author} {\bibfnamefont
  {I.~W.}\ \bibnamefont {Harry}}, \bibinfo {author} {\bibfnamefont
  {I.}~\bibnamefont {Hinder}}, \bibinfo {author} {\bibfnamefont
  {S.}~\bibnamefont {Ossokine}}, \bibinfo {author} {\bibfnamefont
  {M.}~\bibnamefont {P\"urrer}}, \bibinfo {author} {\bibfnamefont
  {V.}~\bibnamefont {Raymond}}, \bibinfo {author} {\bibfnamefont
  {T.}~\bibnamefont {Chu}}, \bibinfo {author} {\bibfnamefont {H.}~\bibnamefont
  {Fong}}, \bibinfo {author} {\bibfnamefont {P.}~\bibnamefont {Kumar}},
  \bibinfo {author} {\bibfnamefont {H.~P.}\ \bibnamefont {Pfeiffer}}, \bibinfo
  {author} {\bibfnamefont {M.}~\bibnamefont {Boyle}}, \bibinfo {author}
  {\bibfnamefont {D.~A.}\ \bibnamefont {Hemberger}}, \bibinfo {author}
  {\bibfnamefont {L.~E.}\ \bibnamefont {Kidder}}, \bibinfo {author}
  {\bibfnamefont {G.}~\bibnamefont {Lovelace}}, \bibinfo {author}
  {\bibfnamefont {M.~A.}\ \bibnamefont {Scheel}}, \ and\ \bibinfo {author}
  {\bibfnamefont {B.}~\bibnamefont {Szil\'agyi}},\ }\href {\doibase
  10.1103/PhysRevD.95.044028} {\bibfield  {journal} {\bibinfo  {journal} {Phys.
  Rev. D}\ }\textbf {\bibinfo {volume} {95}},\ \bibinfo {pages} {044028}
  (\bibinfo {year} {2017})},\ \Eprint {http://arxiv.org/abs/1611.03703}
  {arXiv:1611.03703 [gr-qc]} \BibitemShut {NoStop}%
\bibitem [{\citenamefont {Bini}\ \emph {et~al.}(2012)\citenamefont {Bini},
  \citenamefont {Damour},\ and\ \citenamefont {Faye}}]{Bini:2012gu}%
  \BibitemOpen
  \bibfield  {author} {\bibinfo {author} {\bibfnamefont {D.}~\bibnamefont
  {Bini}}, \bibinfo {author} {\bibfnamefont {T.}~\bibnamefont {Damour}}, \ and\
  \bibinfo {author} {\bibfnamefont {G.}~\bibnamefont {Faye}},\ }\href {\doibase
  10.1103/PhysRevD.85.124034} {\bibfield  {journal} {\bibinfo  {journal}
  {Phys.\ Rev.\ D}\ }\textbf {\bibinfo {volume} {85}},\ \bibinfo {pages}
  {124034} (\bibinfo {year} {2012})},\ \Eprint {http://arxiv.org/abs/1202.3565}
  {arXiv:1202.3565 [gr-qc]} \BibitemShut {NoStop}%
\bibitem [{\citenamefont {{Baiotti}}\ \emph {et~al.}(2011)\citenamefont
  {{Baiotti}}, \citenamefont {{Damour}}, \citenamefont {{Giacomazzo}},
  \citenamefont {{Nagar}},\ and\ \citenamefont {{Rezzolla}}}]{Baiotti2011}%
  \BibitemOpen
  \bibfield  {author} {\bibinfo {author} {\bibfnamefont {L.}~\bibnamefont
  {{Baiotti}}}, \bibinfo {author} {\bibfnamefont {T.}~\bibnamefont {{Damour}}},
  \bibinfo {author} {\bibfnamefont {B.}~\bibnamefont {{Giacomazzo}}}, \bibinfo
  {author} {\bibfnamefont {A.}~\bibnamefont {{Nagar}}}, \ and\ \bibinfo
  {author} {\bibfnamefont {L.}~\bibnamefont {{Rezzolla}}},\ }\href {\doibase
  10.1103/PhysRevD.84.024017} {\bibfield  {journal} {\bibinfo  {journal}
  {Phys.\ Rev.\ D}\ }\textbf {\bibinfo {volume} {84}},\ \bibinfo {eid} {024017}
  (\bibinfo {year} {2011})},\ \Eprint {http://arxiv.org/abs/1103.3874}
  {arXiv:1103.3874 [gr-qc]} \BibitemShut {NoStop}%
\bibitem [{\citenamefont {Dietrich}\ and\ \citenamefont
  {Hinderer}(2017)}]{Dietrich:2017dh}%
  \BibitemOpen
  \bibfield  {author} {\bibinfo {author} {\bibfnamefont {T.}~\bibnamefont
  {Dietrich}}\ and\ \bibinfo {author} {\bibfnamefont {T.}~\bibnamefont
  {Hinderer}},\ }\href {\doibase 10.1103/PhysRevD.95.124006} {\bibfield
  {journal} {\bibinfo  {journal} {Phys. Rev. D}\ }\textbf {\bibinfo {volume}
  {95}},\ \bibinfo {pages} {124006} (\bibinfo {year} {2017})},\ \Eprint
  {http://arxiv.org/abs/1702.02053} {arXiv:1702.02053 [gr-qc]} \BibitemShut
  {NoStop}%
\bibitem [{\citenamefont {Nagar}\ \emph {et~al.}(2018)\citenamefont {Nagar}
  \emph {et~al.}}]{Nagar:2018zoe}%
  \BibitemOpen
  \bibfield  {author} {\bibinfo {author} {\bibfnamefont {A.}~\bibnamefont
  {Nagar}} \emph {et~al.},\ }\href {\doibase 10.1103/PhysRevD.98.104052}
  {\bibfield  {journal} {\bibinfo  {journal} {Phys. Rev. D}\ }\textbf {\bibinfo
  {volume} {98}},\ \bibinfo {pages} {104052} (\bibinfo {year} {2018})},\
  \Eprint {http://arxiv.org/abs/1806.01772} {arXiv:1806.01772 [gr-qc]}
  \BibitemShut {NoStop}%
\bibitem [{\citenamefont {Nagar}\ \emph {et~al.}(2019)\citenamefont {Nagar},
  \citenamefont {Messina}, \citenamefont {Rettegno}, \citenamefont {Bini},
  \citenamefont {Damour}, \citenamefont {Geralico}, \citenamefont {Akcay},\
  and\ \citenamefont {Bernuzzi}}]{Nagar:2018plt}%
  \BibitemOpen
  \bibfield  {author} {\bibinfo {author} {\bibfnamefont {A.}~\bibnamefont
  {Nagar}}, \bibinfo {author} {\bibfnamefont {F.}~\bibnamefont {Messina}},
  \bibinfo {author} {\bibfnamefont {P.}~\bibnamefont {Rettegno}}, \bibinfo
  {author} {\bibfnamefont {D.}~\bibnamefont {Bini}}, \bibinfo {author}
  {\bibfnamefont {T.}~\bibnamefont {Damour}}, \bibinfo {author} {\bibfnamefont
  {A.}~\bibnamefont {Geralico}}, \bibinfo {author} {\bibfnamefont
  {S.}~\bibnamefont {Akcay}}, \ and\ \bibinfo {author} {\bibfnamefont
  {S.}~\bibnamefont {Bernuzzi}},\ }\href {\doibase 10.1103/PhysRevD.99.044007}
  {\bibfield  {journal} {\bibinfo  {journal} {Phys. Rev. D}\ }\textbf {\bibinfo
  {volume} {99}},\ \bibinfo {pages} {044007} (\bibinfo {year} {2019})},\
  \Eprint {http://arxiv.org/abs/1812.07923} {arXiv:1812.07923 [gr-qc]}
  \BibitemShut {NoStop}%
\bibitem [{\citenamefont {Lackey}\ \emph {et~al.}(2014)\citenamefont {Lackey},
  \citenamefont {Kyutoku}, \citenamefont {Shibata}, \citenamefont {Brady},\
  and\ \citenamefont {Friedman}}]{Lackey:2013axa}%
  \BibitemOpen
  \bibfield  {author} {\bibinfo {author} {\bibfnamefont {B.~D.}\ \bibnamefont
  {Lackey}}, \bibinfo {author} {\bibfnamefont {K.}~\bibnamefont {Kyutoku}},
  \bibinfo {author} {\bibfnamefont {M.}~\bibnamefont {Shibata}}, \bibinfo
  {author} {\bibfnamefont {P.~R.}\ \bibnamefont {Brady}}, \ and\ \bibinfo
  {author} {\bibfnamefont {J.~L.}\ \bibnamefont {Friedman}},\ }\href {\doibase
  10.1103/PhysRevD.89.043009} {\bibfield  {journal} {\bibinfo  {journal}
  {Phys.\ Rev.\ D}\ }\textbf {\bibinfo {volume} {89}},\ \bibinfo {pages}
  {043009} (\bibinfo {year} {2014})},\ \Eprint
  {http://arxiv.org/abs/arXiv:1303.6298 [gr-qc]} {arXiv:1303.6298 [gr-qc]}
  \BibitemShut {NoStop}%
\bibitem [{\citenamefont {Taracchini}\ \emph {et~al.}(2014)\citenamefont
  {Taracchini} \emph {et~al.}}]{Taracchini:2013rva}%
  \BibitemOpen
  \bibfield  {author} {\bibinfo {author} {\bibfnamefont {A.}~\bibnamefont
  {Taracchini}} \emph {et~al.},\ }\href {\doibase 10.1103/PhysRevD.89.061502}
  {\bibfield  {journal} {\bibinfo  {journal} {Phys. Rev.}\ }\textbf {\bibinfo
  {volume} {D89}},\ \bibinfo {pages} {061502} (\bibinfo {year} {2014})},\
  \Eprint {http://arxiv.org/abs/1311.2544} {arXiv:1311.2544 [gr-qc]}
  \BibitemShut {NoStop}%
\bibitem [{\citenamefont {Dietrich}\ \emph
  {et~al.}(2019{\natexlab{a}})\citenamefont {Dietrich}, \citenamefont {Khan},
  \citenamefont {Dudi}, \citenamefont {Kapadia}, \citenamefont {Kumar},
  \citenamefont {Nagar}, \citenamefont {Ohme}, \citenamefont {Pannarale},
  \citenamefont {Samajdar}, \citenamefont {Bernuzzi}, \citenamefont {Carullo},
  \citenamefont {Del~Pozzo}, \citenamefont {Haney}, \citenamefont {Markakis},
  \citenamefont {P\"urrer}, \citenamefont {Riemenschneider}, \citenamefont
  {Setyawati}, \citenamefont {Tsang},\ and\ \citenamefont {Van
  Den~Broeck}}]{Dietrich:2018nrt}%
  \BibitemOpen
  \bibfield  {author} {\bibinfo {author} {\bibfnamefont {T.}~\bibnamefont
  {Dietrich}}, \bibinfo {author} {\bibfnamefont {S.}~\bibnamefont {Khan}},
  \bibinfo {author} {\bibfnamefont {R.}~\bibnamefont {Dudi}}, \bibinfo {author}
  {\bibfnamefont {S.~J.}\ \bibnamefont {Kapadia}}, \bibinfo {author}
  {\bibfnamefont {P.}~\bibnamefont {Kumar}}, \bibinfo {author} {\bibfnamefont
  {A.}~\bibnamefont {Nagar}}, \bibinfo {author} {\bibfnamefont
  {F.}~\bibnamefont {Ohme}}, \bibinfo {author} {\bibfnamefont {F.}~\bibnamefont
  {Pannarale}}, \bibinfo {author} {\bibfnamefont {A.}~\bibnamefont {Samajdar}},
  \bibinfo {author} {\bibfnamefont {S.}~\bibnamefont {Bernuzzi}}, \bibinfo
  {author} {\bibfnamefont {G.}~\bibnamefont {Carullo}}, \bibinfo {author}
  {\bibfnamefont {W.}~\bibnamefont {Del~Pozzo}}, \bibinfo {author}
  {\bibfnamefont {M.}~\bibnamefont {Haney}}, \bibinfo {author} {\bibfnamefont
  {C.}~\bibnamefont {Markakis}}, \bibinfo {author} {\bibfnamefont
  {M.}~\bibnamefont {P\"urrer}}, \bibinfo {author} {\bibfnamefont
  {G.}~\bibnamefont {Riemenschneider}}, \bibinfo {author} {\bibfnamefont
  {Y.~E.}\ \bibnamefont {Setyawati}}, \bibinfo {author} {\bibfnamefont {K.~W.}\
  \bibnamefont {Tsang}}, \ and\ \bibinfo {author} {\bibfnamefont
  {C.}~\bibnamefont {Van Den~Broeck}},\ }\href {\doibase
  10.1103/PhysRevD.99.024029} {\bibfield  {journal} {\bibinfo  {journal} {Phys.
  Rev. D}\ }\textbf {\bibinfo {volume} {99}},\ \bibinfo {pages} {024029}
  (\bibinfo {year} {2019}{\natexlab{a}})},\ \Eprint
  {http://arxiv.org/abs/1804.02235} {arXiv:1804.02235 [gr-qc]} \BibitemShut
  {NoStop}%
\bibitem [{\citenamefont {Husa}\ \emph {et~al.}(2016)\citenamefont {Husa},
  \citenamefont {Khan}, \citenamefont {Hannam}, \citenamefont {P{\"u}rrer},
  \citenamefont {Ohme}, \citenamefont {Jim{\'e}nez~Forteza},\ and\
  \citenamefont {Boh{\'e}}}]{Husa:2015iqa}%
  \BibitemOpen
  \bibfield  {author} {\bibinfo {author} {\bibfnamefont {S.}~\bibnamefont
  {Husa}}, \bibinfo {author} {\bibfnamefont {S.}~\bibnamefont {Khan}}, \bibinfo
  {author} {\bibfnamefont {M.}~\bibnamefont {Hannam}}, \bibinfo {author}
  {\bibfnamefont {M.}~\bibnamefont {P{\"u}rrer}}, \bibinfo {author}
  {\bibfnamefont {F.}~\bibnamefont {Ohme}}, \bibinfo {author} {\bibfnamefont
  {X.}~\bibnamefont {Jim{\'e}nez~Forteza}}, \ and\ \bibinfo {author}
  {\bibfnamefont {A.}~\bibnamefont {Boh{\'e}}},\ }\href {\doibase
  10.1103/PhysRevD.93.044006} {\bibfield  {journal} {\bibinfo  {journal} {Phys.
  Rev.}\ }\textbf {\bibinfo {volume} {D93}},\ \bibinfo {pages} {044006}
  (\bibinfo {year} {2016})},\ \Eprint {http://arxiv.org/abs/1508.07250}
  {arXiv:1508.07250 [gr-qc]} \BibitemShut {NoStop}%
\bibitem [{\citenamefont {Khan}\ \emph {et~al.}(2016)\citenamefont {Khan},
  \citenamefont {Husa}, \citenamefont {Hannam}, \citenamefont {Ohme},
  \citenamefont {P{\"u}rrer}, \citenamefont {Jim{\'e}nez~Forteza},\ and\
  \citenamefont {Boh{\'e}}}]{Khan:2015jqa}%
  \BibitemOpen
  \bibfield  {author} {\bibinfo {author} {\bibfnamefont {S.}~\bibnamefont
  {Khan}}, \bibinfo {author} {\bibfnamefont {S.}~\bibnamefont {Husa}}, \bibinfo
  {author} {\bibfnamefont {M.}~\bibnamefont {Hannam}}, \bibinfo {author}
  {\bibfnamefont {F.}~\bibnamefont {Ohme}}, \bibinfo {author} {\bibfnamefont
  {M.}~\bibnamefont {P{\"u}rrer}}, \bibinfo {author} {\bibfnamefont
  {X.}~\bibnamefont {Jim{\'e}nez~Forteza}}, \ and\ \bibinfo {author}
  {\bibfnamefont {A.}~\bibnamefont {Boh{\'e}}},\ }\href {\doibase
  10.1103/PhysRevD.93.044007} {\bibfield  {journal} {\bibinfo  {journal} {Phys.
  Rev.}\ }\textbf {\bibinfo {volume} {D93}},\ \bibinfo {pages} {044007}
  (\bibinfo {year} {2016})},\ \Eprint {http://arxiv.org/abs/1508.07253}
  {arXiv:1508.07253 [gr-qc]} \BibitemShut {NoStop}%
\bibitem [{\citenamefont {Schmidt}\ \emph {et~al.}(2012)\citenamefont
  {Schmidt}, \citenamefont {Hannam},\ and\ \citenamefont
  {Husa}}]{Schmidt:2012rh}%
  \BibitemOpen
  \bibfield  {author} {\bibinfo {author} {\bibfnamefont {P.}~\bibnamefont
  {Schmidt}}, \bibinfo {author} {\bibfnamefont {M.}~\bibnamefont {Hannam}}, \
  and\ \bibinfo {author} {\bibfnamefont {S.}~\bibnamefont {Husa}},\ }\href
  {\doibase 10.1103/PhysRevD.86.104063} {\bibfield  {journal} {\bibinfo
  {journal} {Phys.\ Rev.\ D}\ }\textbf {\bibinfo {volume} {86}},\ \bibinfo
  {pages} {104063} (\bibinfo {year} {2012})},\ \Eprint
  {http://arxiv.org/abs/1207.3088} {arXiv:1207.3088 [gr-qc]} \BibitemShut
  {NoStop}%
\bibitem [{\citenamefont {Schmidt}\ \emph {et~al.}(2015)\citenamefont
  {Schmidt}, \citenamefont {Ohme},\ and\ \citenamefont
  {Hannam}}]{Schmidt:2014iyl}%
  \BibitemOpen
  \bibfield  {author} {\bibinfo {author} {\bibfnamefont {P.}~\bibnamefont
  {Schmidt}}, \bibinfo {author} {\bibfnamefont {F.}~\bibnamefont {Ohme}}, \
  and\ \bibinfo {author} {\bibfnamefont {M.}~\bibnamefont {Hannam}},\ }\href
  {\doibase 10.1103/PhysRevD.91.024043} {\bibfield  {journal} {\bibinfo
  {journal} {Phys.\ Rev.\ D}\ }\textbf {\bibinfo {volume} {91}},\ \bibinfo
  {pages} {024043} (\bibinfo {year} {2015})},\ \Eprint
  {http://arxiv.org/abs/1408.1810} {arXiv:1408.1810 [gr-qc]} \BibitemShut
  {NoStop}%
\bibitem [{\citenamefont {Dietrich}\ \emph {et~al.}(2017)\citenamefont
  {Dietrich}, \citenamefont {Bernuzzi},\ and\ \citenamefont
  {Tichy}}]{Dietrich:2017aum}%
  \BibitemOpen
  \bibfield  {author} {\bibinfo {author} {\bibfnamefont {T.}~\bibnamefont
  {Dietrich}}, \bibinfo {author} {\bibfnamefont {S.}~\bibnamefont {Bernuzzi}},
  \ and\ \bibinfo {author} {\bibfnamefont {W.}~\bibnamefont {Tichy}},\ }\href
  {\doibase 10.1103/PhysRevD.96.121501} {\bibfield  {journal} {\bibinfo
  {journal} {Phys. Rev.}\ }\textbf {\bibinfo {volume} {D96}},\ \bibinfo {pages}
  {121501} (\bibinfo {year} {2017})},\ \Eprint
  {http://arxiv.org/abs/1706.02969} {arXiv:1706.02969 [gr-qc]} \BibitemShut
  {NoStop}%
\bibitem [{\citenamefont {Dietrich}\ \emph
  {et~al.}(2019{\natexlab{b}})\citenamefont {Dietrich}, \citenamefont
  {Samajdar}, \citenamefont {Khan}, \citenamefont {Johnson-McDaniel},
  \citenamefont {Dudi},\ and\ \citenamefont {Tichy}}]{Dietrich:2019nrt}%
  \BibitemOpen
  \bibfield  {author} {\bibinfo {author} {\bibfnamefont {T.}~\bibnamefont
  {Dietrich}}, \bibinfo {author} {\bibfnamefont {A.}~\bibnamefont {Samajdar}},
  \bibinfo {author} {\bibfnamefont {S.}~\bibnamefont {Khan}}, \bibinfo {author}
  {\bibfnamefont {N.~K.}\ \bibnamefont {Johnson-McDaniel}}, \bibinfo {author}
  {\bibfnamefont {R.}~\bibnamefont {Dudi}}, \ and\ \bibinfo {author}
  {\bibfnamefont {W.}~\bibnamefont {Tichy}},\ }\href {\doibase
  10.1103/PhysRevD.100.044003} {\bibfield  {journal} {\bibinfo  {journal}
  {Phys. Rev. D}\ }\textbf {\bibinfo {volume} {100}},\ \bibinfo {pages}
  {044003} (\bibinfo {year} {2019}{\natexlab{b}})},\ \Eprint
  {http://arxiv.org/abs/1905.06011} {arXiv:1905.06011 [gr-qc]} \BibitemShut
  {NoStop}%
\bibitem [{\citenamefont {{Foucart}}\ \emph {et~al.}(2013)\citenamefont
  {{Foucart}}, \citenamefont {{Buchman}}, \citenamefont {{Duez}}, \citenamefont
  {{Grudich}}, \citenamefont {{Kidder}}, \citenamefont {{MacDonald}},
  \citenamefont {{Mroue}}, \citenamefont {{Pfeiffer}}, \citenamefont
  {{Scheel}},\ and\ \citenamefont {{Szil{\'a}gyi}}}]{Foucart:2013psa}%
  \BibitemOpen
  \bibfield  {author} {\bibinfo {author} {\bibfnamefont {F.}~\bibnamefont
  {{Foucart}}}, \bibinfo {author} {\bibfnamefont {L.}~\bibnamefont
  {{Buchman}}}, \bibinfo {author} {\bibfnamefont {M.~D.}\ \bibnamefont
  {{Duez}}}, \bibinfo {author} {\bibfnamefont {M.}~\bibnamefont {{Grudich}}},
  \bibinfo {author} {\bibfnamefont {L.~E.}\ \bibnamefont {{Kidder}}}, \bibinfo
  {author} {\bibfnamefont {I.}~\bibnamefont {{MacDonald}}}, \bibinfo {author}
  {\bibfnamefont {A.}~\bibnamefont {{Mroue}}}, \bibinfo {author} {\bibfnamefont
  {H.~P.}\ \bibnamefont {{Pfeiffer}}}, \bibinfo {author} {\bibfnamefont
  {M.~A.}\ \bibnamefont {{Scheel}}}, \ and\ \bibinfo {author} {\bibfnamefont
  {B.}~\bibnamefont {{Szil{\'a}gyi}}},\ }\href@noop {} {\bibfield  {journal}
  {\bibinfo  {journal} {Phys.\ Rev.\ D}\ }\textbf {\bibinfo {volume} {88}},\
  \bibinfo {pages} {064017} (\bibinfo {year} {2013})},\ \Eprint
  {http://arxiv.org/abs/1307.7685} {arXiv:1307.7685 [gr-qc]} \BibitemShut
  {NoStop}%
\bibitem [{\citenamefont {{Lovelace}}\ \emph {et~al.}(2013)\citenamefont
  {{Lovelace}}, \citenamefont {{Duez}}, \citenamefont {{Foucart}},
  \citenamefont {{Kidder}}, \citenamefont {{Pfeiffer}}, \citenamefont
  {{Scheel}},\ and\ \citenamefont {{Szil{\'a}gyi}}}]{Lovelace:2013vma}%
  \BibitemOpen
  \bibfield  {author} {\bibinfo {author} {\bibfnamefont {G.}~\bibnamefont
  {{Lovelace}}}, \bibinfo {author} {\bibfnamefont {M.~D.}\ \bibnamefont
  {{Duez}}}, \bibinfo {author} {\bibfnamefont {F.}~\bibnamefont {{Foucart}}},
  \bibinfo {author} {\bibfnamefont {L.~E.}\ \bibnamefont {{Kidder}}}, \bibinfo
  {author} {\bibfnamefont {H.~P.}\ \bibnamefont {{Pfeiffer}}}, \bibinfo
  {author} {\bibfnamefont {M.~A.}\ \bibnamefont {{Scheel}}}, \ and\ \bibinfo
  {author} {\bibfnamefont {B.}~\bibnamefont {{Szil{\'a}gyi}}},\ }\href
  {\doibase 10.1088/0264-9381/30/13/135004} {\bibfield  {journal} {\bibinfo
  {journal} {Class.\ Quantum Grav.}\ }\textbf {\bibinfo {volume} {30}},\
  \bibinfo {eid} {135004} (\bibinfo {year} {2013})},\ \Eprint
  {http://arxiv.org/abs/1302.6297} {arXiv:1302.6297 [gr-qc]} \BibitemShut
  {NoStop}%
\bibitem [{\citenamefont {Kawaguchi}\ \emph {et~al.}(2015)\citenamefont
  {Kawaguchi}, \citenamefont {Kyutoku}, \citenamefont {Nakano}, \citenamefont
  {Okawa}, \citenamefont {Shibata},\ and\ \citenamefont
  {Taniguchi}}]{Kawaguchi:2015}%
  \BibitemOpen
  \bibfield  {author} {\bibinfo {author} {\bibfnamefont {K.}~\bibnamefont
  {Kawaguchi}}, \bibinfo {author} {\bibfnamefont {K.}~\bibnamefont {Kyutoku}},
  \bibinfo {author} {\bibfnamefont {H.}~\bibnamefont {Nakano}}, \bibinfo
  {author} {\bibfnamefont {H.}~\bibnamefont {Okawa}}, \bibinfo {author}
  {\bibfnamefont {M.}~\bibnamefont {Shibata}}, \ and\ \bibinfo {author}
  {\bibfnamefont {K.}~\bibnamefont {Taniguchi}},\ }\href {\doibase
  10.1103/PhysRevD.92.024014} {\bibfield  {journal} {\bibinfo  {journal}
  {Phys.\ Rev.\ D}\ }\textbf {\bibinfo {volume} {92}},\ \bibinfo {eid} {024014}
  (\bibinfo {year} {2015})},\ \Eprint {http://arxiv.org/abs/1506.05473}
  {arXiv:1506.05473 [astro-ph.HE]} \BibitemShut {NoStop}%
\bibitem [{\citenamefont {{Haas}}\ \emph {et~al.}(2016)\citenamefont {{Haas}},
  \citenamefont {{Ott}}, \citenamefont {{Szil{\'a}gyi}}, \citenamefont
  {{Kaplan}}, \citenamefont {{Lippuner}}, \citenamefont {{Scheel}},
  \citenamefont {{Barkett}}, \citenamefont {{Muhlberger}}, \citenamefont
  {{Dietrich}}, \citenamefont {{Duez}}, \citenamefont {{Foucart}},
  \citenamefont {{Pfeiffer}}, \citenamefont {{Kidder}},\ and\ \citenamefont
  {{Teukolsky}}}]{Haas:2016}%
  \BibitemOpen
  \bibfield  {author} {\bibinfo {author} {\bibfnamefont {R.}~\bibnamefont
  {{Haas}}}, \bibinfo {author} {\bibfnamefont {C.~D.}\ \bibnamefont {{Ott}}},
  \bibinfo {author} {\bibfnamefont {B.}~\bibnamefont {{Szil{\'a}gyi}}},
  \bibinfo {author} {\bibfnamefont {J.~D.}\ \bibnamefont {{Kaplan}}}, \bibinfo
  {author} {\bibfnamefont {J.}~\bibnamefont {{Lippuner}}}, \bibinfo {author}
  {\bibfnamefont {M.~A.}\ \bibnamefont {{Scheel}}}, \bibinfo {author}
  {\bibfnamefont {K.}~\bibnamefont {{Barkett}}}, \bibinfo {author}
  {\bibfnamefont {C.~D.}\ \bibnamefont {{Muhlberger}}}, \bibinfo {author}
  {\bibfnamefont {T.}~\bibnamefont {{Dietrich}}}, \bibinfo {author}
  {\bibfnamefont {M.~D.}\ \bibnamefont {{Duez}}}, \bibinfo {author}
  {\bibfnamefont {F.}~\bibnamefont {{Foucart}}}, \bibinfo {author}
  {\bibfnamefont {H.~P.}\ \bibnamefont {{Pfeiffer}}}, \bibinfo {author}
  {\bibfnamefont {L.~E.}\ \bibnamefont {{Kidder}}}, \ and\ \bibinfo {author}
  {\bibfnamefont {S.~A.}\ \bibnamefont {{Teukolsky}}},\ }\href {\doibase
  10.1103/PhysRevD.93.124062} {\bibfield  {journal} {\bibinfo  {journal}
  {Phys.\ Rev.\ D}\ }\textbf {\bibinfo {volume} {D93}},\ \bibinfo {pages}
  {124062} (\bibinfo {year} {2016})},\ \Eprint
  {http://arxiv.org/abs/1604.00782} {arXiv:1604.00782 [gr-qc]} \BibitemShut
  {NoStop}%
\bibitem [{\citenamefont {{Hinderer}}\ \emph {et~al.}(2016)\citenamefont
  {{Hinderer}}, \citenamefont {{Taracchini}}, \citenamefont {{Foucart}},
  \citenamefont {{Buonanno}}, \citenamefont {{Steinhoff}}, \citenamefont
  {{Duez}}, \citenamefont {{Kidder}}, \citenamefont {{Pfeiffer}}, \citenamefont
  {{Scheel}}, \citenamefont {{Szil{\'a}gyi}}, \citenamefont {{Hotokezaka}},
  \citenamefont {{Kyutoku}}, \citenamefont {{Shibata}},\ and\ \citenamefont
  {{Carpenter}}}]{Hinderer:2016a}%
  \BibitemOpen
  \bibfield  {author} {\bibinfo {author} {\bibfnamefont {T.}~\bibnamefont
  {{Hinderer}}}, \bibinfo {author} {\bibfnamefont {A.}~\bibnamefont
  {{Taracchini}}}, \bibinfo {author} {\bibfnamefont {F.}~\bibnamefont
  {{Foucart}}}, \bibinfo {author} {\bibfnamefont {A.}~\bibnamefont
  {{Buonanno}}}, \bibinfo {author} {\bibfnamefont {J.}~\bibnamefont
  {{Steinhoff}}}, \bibinfo {author} {\bibfnamefont {M.}~\bibnamefont {{Duez}}},
  \bibinfo {author} {\bibfnamefont {L.~E.}\ \bibnamefont {{Kidder}}}, \bibinfo
  {author} {\bibfnamefont {H.~P.}\ \bibnamefont {{Pfeiffer}}}, \bibinfo
  {author} {\bibfnamefont {M.~A.}\ \bibnamefont {{Scheel}}}, \bibinfo {author}
  {\bibfnamefont {B.}~\bibnamefont {{Szil{\'a}gyi}}}, \bibinfo {author}
  {\bibfnamefont {K.}~\bibnamefont {{Hotokezaka}}}, \bibinfo {author}
  {\bibfnamefont {K.}~\bibnamefont {{Kyutoku}}}, \bibinfo {author}
  {\bibfnamefont {M.}~\bibnamefont {{Shibata}}}, \ and\ \bibinfo {author}
  {\bibfnamefont {C.~W.}\ \bibnamefont {{Carpenter}}},\ }\href {\doibase
  10.1103/PhysRevLett.116.181101} {\bibfield  {journal} {\bibinfo  {journal}
  {Phys. Rev. Lett.}\ }\textbf {\bibinfo {volume} {116}},\ \bibinfo {eid}
  {181101} (\bibinfo {year} {2016})},\ \Eprint
  {http://arxiv.org/abs/1602.00599} {arXiv:1602.00599 [gr-qc]} \BibitemShut
  {NoStop}%
\bibitem [{\citenamefont {Dietrich}\ \emph {et~al.}(2018)\citenamefont
  {Dietrich}, \citenamefont {Radice}, \citenamefont {Bernuzzi}, \citenamefont
  {Zappa}, \citenamefont {Perego}, \citenamefont {Brügmann}, \citenamefont
  {Chaurasia}, \citenamefont {Dudi}, \citenamefont {Tichy},\ and\ \citenamefont
  {Ujevic}}]{2018arXiv180601625D}%
  \BibitemOpen
  \bibfield  {author} {\bibinfo {author} {\bibfnamefont {T.}~\bibnamefont
  {Dietrich}}, \bibinfo {author} {\bibfnamefont {D.}~\bibnamefont {Radice}},
  \bibinfo {author} {\bibfnamefont {S.}~\bibnamefont {Bernuzzi}}, \bibinfo
  {author} {\bibfnamefont {F.}~\bibnamefont {Zappa}}, \bibinfo {author}
  {\bibfnamefont {A.}~\bibnamefont {Perego}}, \bibinfo {author} {\bibfnamefont
  {B.}~\bibnamefont {Brügmann}}, \bibinfo {author} {\bibfnamefont {S.~V.}\
  \bibnamefont {Chaurasia}}, \bibinfo {author} {\bibfnamefont {R.}~\bibnamefont
  {Dudi}}, \bibinfo {author} {\bibfnamefont {W.}~\bibnamefont {Tichy}}, \ and\
  \bibinfo {author} {\bibfnamefont {M.}~\bibnamefont {Ujevic}},\ }\href
  {\doibase 10.1088/1361-6382/aaebc0} {\bibfield  {journal} {\bibinfo
  {journal} {Class. Quant. Grav.}\ }\textbf {\bibinfo {volume} {35}},\ \bibinfo
  {pages} {24LT01} (\bibinfo {year} {2018})},\ \Eprint
  {http://arxiv.org/abs/1806.01625} {arXiv:1806.01625 [gr-qc]} \BibitemShut
  {NoStop}%
\bibitem [{\citenamefont {Foucart}\ \emph {et~al.}(2019)\citenamefont {Foucart}
  \emph {et~al.}}]{Foucart:2018inp}%
  \BibitemOpen
  \bibfield  {author} {\bibinfo {author} {\bibfnamefont {F.}~\bibnamefont
  {Foucart}} \emph {et~al.},\ }\href {\doibase 10.1103/PhysRevD.99.044008}
  {\bibfield  {journal} {\bibinfo  {journal} {Phys. Rev.}\ }\textbf {\bibinfo
  {volume} {D99}},\ \bibinfo {pages} {044008} (\bibinfo {year} {2019})},\
  \Eprint {http://arxiv.org/abs/1812.06988} {arXiv:1812.06988 [gr-qc]}
  \BibitemShut {NoStop}%
\bibitem [{\citenamefont {Kiuchi}\ \emph {et~al.}(2020)\citenamefont {Kiuchi},
  \citenamefont {Kawaguchi}, \citenamefont {Kyutoku}, \citenamefont
  {Sekiguchi},\ and\ \citenamefont {Shibata}}]{Kiuchi:2019kzt}%
  \BibitemOpen
  \bibfield  {author} {\bibinfo {author} {\bibfnamefont {K.}~\bibnamefont
  {Kiuchi}}, \bibinfo {author} {\bibfnamefont {K.}~\bibnamefont {Kawaguchi}},
  \bibinfo {author} {\bibfnamefont {K.}~\bibnamefont {Kyutoku}}, \bibinfo
  {author} {\bibfnamefont {Y.}~\bibnamefont {Sekiguchi}}, \ and\ \bibinfo
  {author} {\bibfnamefont {M.}~\bibnamefont {Shibata}},\ }\href {\doibase
  10.1103/PhysRevD.101.084006} {\bibfield  {journal} {\bibinfo  {journal}
  {Phys. Rev. D}\ }\textbf {\bibinfo {volume} {101}},\ \bibinfo {pages}
  {084006} (\bibinfo {year} {2020})},\ \Eprint
  {http://arxiv.org/abs/1907.03790} {arXiv:1907.03790 [astro-ph.HE]}
  \BibitemShut {NoStop}%
\bibitem [{\citenamefont {Aylott}\ \emph {et~al.}(2009)\citenamefont {Aylott}
  \emph {et~al.}}]{Aylott:2009tn}%
  \BibitemOpen
  \bibfield  {author} {\bibinfo {author} {\bibfnamefont {B.}~\bibnamefont
  {Aylott}} \emph {et~al.},\ }\href {\doibase 10.1088/0264-9381/26/11/114008}
  {\bibfield  {journal} {\bibinfo  {journal} {Class.\ Quantum Grav.}\ }\textbf
  {\bibinfo {volume} {26}},\ \bibinfo {pages} {114008} (\bibinfo {year}
  {2009})},\ \Eprint {http://arxiv.org/abs/0905.4227} {arXiv:0905.4227 [gr-qc]}
  \BibitemShut {NoStop}%
\bibitem [{\citenamefont {Ajith}\ \emph {et~al.}(2012)\citenamefont {Ajith}
  \emph {et~al.}}]{Ajith:2012az}%
  \BibitemOpen
  \bibfield  {author} {\bibinfo {author} {\bibfnamefont {P.}~\bibnamefont
  {Ajith}} \emph {et~al.},\ }\href {\doibase 10.1088/0264-9381/30/19/199401,
  10.1088/0264-9381/29/12/124001} {\bibfield  {journal} {\bibinfo  {journal}
  {Class. Quant. Grav.}\ }\textbf {\bibinfo {volume} {29}},\ \bibinfo {pages}
  {124001} (\bibinfo {year} {2012})},\ \bibinfo {note} {[Erratum: Class. Quant.
  Grav.30,199401(2013)]},\ \Eprint {http://arxiv.org/abs/1201.5319}
  {arXiv:1201.5319 [gr-qc]} \BibitemShut {NoStop}%
\bibitem [{\citenamefont {Hinder}\ \emph {et~al.}(2014)\citenamefont {Hinder}
  \emph {et~al.}}]{Hinder:2013oqa}%
  \BibitemOpen
  \bibfield  {author} {\bibinfo {author} {\bibfnamefont {I.}~\bibnamefont
  {Hinder}} \emph {et~al.} (\bibinfo {collaboration} {The NRAR
  Collaboration}),\ }\href@noop {} {\bibfield  {journal} {\bibinfo  {journal}
  {Class.\ Quantum Grav.}\ }\textbf {\bibinfo {volume} {31}},\ \bibinfo {pages}
  {025012} (\bibinfo {year} {2014})},\ \Eprint {http://arxiv.org/abs/1307.5307}
  {arXiv:1307.5307 [gr-qc]} \BibitemShut {NoStop}%
\bibitem [{\citenamefont {{Mrou{\'e}}}\ \emph {et~al.}(2013)\citenamefont
  {{Mrou{\'e}}}, \citenamefont {{Scheel}}, \citenamefont {{Szil{\'a}gyi}},
  \citenamefont {{Pfeiffer}}, \citenamefont {{Boyle}}, \citenamefont
  {{Hemberger}}, \citenamefont {{Kidder}}, \citenamefont {{Lovelace}},
  \citenamefont {{Ossokine}}, \citenamefont {{Taylor}}, \citenamefont
  {{Zengino{\u g}lu}}, \citenamefont {{Buchman}}, \citenamefont {{Chu}},
  \citenamefont {{Foley}}, \citenamefont {{Giesler}}, \citenamefont {{Owen}},\
  and\ \citenamefont {{Teukolsky}}}]{Mroue:2013PRL}%
  \BibitemOpen
  \bibfield  {author} {\bibinfo {author} {\bibfnamefont {A.~H.}\ \bibnamefont
  {{Mrou{\'e}}}}, \bibinfo {author} {\bibfnamefont {M.~A.}\ \bibnamefont
  {{Scheel}}}, \bibinfo {author} {\bibfnamefont {B.}~\bibnamefont
  {{Szil{\'a}gyi}}}, \bibinfo {author} {\bibfnamefont {H.~P.}\ \bibnamefont
  {{Pfeiffer}}}, \bibinfo {author} {\bibfnamefont {M.}~\bibnamefont {{Boyle}}},
  \bibinfo {author} {\bibfnamefont {D.~A.}\ \bibnamefont {{Hemberger}}},
  \bibinfo {author} {\bibfnamefont {L.~E.}\ \bibnamefont {{Kidder}}}, \bibinfo
  {author} {\bibfnamefont {G.}~\bibnamefont {{Lovelace}}}, \bibinfo {author}
  {\bibfnamefont {S.}~\bibnamefont {{Ossokine}}}, \bibinfo {author}
  {\bibfnamefont {N.~W.}\ \bibnamefont {{Taylor}}}, \bibinfo {author}
  {\bibfnamefont {A.}~\bibnamefont {{Zengino{\u g}lu}}}, \bibinfo {author}
  {\bibfnamefont {L.~T.}\ \bibnamefont {{Buchman}}}, \bibinfo {author}
  {\bibfnamefont {T.}~\bibnamefont {{Chu}}}, \bibinfo {author} {\bibfnamefont
  {E.}~\bibnamefont {{Foley}}}, \bibinfo {author} {\bibfnamefont
  {M.}~\bibnamefont {{Giesler}}}, \bibinfo {author} {\bibfnamefont
  {R.}~\bibnamefont {{Owen}}}, \ and\ \bibinfo {author} {\bibfnamefont {S.~A.}\
  \bibnamefont {{Teukolsky}}},\ }\href@noop {} {\bibfield  {journal} {\bibinfo
  {journal} {Phys.\ Rev.\ Lett.}\ }\textbf {\bibinfo {volume} {111}},\ \bibinfo
  {pages} {241104} (\bibinfo {year} {2013})},\ \Eprint
  {http://arxiv.org/abs/1304.6077} {arXiv:1304.6077 [gr-qc]} \BibitemShut
  {NoStop}%
\bibitem [{\citenamefont {Jani}\ \emph {et~al.}(2016)\citenamefont {Jani},
  \citenamefont {Healy}, \citenamefont {Clark}, \citenamefont {London},
  \citenamefont {Laguna},\ and\ \citenamefont {Shoemaker}}]{Jani:2016wkt}%
  \BibitemOpen
  \bibfield  {author} {\bibinfo {author} {\bibfnamefont {K.}~\bibnamefont
  {Jani}}, \bibinfo {author} {\bibfnamefont {J.}~\bibnamefont {Healy}},
  \bibinfo {author} {\bibfnamefont {J.~A.}\ \bibnamefont {Clark}}, \bibinfo
  {author} {\bibfnamefont {L.}~\bibnamefont {London}}, \bibinfo {author}
  {\bibfnamefont {P.}~\bibnamefont {Laguna}}, \ and\ \bibinfo {author}
  {\bibfnamefont {D.}~\bibnamefont {Shoemaker}},\ }\href {\doibase
  10.1088/0264-9381/33/20/204001} {\bibfield  {journal} {\bibinfo  {journal}
  {Class. Quant. Grav.}\ }\textbf {\bibinfo {volume} {33}},\ \bibinfo {pages}
  {204001} (\bibinfo {year} {2016})},\ \Eprint
  {http://arxiv.org/abs/1605.03204} {arXiv:1605.03204 [gr-qc]} \BibitemShut
  {NoStop}%
\bibitem [{\citenamefont {Healy}\ \emph {et~al.}(2017)\citenamefont {Healy},
  \citenamefont {Lousto}, \citenamefont {Zlochower},\ and\ \citenamefont
  {Campanelli}}]{Healy:2017bsc}%
  \BibitemOpen
  \bibfield  {author} {\bibinfo {author} {\bibfnamefont {J.}~\bibnamefont
  {Healy}}, \bibinfo {author} {\bibfnamefont {C.~O.}\ \bibnamefont {Lousto}},
  \bibinfo {author} {\bibfnamefont {Y.}~\bibnamefont {Zlochower}}, \ and\
  \bibinfo {author} {\bibfnamefont {M.}~\bibnamefont {Campanelli}},\ }\href
  {\doibase 10.1088/1361-6382/aa91b1} {\bibfield  {journal} {\bibinfo
  {journal} {Classical and Quantum Gravity}\ }\textbf {\bibinfo {volume}
  {34}},\ \bibinfo {pages} {224001} (\bibinfo {year} {2017})},\ \Eprint
  {http://arxiv.org/abs/1703.03423} {arXiv:1703.03423 [gr-qc]} \BibitemShut
  {NoStop}%
\bibitem [{\citenamefont {Healy}\ \emph {et~al.}(2019)\citenamefont {Healy},
  \citenamefont {Lousto}, \citenamefont {Lange}, \citenamefont {O'Shaughnessy},
  \citenamefont {Zlochower},\ and\ \citenamefont {Campanelli}}]{Healy:2019bsc}%
  \BibitemOpen
  \bibfield  {author} {\bibinfo {author} {\bibfnamefont {J.}~\bibnamefont
  {Healy}}, \bibinfo {author} {\bibfnamefont {C.~O.}\ \bibnamefont {Lousto}},
  \bibinfo {author} {\bibfnamefont {J.}~\bibnamefont {Lange}}, \bibinfo
  {author} {\bibfnamefont {R.}~\bibnamefont {O'Shaughnessy}}, \bibinfo {author}
  {\bibfnamefont {Y.}~\bibnamefont {Zlochower}}, \ and\ \bibinfo {author}
  {\bibfnamefont {M.}~\bibnamefont {Campanelli}},\ }\href {\doibase
  10.1103/PhysRevD.100.024021} {\bibfield  {journal} {\bibinfo  {journal}
  {Phys. Rev. D}\ }\textbf {\bibinfo {volume} {100}},\ \bibinfo {pages}
  {024021} (\bibinfo {year} {2019})},\ \Eprint
  {http://arxiv.org/abs/1901.02553} {arXiv:1901.02553 [gr-qc]} \BibitemShut
  {NoStop}%
\bibitem [{\citenamefont {Huerta}\ \emph {et~al.}(2019)\citenamefont {Huerta},
  \citenamefont {Haas}, \citenamefont {Habib}, \citenamefont {Gupta},
  \citenamefont {Rebei}, \citenamefont {Chavva}, \citenamefont {Johnson},
  \citenamefont {Rosofsky}, \citenamefont {Wessel}, \citenamefont {Agarwal},
  \citenamefont {Luo},\ and\ \citenamefont {Ren}}]{Huerta:2019hrh}%
  \BibitemOpen
  \bibfield  {author} {\bibinfo {author} {\bibfnamefont {E.~A.}\ \bibnamefont
  {Huerta}}, \bibinfo {author} {\bibfnamefont {R.}~\bibnamefont {Haas}},
  \bibinfo {author} {\bibfnamefont {S.}~\bibnamefont {Habib}}, \bibinfo
  {author} {\bibfnamefont {A.}~\bibnamefont {Gupta}}, \bibinfo {author}
  {\bibfnamefont {A.}~\bibnamefont {Rebei}}, \bibinfo {author} {\bibfnamefont
  {V.}~\bibnamefont {Chavva}}, \bibinfo {author} {\bibfnamefont
  {D.}~\bibnamefont {Johnson}}, \bibinfo {author} {\bibfnamefont
  {S.}~\bibnamefont {Rosofsky}}, \bibinfo {author} {\bibfnamefont
  {E.}~\bibnamefont {Wessel}}, \bibinfo {author} {\bibfnamefont
  {B.}~\bibnamefont {Agarwal}}, \bibinfo {author} {\bibfnamefont
  {D.}~\bibnamefont {Luo}}, \ and\ \bibinfo {author} {\bibfnamefont
  {W.}~\bibnamefont {Ren}},\ }\href {\doibase 10.1103/PhysRevD.100.064003}
  {\bibfield  {journal} {\bibinfo  {journal} {Phys. Rev. D}\ }\textbf {\bibinfo
  {volume} {100}},\ \bibinfo {pages} {064003} (\bibinfo {year} {2019})},\
  \Eprint {http://arxiv.org/abs/1901.07038} {arXiv:1901.07038 [gr-qc]}
  \BibitemShut {NoStop}%
\bibitem [{\citenamefont {Boyle}\ \emph {et~al.}(2019)\citenamefont {Boyle}
  \emph {et~al.}}]{Boyle:2019kee}%
  \BibitemOpen
  \bibfield  {author} {\bibinfo {author} {\bibfnamefont {M.}~\bibnamefont
  {Boyle}} \emph {et~al.},\ }\href@noop {} {\bibfield  {journal} {\bibinfo
  {journal} {Class.\ Quantum Grav.}\ }\textbf {\bibinfo {volume} {36}},\
  \bibinfo {pages} {195006} (\bibinfo {year} {2019})},\ \Eprint
  {http://arxiv.org/abs/1904.04831} {arXiv:1904.04831 [gr-qc]} \BibitemShut
  {NoStop}%
\bibitem [{\citenamefont {{Field}}\ \emph {et~al.}(2014)\citenamefont
  {{Field}}, \citenamefont {{Galley}}, \citenamefont {{Hesthaven}},
  \citenamefont {{Kaye}},\ and\ \citenamefont {{Tiglio}}}]{Field:2013cfa}%
  \BibitemOpen
  \bibfield  {author} {\bibinfo {author} {\bibfnamefont {S.~E.}\ \bibnamefont
  {{Field}}}, \bibinfo {author} {\bibfnamefont {C.~R.}\ \bibnamefont
  {{Galley}}}, \bibinfo {author} {\bibfnamefont {J.~S.}\ \bibnamefont
  {{Hesthaven}}}, \bibinfo {author} {\bibfnamefont {J.}~\bibnamefont {{Kaye}}},
  \ and\ \bibinfo {author} {\bibfnamefont {M.}~\bibnamefont {{Tiglio}}},\
  }\href {\doibase 10.1103/PhysRevX.4.031006} {\bibfield  {journal} {\bibinfo
  {journal} {Phys.\ Rev.\ X}\ }\textbf {\bibinfo {volume} {4}},\ \bibinfo {eid}
  {031006} (\bibinfo {year} {2014})},\ \Eprint {http://arxiv.org/abs/1308.3565}
  {arXiv:1308.3565 [gr-qc]} \BibitemShut {NoStop}%
\bibitem [{\citenamefont {P{\"u}rrer}(2014)}]{Purrer:2014fza}%
  \BibitemOpen
  \bibfield  {author} {\bibinfo {author} {\bibfnamefont {M.}~\bibnamefont
  {P{\"u}rrer}},\ }\href {\doibase 10.1088/0264-9381/31/19/195010} {\bibfield
  {journal} {\bibinfo  {journal} {Class.\ Quantum Grav.}\ }\textbf {\bibinfo
  {volume} {31}},\ \bibinfo {pages} {195010} (\bibinfo {year} {2014})},\
  \Eprint {http://arxiv.org/abs/1402.4146} {arXiv:1402.4146 [gr-qc]}
  \BibitemShut {NoStop}%
\bibitem [{\citenamefont {{Blackman}}\ \emph {et~al.}(2015)\citenamefont
  {{Blackman}}, \citenamefont {{Field}}, \citenamefont {{Galley}},
  \citenamefont {{Szil{\'a}gyi}}, \citenamefont {{Scheel}}, \citenamefont
  {{Tiglio}},\ and\ \citenamefont {{Hemberger}}}]{Blackman:2015pia}%
  \BibitemOpen
  \bibfield  {author} {\bibinfo {author} {\bibfnamefont {J.}~\bibnamefont
  {{Blackman}}}, \bibinfo {author} {\bibfnamefont {S.~E.}\ \bibnamefont
  {{Field}}}, \bibinfo {author} {\bibfnamefont {C.~R.}\ \bibnamefont
  {{Galley}}}, \bibinfo {author} {\bibfnamefont {B.}~\bibnamefont
  {{Szil{\'a}gyi}}}, \bibinfo {author} {\bibfnamefont {M.~A.}\ \bibnamefont
  {{Scheel}}}, \bibinfo {author} {\bibfnamefont {M.}~\bibnamefont {{Tiglio}}},
  \ and\ \bibinfo {author} {\bibfnamefont {D.~A.}\ \bibnamefont
  {{Hemberger}}},\ }\href {\doibase 10.1103/PhysRevLett.115.121102} {\bibfield
  {journal} {\bibinfo  {journal} {Phys.\ Rev.\ Lett.}\ }\textbf {\bibinfo
  {volume} {115}},\ \bibinfo {eid} {121102} (\bibinfo {year} {2015})},\ \Eprint
  {http://arxiv.org/abs/1502.07758} {arXiv:1502.07758 [gr-qc]} \BibitemShut
  {NoStop}%
\bibitem [{\citenamefont {Blackman}\ \emph
  {et~al.}(2017{\natexlab{a}})\citenamefont {Blackman}, \citenamefont {Field},
  \citenamefont {Scheel}, \citenamefont {Galley}, \citenamefont {Hemberger},
  \citenamefont {Schmidt},\ and\ \citenamefont {Smith}}]{Blackman:2017dfb}%
  \BibitemOpen
  \bibfield  {author} {\bibinfo {author} {\bibfnamefont {J.}~\bibnamefont
  {Blackman}}, \bibinfo {author} {\bibfnamefont {S.~E.}\ \bibnamefont {Field}},
  \bibinfo {author} {\bibfnamefont {M.~A.}\ \bibnamefont {Scheel}}, \bibinfo
  {author} {\bibfnamefont {C.~R.}\ \bibnamefont {Galley}}, \bibinfo {author}
  {\bibfnamefont {D.~A.}\ \bibnamefont {Hemberger}}, \bibinfo {author}
  {\bibfnamefont {P.}~\bibnamefont {Schmidt}}, \ and\ \bibinfo {author}
  {\bibfnamefont {R.}~\bibnamefont {Smith}},\ }\href {\doibase
  10.1103/PhysRevD.95.104023} {\bibfield  {journal} {\bibinfo  {journal} {Phys.
  Rev.}\ }\textbf {\bibinfo {volume} {D95}},\ \bibinfo {pages} {104023}
  (\bibinfo {year} {2017}{\natexlab{a}})},\ \Eprint
  {http://arxiv.org/abs/1701.00550} {arXiv:1701.00550 [gr-qc]} \BibitemShut
  {NoStop}%
\bibitem [{\citenamefont {Blackman}\ \emph
  {et~al.}(2017{\natexlab{b}})\citenamefont {Blackman}, \citenamefont {Field},
  \citenamefont {Scheel}, \citenamefont {Galley}, \citenamefont {Ott},
  \citenamefont {Boyle}, \citenamefont {Kidder}, \citenamefont {Pfeiffer},\
  and\ \citenamefont {Szilágyi}}]{Blackman:2017pcm}%
  \BibitemOpen
  \bibfield  {author} {\bibinfo {author} {\bibfnamefont {J.}~\bibnamefont
  {Blackman}}, \bibinfo {author} {\bibfnamefont {S.~E.}\ \bibnamefont {Field}},
  \bibinfo {author} {\bibfnamefont {M.~A.}\ \bibnamefont {Scheel}}, \bibinfo
  {author} {\bibfnamefont {C.~R.}\ \bibnamefont {Galley}}, \bibinfo {author}
  {\bibfnamefont {C.~D.}\ \bibnamefont {Ott}}, \bibinfo {author} {\bibfnamefont
  {M.}~\bibnamefont {Boyle}}, \bibinfo {author} {\bibfnamefont {L.~E.}\
  \bibnamefont {Kidder}}, \bibinfo {author} {\bibfnamefont {H.~P.}\
  \bibnamefont {Pfeiffer}}, \ and\ \bibinfo {author} {\bibfnamefont
  {B.}~\bibnamefont {Szilágyi}},\ }\href {\doibase 10.1103/PhysRevD.96.024058}
  {\bibfield  {journal} {\bibinfo  {journal} {Phys. Rev.}\ }\textbf {\bibinfo
  {volume} {D96}},\ \bibinfo {pages} {024058} (\bibinfo {year}
  {2017}{\natexlab{b}})},\ \Eprint {http://arxiv.org/abs/1705.07089}
  {arXiv:1705.07089 [gr-qc]} \BibitemShut {NoStop}%
\bibitem [{\citenamefont {Varma}\ \emph
  {et~al.}(2019{\natexlab{a}})\citenamefont {Varma}, \citenamefont {Field},
  \citenamefont {Scheel}, \citenamefont {Blackman}, \citenamefont {Kidder},\
  and\ \citenamefont {Pfeiffer}}]{Varma:2018mmi}%
  \BibitemOpen
  \bibfield  {author} {\bibinfo {author} {\bibfnamefont {V.}~\bibnamefont
  {Varma}}, \bibinfo {author} {\bibfnamefont {S.~E.}\ \bibnamefont {Field}},
  \bibinfo {author} {\bibfnamefont {M.~A.}\ \bibnamefont {Scheel}}, \bibinfo
  {author} {\bibfnamefont {J.}~\bibnamefont {Blackman}}, \bibinfo {author}
  {\bibfnamefont {L.~E.}\ \bibnamefont {Kidder}}, \ and\ \bibinfo {author}
  {\bibfnamefont {H.~P.}\ \bibnamefont {Pfeiffer}},\ }\href {\doibase
  10.1103/PhysRevD.99.064045} {\bibfield  {journal} {\bibinfo  {journal} {Phys.
  Rev.}\ }\textbf {\bibinfo {volume} {D99}},\ \bibinfo {pages} {064045}
  (\bibinfo {year} {2019}{\natexlab{a}})},\ \Eprint
  {http://arxiv.org/abs/1812.07865} {arXiv:1812.07865 [gr-qc]} \BibitemShut
  {NoStop}%
\bibitem [{\citenamefont {Varma}\ \emph
  {et~al.}(2019{\natexlab{b}})\citenamefont {Varma}, \citenamefont {Field},
  \citenamefont {Scheel}, \citenamefont {Blackman}, \citenamefont {Gerosa},
  \citenamefont {Stein}, \citenamefont {Kidder},\ and\ \citenamefont
  {Pfeiffer}}]{Varma:2019csw}%
  \BibitemOpen
  \bibfield  {author} {\bibinfo {author} {\bibfnamefont {V.}~\bibnamefont
  {Varma}}, \bibinfo {author} {\bibfnamefont {S.~E.}\ \bibnamefont {Field}},
  \bibinfo {author} {\bibfnamefont {M.~A.}\ \bibnamefont {Scheel}}, \bibinfo
  {author} {\bibfnamefont {J.}~\bibnamefont {Blackman}}, \bibinfo {author}
  {\bibfnamefont {D.}~\bibnamefont {Gerosa}}, \bibinfo {author} {\bibfnamefont
  {L.~C.}\ \bibnamefont {Stein}}, \bibinfo {author} {\bibfnamefont {L.~E.}\
  \bibnamefont {Kidder}}, \ and\ \bibinfo {author} {\bibfnamefont {H.~P.}\
  \bibnamefont {Pfeiffer}},\ }\href {\doibase 10.1103/PhysRevResearch.1.033015}
  {\bibfield  {journal} {\bibinfo  {journal} {Phys. Rev. Research}\ }\textbf
  {\bibinfo {volume} {1}},\ \bibinfo {pages} {033015} (\bibinfo {year}
  {2019}{\natexlab{b}})},\ \Eprint {http://arxiv.org/abs/1905.09300}
  {arXiv:1905.09300 [gr-qc]} \BibitemShut {NoStop}%
\bibitem [{\citenamefont {{Barkett}}\ \emph {et~al.}(2016)\citenamefont
  {{Barkett}}, \citenamefont {{Scheel}}, \citenamefont {{Haas}}, \citenamefont
  {{Ott}}, \citenamefont {{Bernuzzi}}, \citenamefont {{Brown}}, \citenamefont
  {{Szil{\'a}gyi}}, \citenamefont {{Kaplan}}, \citenamefont {{Lippuner}},
  \citenamefont {{Muhlberger}}, \citenamefont {{Foucart}},\ and\ \citenamefont
  {{Duez}}}]{Barkett2015}%
  \BibitemOpen
  \bibfield  {author} {\bibinfo {author} {\bibfnamefont {K.}~\bibnamefont
  {{Barkett}}}, \bibinfo {author} {\bibfnamefont {M.~A.}\ \bibnamefont
  {{Scheel}}}, \bibinfo {author} {\bibfnamefont {R.}~\bibnamefont {{Haas}}},
  \bibinfo {author} {\bibfnamefont {C.~D.}\ \bibnamefont {{Ott}}}, \bibinfo
  {author} {\bibfnamefont {S.}~\bibnamefont {{Bernuzzi}}}, \bibinfo {author}
  {\bibfnamefont {D.~A.}\ \bibnamefont {{Brown}}}, \bibinfo {author}
  {\bibfnamefont {B.}~\bibnamefont {{Szil{\'a}gyi}}}, \bibinfo {author}
  {\bibfnamefont {J.~D.}\ \bibnamefont {{Kaplan}}}, \bibinfo {author}
  {\bibfnamefont {J.}~\bibnamefont {{Lippuner}}}, \bibinfo {author}
  {\bibfnamefont {C.~D.}\ \bibnamefont {{Muhlberger}}}, \bibinfo {author}
  {\bibfnamefont {F.}~\bibnamefont {{Foucart}}}, \ and\ \bibinfo {author}
  {\bibfnamefont {M.~D.}\ \bibnamefont {{Duez}}},\ }\href {\doibase
  10.1103/PhysRevD.93.044064} {\bibfield  {journal} {\bibinfo  {journal}
  {Phys.\ Rev.\ D}\ }\textbf {\bibinfo {volume} {93}},\ \bibinfo {pages}
  {044064} (\bibinfo {year} {2016})},\ \Eprint
  {http://arxiv.org/abs/1509.05782} {arXiv:1509.05782 [gr-qc]} \BibitemShut
  {NoStop}%
\bibitem [{\citenamefont {Blanchet}(2014)}]{Blanchet:2013haa}%
  \BibitemOpen
  \bibfield  {author} {\bibinfo {author} {\bibfnamefont {L.}~\bibnamefont
  {Blanchet}},\ }\href {\doibase 10.12942/lrr-2014-2} {\bibfield  {journal}
  {\bibinfo  {journal} {Living Rev. Rel.}\ }\textbf {\bibinfo {volume} {17}},\
  \bibinfo {pages} {2} (\bibinfo {year} {2014})},\ \Eprint
  {http://arxiv.org/abs/1310.1528} {arXiv:1310.1528 [gr-qc]} \BibitemShut
  {NoStop}%
\bibitem [{\citenamefont {Fang}\ and\ \citenamefont
  {Lovelace}(2005)}]{FangLovelace:2005}%
  \BibitemOpen
  \bibfield  {author} {\bibinfo {author} {\bibfnamefont {H.}~\bibnamefont
  {Fang}}\ and\ \bibinfo {author} {\bibfnamefont {G.}~\bibnamefont
  {Lovelace}},\ }\href@noop {} {\bibfield  {journal} {\bibinfo  {journal}
  {Phys.\ Rev.\ D}\ }\textbf {\bibinfo {volume} {72}},\ \bibinfo {pages}
  {124016} (\bibinfo {year} {2005})}\BibitemShut {NoStop}%
\bibitem [{\citenamefont {Alvi}(2001)}]{Alvi:2001mx}%
  \BibitemOpen
  \bibfield  {author} {\bibinfo {author} {\bibfnamefont {K.}~\bibnamefont
  {Alvi}},\ }\href {http://link.aps.org/doi/10.1103/PhysRevD.64.104020}
  {\bibfield  {journal} {\bibinfo  {journal} {Phys.\ Rev.\ D}\ }\textbf
  {\bibinfo {volume} {64}},\ \bibinfo {pages} {104020} (\bibinfo {year}
  {2001})}\BibitemShut {NoStop}%
\bibitem [{\citenamefont {{Flanagan}}\ and\ \citenamefont
  {{Hinderer}}(2008)}]{Flanagan2008}%
  \BibitemOpen
  \bibfield  {author} {\bibinfo {author} {\bibfnamefont {{\'E}.~{\'E}.}\
  \bibnamefont {{Flanagan}}}\ and\ \bibinfo {author} {\bibfnamefont
  {T.}~\bibnamefont {{Hinderer}}},\ }\href {\doibase
  10.1103/PhysRevD.77.021502} {\bibfield  {journal} {\bibinfo  {journal}
  {Phys.\ Rev.\ D}\ }\textbf {\bibinfo {volume} {77}},\ \bibinfo {eid} {021502}
  (\bibinfo {year} {2008})},\ \Eprint {http://arxiv.org/abs/0709.1915}
  {arXiv:0709.1915} \BibitemShut {NoStop}%
\bibitem [{\citenamefont {Boyle}\ \emph {et~al.}(2007)\citenamefont {Boyle},
  \citenamefont {Brown}, \citenamefont {Kidder}, \citenamefont {Mroue},
  \citenamefont {Pfeiffer}, \citenamefont {Scheel}, \citenamefont {Cook},\ and\
  \citenamefont {Teukolsky}}]{Boyle2007}%
  \BibitemOpen
  \bibfield  {author} {\bibinfo {author} {\bibfnamefont {M.}~\bibnamefont
  {Boyle}}, \bibinfo {author} {\bibfnamefont {D.~A.}\ \bibnamefont {Brown}},
  \bibinfo {author} {\bibfnamefont {L.~E.}\ \bibnamefont {Kidder}}, \bibinfo
  {author} {\bibfnamefont {A.~H.}\ \bibnamefont {Mroue}}, \bibinfo {author}
  {\bibfnamefont {H.~P.}\ \bibnamefont {Pfeiffer}}, \bibinfo {author}
  {\bibfnamefont {M.~A.}\ \bibnamefont {Scheel}}, \bibinfo {author}
  {\bibfnamefont {G.~B.}\ \bibnamefont {Cook}}, \ and\ \bibinfo {author}
  {\bibfnamefont {S.~A.}\ \bibnamefont {Teukolsky}},\ }\href {\doibase
  10.1103/PhysRevD.76.124038} {\bibfield  {journal} {\bibinfo  {journal} {Phys.
  Rev.}\ }\textbf {\bibinfo {volume} {D76}},\ \bibinfo {pages} {124038}
  (\bibinfo {year} {2007})},\ \Eprint {http://arxiv.org/abs/0710.0158}
  {arXiv:0710.0158 [gr-qc]} \BibitemShut {NoStop}%
\bibitem [{\citenamefont {Damour}\ \emph {et~al.}(2001)\citenamefont {Damour},
  \citenamefont {Iyer},\ and\ \citenamefont {Sathyaprakash}}]{Damour:2000zb}%
  \BibitemOpen
  \bibfield  {author} {\bibinfo {author} {\bibfnamefont {T.}~\bibnamefont
  {Damour}}, \bibinfo {author} {\bibfnamefont {B.~R.}\ \bibnamefont {Iyer}}, \
  and\ \bibinfo {author} {\bibfnamefont {B.}~\bibnamefont {Sathyaprakash}},\
  }\href@noop {} {\bibfield  {journal} {\bibinfo  {journal} {Phys.\ Rev.\ D}\
  }\textbf {\bibinfo {volume} {63}},\ \bibinfo {pages} {044023} (\bibinfo
  {year} {2001})},\ \Eprint {http://arxiv.org/abs/gr-qc/0010009}
  {arXiv:gr-qc/0010009 [gr-qc]} \BibitemShut {NoStop}%
\bibitem [{\citenamefont {Blanchet}\ \emph {et~al.}(2008)\citenamefont
  {Blanchet}, \citenamefont {Faye}, \citenamefont {Iyer},\ and\ \citenamefont
  {Sinha}}]{BFIS}%
  \BibitemOpen
  \bibfield  {author} {\bibinfo {author} {\bibfnamefont {L.}~\bibnamefont
  {Blanchet}}, \bibinfo {author} {\bibfnamefont {G.}~\bibnamefont {Faye}},
  \bibinfo {author} {\bibfnamefont {B.~R.}\ \bibnamefont {Iyer}}, \ and\
  \bibinfo {author} {\bibfnamefont {S.}~\bibnamefont {Sinha}},\ }\href
  {\doibase 10.1088/0264-9381/25/16/165003} {\bibfield  {journal} {\bibinfo
  {journal} {Class.\ Quantum Grav.}\ }\textbf {\bibinfo {volume} {25}},\
  \bibinfo {pages} {165003} (\bibinfo {year} {2008})},\ \Eprint
  {http://arxiv.org/abs/0802.1249} {arXiv:0802.1249 [gr-qc]} \BibitemShut
  {NoStop}%
\bibitem [{\citenamefont {Blanchet}\ and\ \citenamefont
  {Sch\"afer}(1993)}]{Blanchet93}%
  \BibitemOpen
  \bibfield  {author} {\bibinfo {author} {\bibfnamefont {L.}~\bibnamefont
  {Blanchet}}\ and\ \bibinfo {author} {\bibfnamefont {G.}~\bibnamefont
  {Sch\"afer}},\ }\href {http://stacks.iop.org/0264-9381/10/2699} {\bibfield
  {journal} {\bibinfo  {journal} {Class.\ Quantum Grav.}\ }\textbf {\bibinfo
  {volume} {10}},\ \bibinfo {pages} {2699} (\bibinfo {year}
  {1993})}\BibitemShut {NoStop}%
\bibitem [{\citenamefont {{Arun}}\ \emph {et~al.}(2004)\citenamefont {{Arun}},
  \citenamefont {{Blanchet}}, \citenamefont {{Iyer}},\ and\ \citenamefont
  {{Qusailah}}}]{Arun:2004}%
  \BibitemOpen
  \bibfield  {author} {\bibinfo {author} {\bibfnamefont {K.~G.}\ \bibnamefont
  {{Arun}}}, \bibinfo {author} {\bibfnamefont {L.}~\bibnamefont {{Blanchet}}},
  \bibinfo {author} {\bibfnamefont {B.~R.}\ \bibnamefont {{Iyer}}}, \ and\
  \bibinfo {author} {\bibfnamefont {M.~S.~S.}\ \bibnamefont {{Qusailah}}},\
  }\href {\doibase 10.1088/0264-9381/21/15/010} {\bibfield  {journal} {\bibinfo
   {journal} {Class.\ Quantum Grav.}\ }\textbf {\bibinfo {volume} {21}},\
  \bibinfo {pages} {3771} (\bibinfo {year} {2004})},\ \Eprint
  {http://arxiv.org/abs/arXiv:gr-qc/0404085} {arXiv:gr-qc/0404085} \BibitemShut
  {NoStop}%
\bibitem [{\citenamefont {Kidder}(2008)}]{Kidder2008}%
  \BibitemOpen
  \bibfield  {author} {\bibinfo {author} {\bibfnamefont {L.~E.}\ \bibnamefont
  {Kidder}},\ }\href {\doibase 10.1103/PhysRevD.77.044016} {\bibfield
  {journal} {\bibinfo  {journal} {Phys.\ Rev.\ D}\ }\textbf {\bibinfo {volume}
  {77}},\ \bibinfo {pages} {044016} (\bibinfo {year} {2008})},\ \Eprint
  {http://arxiv.org/abs/0710.0614} {arXiv:0710.0614 [gr-qc]} \BibitemShut
  {NoStop}%
\bibitem [{\citenamefont {{Damour}}\ \emph {et~al.}(2012)\citenamefont
  {{Damour}}, \citenamefont {{Nagar}},\ and\ \citenamefont
  {{Villain}}}]{damour:12}%
  \BibitemOpen
  \bibfield  {author} {\bibinfo {author} {\bibfnamefont {T.}~\bibnamefont
  {{Damour}}}, \bibinfo {author} {\bibfnamefont {A.}~\bibnamefont {{Nagar}}}, \
  and\ \bibinfo {author} {\bibfnamefont {L.}~\bibnamefont {{Villain}}},\ }\href
  {\doibase 10.1103/PhysRevD.85.123007} {\bibfield  {journal} {\bibinfo
  {journal} {Phys.\ Rev.\ D}\ }\textbf {\bibinfo {volume} {85}},\ \bibinfo
  {eid} {123007} (\bibinfo {year} {2012})},\ \Eprint
  {http://arxiv.org/abs/1203.4352} {arXiv:1203.4352 [gr-qc]} \BibitemShut
  {NoStop}%
\bibitem [{\citenamefont {Jaranowski}\ and\ \citenamefont
  {Sch\"afer}(1999)}]{Jaranowski99a}%
  \BibitemOpen
  \bibfield  {author} {\bibinfo {author} {\bibfnamefont {P.}~\bibnamefont
  {Jaranowski}}\ and\ \bibinfo {author} {\bibfnamefont {G.}~\bibnamefont
  {Sch\"afer}},\ }\href {\doibase 10.1103/PhysRevD.60.124003} {\bibfield
  {journal} {\bibinfo  {journal} {Phys.\ Rev.\ D}\ }\textbf {\bibinfo {volume}
  {60}},\ \bibinfo {pages} {124003} (\bibinfo {year} {1999})}\BibitemShut
  {NoStop}%
\bibitem [{\citenamefont {de~Andrade}\ \emph {et~al.}(2001)\citenamefont
  {de~Andrade}, \citenamefont {Blanchet},\ and\ \citenamefont
  {Faye}}]{Andrade01}%
  \BibitemOpen
  \bibfield  {author} {\bibinfo {author} {\bibfnamefont {V.~C.}\ \bibnamefont
  {de~Andrade}}, \bibinfo {author} {\bibfnamefont {L.}~\bibnamefont
  {Blanchet}}, \ and\ \bibinfo {author} {\bibfnamefont {G.}~\bibnamefont
  {Faye}},\ }\href {http://stacks.iop.org/0264-9381/18/753} {\bibfield
  {journal} {\bibinfo  {journal} {Class.\ Quantum Grav.}\ }\textbf {\bibinfo
  {volume} {18}},\ \bibinfo {pages} {753} (\bibinfo {year} {2001})},\ \Eprint
  {http://arxiv.org/abs/gr-qc/0011063} {arXiv:gr-qc/0011063 [gr-qc]}
  \BibitemShut {NoStop}%
\bibitem [{\citenamefont {Blanchet}\ and\ \citenamefont
  {Faye}(2001)}]{Blanchet:2000ub}%
  \BibitemOpen
  \bibfield  {author} {\bibinfo {author} {\bibfnamefont {L.}~\bibnamefont
  {Blanchet}}\ and\ \bibinfo {author} {\bibfnamefont {G.}~\bibnamefont
  {Faye}},\ }\href {\doibase 10.1103/PhysRevD.63.062005} {\bibfield  {journal}
  {\bibinfo  {journal} {Phys. Rev.}\ }\textbf {\bibinfo {volume} {D63}},\
  \bibinfo {pages} {062005} (\bibinfo {year} {2001})},\ \Eprint
  {http://arxiv.org/abs/gr-qc/0007051} {arXiv:gr-qc/0007051 [gr-qc]}
  \BibitemShut {NoStop}%
\bibitem [{\citenamefont {{Damour}}\ \emph {et~al.}(2001)\citenamefont
  {{Damour}}, \citenamefont {{Jaranowski}},\ and\ \citenamefont
  {{Sch{\"a}fer}}}]{Damour01}%
  \BibitemOpen
  \bibfield  {author} {\bibinfo {author} {\bibfnamefont {T.}~\bibnamefont
  {{Damour}}}, \bibinfo {author} {\bibfnamefont {P.}~\bibnamefont
  {{Jaranowski}}}, \ and\ \bibinfo {author} {\bibfnamefont {G.}~\bibnamefont
  {{Sch{\"a}fer}}},\ }\href {\doibase 10.1016/S0370-2693(01)00642-6} {\bibfield
   {journal} {\bibinfo  {journal} {Physics Letters B}\ }\textbf {\bibinfo
  {volume} {513}},\ \bibinfo {pages} {147} (\bibinfo {year} {2001})},\ \Eprint
  {http://arxiv.org/abs/arXiv:gr-qc/0105038} {arXiv:gr-qc/0105038} \BibitemShut
  {NoStop}%
\bibitem [{\citenamefont {Blanchet}\ \emph {et~al.}(2002)\citenamefont
  {Blanchet}, \citenamefont {Faye}, \citenamefont {Iyer},\ and\ \citenamefont
  {Joguet}}]{Blanchet02a}%
  \BibitemOpen
  \bibfield  {author} {\bibinfo {author} {\bibfnamefont {L.}~\bibnamefont
  {Blanchet}}, \bibinfo {author} {\bibfnamefont {G.}~\bibnamefont {Faye}},
  \bibinfo {author} {\bibfnamefont {B.~R.}\ \bibnamefont {Iyer}}, \ and\
  \bibinfo {author} {\bibfnamefont {B.}~\bibnamefont {Joguet}},\ }\href
  {\doibase 10.1103/PhysRevD.65.061501} {\bibfield  {journal} {\bibinfo
  {journal} {Phys.\ Rev.\ D}\ }\textbf {\bibinfo {volume} {65}},\ \bibinfo
  {pages} {061501} (\bibinfo {year} {2002})},\ \bibinfo {note} {erratum:
  \cite{BlanchetEtAl2005a}},\ \Eprint {http://arxiv.org/abs/gr-qc/0105099}
  {arXiv:gr-qc/0105099 [gr-qc]} \BibitemShut {NoStop}%
\bibitem [{\citenamefont {Blanchet}\ \emph {et~al.}(2004)\citenamefont
  {Blanchet}, \citenamefont {Damour}, \citenamefont {Esposito-Farese},\ and\
  \citenamefont {Iyer}}]{Blanchet04a}%
  \BibitemOpen
  \bibfield  {author} {\bibinfo {author} {\bibfnamefont {L.}~\bibnamefont
  {Blanchet}}, \bibinfo {author} {\bibfnamefont {T.}~\bibnamefont {Damour}},
  \bibinfo {author} {\bibfnamefont {G.}~\bibnamefont {Esposito-Farese}}, \ and\
  \bibinfo {author} {\bibfnamefont {B.~R.}\ \bibnamefont {Iyer}},\ }\href
  {\doibase 10.1103/PhysRevLett.93.091101} {\bibfield  {journal} {\bibinfo
  {journal} {Phys. Rev. Lett.}\ }\textbf {\bibinfo {volume} {93}},\ \bibinfo
  {pages} {091101} (\bibinfo {year} {2004})},\ \Eprint
  {http://arxiv.org/abs/gr-qc/0406012} {arXiv:gr-qc/0406012 [gr-qc]}
  \BibitemShut {NoStop}%
\bibitem [{\citenamefont {Blanchet}\ \emph {et~al.}(2006)\citenamefont
  {Blanchet}, \citenamefont {Buonanno},\ and\ \citenamefont
  {Faye}}]{Blanchet-Buonanno-Faye:2006}%
  \BibitemOpen
  \bibfield  {author} {\bibinfo {author} {\bibfnamefont {L.}~\bibnamefont
  {Blanchet}}, \bibinfo {author} {\bibfnamefont {A.}~\bibnamefont {Buonanno}},
  \ and\ \bibinfo {author} {\bibfnamefont {G.}~\bibnamefont {Faye}},\ }\href
  {\doibase 10.1103/PhysRevD.74.104034} {\bibfield  {journal} {\bibinfo
  {journal} {Phys.\ Rev.\ D}\ }\textbf {\bibinfo {volume} {74}},\ \bibinfo
  {eid} {104034} (\bibinfo {year} {2006})}\BibitemShut {NoStop}%
\bibitem [{\citenamefont {Kidder}(1995)}]{Kidder:1995zr}%
  \BibitemOpen
  \bibfield  {author} {\bibinfo {author} {\bibfnamefont {L.~E.}\ \bibnamefont
  {Kidder}},\ }\href {\doibase 10.1103/PhysRevD.52.821} {\bibfield  {journal}
  {\bibinfo  {journal} {Phys.\ Rev.}\ }\textbf {\bibinfo {volume} {D52}},\
  \bibinfo {pages} {821} (\bibinfo {year} {1995})},\ \Eprint
  {http://arxiv.org/abs/gr-qc/9506022} {arXiv:gr-qc/9506022} \BibitemShut
  {NoStop}%
\bibitem [{\citenamefont {Will}\ and\ \citenamefont {Wiseman}(1996)}]{Will96}%
  \BibitemOpen
  \bibfield  {author} {\bibinfo {author} {\bibfnamefont {C.~M.}\ \bibnamefont
  {Will}}\ and\ \bibinfo {author} {\bibfnamefont {A.~G.}\ \bibnamefont
  {Wiseman}},\ }\href {\doibase 10.1103/PhysRevD.54.4813} {\bibfield  {journal}
  {\bibinfo  {journal} {Phys. Rev.}\ }\textbf {\bibinfo {volume} {D54}},\
  \bibinfo {pages} {4813} (\bibinfo {year} {1996})},\ \Eprint
  {http://arxiv.org/abs/gr-qc/9608012} {arXiv:gr-qc/9608012 [gr-qc]}
  \BibitemShut {NoStop}%
\bibitem [{\citenamefont {Poisson}(1998)}]{Poisson:1997ha}%
  \BibitemOpen
  \bibfield  {author} {\bibinfo {author} {\bibfnamefont {E.}~\bibnamefont
  {Poisson}},\ }\href {\doibase 10.1103/PhysRevD.57.5287} {\bibfield  {journal}
  {\bibinfo  {journal} {Phys.\ Rev.}\ }\textbf {\bibinfo {volume} {D57}},\
  \bibinfo {pages} {5287} (\bibinfo {year} {1998})},\ \Eprint
  {http://arxiv.org/abs/gr-qc/9709032} {arXiv:gr-qc/9709032} \BibitemShut
  {NoStop}%
\bibitem [{\citenamefont {Bernuzzi}\ \emph {et~al.}(2014)\citenamefont
  {Bernuzzi}, \citenamefont {Nagar}, \citenamefont {Balmelli}, \citenamefont
  {Dietrich},\ and\ \citenamefont {Ujevic}}]{Bernuzzi:2014kca}%
  \BibitemOpen
  \bibfield  {author} {\bibinfo {author} {\bibfnamefont {S.}~\bibnamefont
  {Bernuzzi}}, \bibinfo {author} {\bibfnamefont {A.}~\bibnamefont {Nagar}},
  \bibinfo {author} {\bibfnamefont {S.}~\bibnamefont {Balmelli}}, \bibinfo
  {author} {\bibfnamefont {T.}~\bibnamefont {Dietrich}}, \ and\ \bibinfo
  {author} {\bibfnamefont {M.}~\bibnamefont {Ujevic}},\ }\href {\doibase
  10.1103/PhysRevLett.112.201101} {\bibfield  {journal} {\bibinfo  {journal}
  {Phys.\ Rev.\ Lett.}\ }\textbf {\bibinfo {volume} {112}},\ \bibinfo {pages}
  {201101} (\bibinfo {year} {2014})},\ \Eprint
  {http://arxiv.org/abs/arXiv:1402.6244 [gr-qc]} {arXiv:1402.6244 [gr-qc]}
  \BibitemShut {NoStop}%
\bibitem [{\citenamefont {Buonanno}\ \emph {et~al.}(2009)\citenamefont
  {Buonanno}, \citenamefont {Iyer}, \citenamefont {Ochsner}, \citenamefont
  {Pan},\ and\ \citenamefont {Sathyaprakash}}]{Buonanno:2009}%
  \BibitemOpen
  \bibfield  {author} {\bibinfo {author} {\bibfnamefont {A.}~\bibnamefont
  {Buonanno}}, \bibinfo {author} {\bibfnamefont {B.}~\bibnamefont {Iyer}},
  \bibinfo {author} {\bibfnamefont {E.}~\bibnamefont {Ochsner}}, \bibinfo
  {author} {\bibfnamefont {Y.}~\bibnamefont {Pan}}, \ and\ \bibinfo {author}
  {\bibfnamefont {B.~S.}\ \bibnamefont {Sathyaprakash}},\ }\href {\doibase
  10.1103/PhysRevD.80.084043} {\bibfield  {journal} {\bibinfo  {journal} {Phys.
  Rev.}\ }\textbf {\bibinfo {volume} {D80}},\ \bibinfo {pages} {084043}
  (\bibinfo {year} {2009})},\ \Eprint {http://arxiv.org/abs/0907.0700}
  {arXiv:0907.0700 [gr-qc]} \BibitemShut {NoStop}%
\bibitem [{\citenamefont {Damour}\ \emph {et~al.}(2000)\citenamefont {Damour},
  \citenamefont {Iyer},\ and\ \citenamefont {Sathyaprakash}}]{DIS00}%
  \BibitemOpen
  \bibfield  {author} {\bibinfo {author} {\bibfnamefont {T.}~\bibnamefont
  {Damour}}, \bibinfo {author} {\bibfnamefont {B.~R.}\ \bibnamefont {Iyer}}, \
  and\ \bibinfo {author} {\bibfnamefont {B.~S.}\ \bibnamefont
  {Sathyaprakash}},\ }\href {\doibase 10.1103/PhysRevD.62.084036} {\bibfield
  {journal} {\bibinfo  {journal} {Phys.\ Rev.\ D}\ }\textbf {\bibinfo {volume}
  {62}},\ \bibinfo {pages} {084036} (\bibinfo {year} {2000})}\BibitemShut
  {NoStop}%
\bibitem [{\citenamefont {McKechan}\ \emph {et~al.}(2010)\citenamefont
  {McKechan}, \citenamefont {Robinson},\ and\ \citenamefont
  {Sathyaprakash}}]{McKechan:2010kp}%
  \BibitemOpen
  \bibfield  {author} {\bibinfo {author} {\bibfnamefont {D.}~\bibnamefont
  {McKechan}}, \bibinfo {author} {\bibfnamefont {C.}~\bibnamefont {Robinson}},
  \ and\ \bibinfo {author} {\bibfnamefont {B.}~\bibnamefont {Sathyaprakash}},\
  }\href {\doibase 10.1088/0264-9381/27/8/084020} {\bibfield  {journal}
  {\bibinfo  {journal} {Class.\ Quantum Grav.}\ }\textbf {\bibinfo {volume}
  {27}},\ \bibinfo {pages} {084020} (\bibinfo {year} {2010})},\ \Eprint
  {http://arxiv.org/abs/1003.2939} {arXiv:1003.2939 [gr-qc]} \BibitemShut
  {NoStop}%
\bibitem [{\citenamefont {Varma}\ and\ \citenamefont
  {Ajith}(2017)}]{Varma:2016dnf}%
  \BibitemOpen
  \bibfield  {author} {\bibinfo {author} {\bibfnamefont {V.}~\bibnamefont
  {Varma}}\ and\ \bibinfo {author} {\bibfnamefont {P.}~\bibnamefont {Ajith}},\
  }\href {\doibase 10.1103/PhysRevD.96.124024} {\bibfield  {journal} {\bibinfo
  {journal} {Phys. Rev.}\ }\textbf {\bibinfo {volume} {D96}},\ \bibinfo {pages}
  {124024} (\bibinfo {year} {2017})},\ \Eprint
  {http://arxiv.org/abs/1612.05608} {arXiv:1612.05608 [gr-qc]} \BibitemShut
  {NoStop}%
\bibitem [{\citenamefont {Yagi}\ and\ \citenamefont
  {Yunes}(2013{\natexlab{a}})}]{Yagi:2013ilq}%
  \BibitemOpen
  \bibfield  {author} {\bibinfo {author} {\bibfnamefont {K.}~\bibnamefont
  {Yagi}}\ and\ \bibinfo {author} {\bibfnamefont {N.}~\bibnamefont {Yunes}},\
  }\href {\doibase 10.1103/PhysRevD.88.023009} {\bibfield  {journal} {\bibinfo
  {journal} {Phys. Rev. D}\ }\textbf {\bibinfo {volume} {88}},\ \bibinfo
  {pages} {023009} (\bibinfo {year} {2013}{\natexlab{a}})},\ \Eprint
  {http://arxiv.org/abs/arXiv:1303.1528 [gr-qc]} {arXiv:1303.1528 [gr-qc]}
  \BibitemShut {NoStop}%
\bibitem [{\citenamefont {Yagi}\ and\ \citenamefont
  {Yunes}(2013{\natexlab{b}})}]{Yagi:2013bca}%
  \BibitemOpen
  \bibfield  {author} {\bibinfo {author} {\bibfnamefont {K.}~\bibnamefont
  {Yagi}}\ and\ \bibinfo {author} {\bibfnamefont {N.}~\bibnamefont {Yunes}},\
  }\href {\doibase 10.1126/science.1236462} {\bibfield  {journal} {\bibinfo
  {journal} {Science}\ }\textbf {\bibinfo {volume} {341}},\ \bibinfo {pages}
  {365} (\bibinfo {year} {2013}{\natexlab{b}})},\ \Eprint
  {http://arxiv.org/abs/arXiv:1302.4499 [gr-qc]} {arXiv:1302.4499 [gr-qc]}
  \BibitemShut {NoStop}%
\bibitem [{\citenamefont {Yagi}(2014)}]{Yagi:2014mlr}%
  \BibitemOpen
  \bibfield  {author} {\bibinfo {author} {\bibfnamefont {K.}~\bibnamefont
  {Yagi}},\ }\href {\doibase 10.1103/PhysRevD.89.043011} {\bibfield  {journal}
  {\bibinfo  {journal} {Phys. Rev. D}\ }\textbf {\bibinfo {volume} {89}},\
  \bibinfo {pages} {043011} (\bibinfo {year} {2014})},\ \bibinfo {note}
  {errata: \cite{Yagi:2017erratum},\cite{Yagi:2018erratum}},\ \Eprint
  {http://arxiv.org/abs/arXiv:1311.0872 [gr-qc]} {arXiv:1311.0872 [gr-qc]}
  \BibitemShut {NoStop}%
\bibitem [{\citenamefont {Chan}\ \emph {et~al.}(2014)\citenamefont {Chan},
  \citenamefont {Sham}, \citenamefont {Leung},\ and\ \citenamefont
  {Lin}}]{Chan:2014a}%
  \BibitemOpen
  \bibfield  {author} {\bibinfo {author} {\bibfnamefont {T.~K.}\ \bibnamefont
  {Chan}}, \bibinfo {author} {\bibfnamefont {Y.-H.}\ \bibnamefont {Sham}},
  \bibinfo {author} {\bibfnamefont {P.~T.}\ \bibnamefont {Leung}}, \ and\
  \bibinfo {author} {\bibfnamefont {L.-M.}\ \bibnamefont {Lin}},\ }\href
  {\doibase 10.1103/PhysRevD.90.124023} {\bibfield  {journal} {\bibinfo
  {journal} {Phys. Rev. D}\ }\textbf {\bibinfo {volume} {90}},\ \bibinfo
  {pages} {124023} (\bibinfo {year} {2014})},\ \Eprint
  {http://arxiv.org/abs/arXiv:1408.3789 [gr-qc]} {arXiv:1408.3789 [gr-qc]}
  \BibitemShut {NoStop}%
\bibitem [{\citenamefont {Abdelsalhin}\ \emph {et~al.}(2018)\citenamefont
  {Abdelsalhin}, \citenamefont {Gualtieri},\ and\ \citenamefont
  {Pani}}]{Abdelsalhin:2018a}%
  \BibitemOpen
  \bibfield  {author} {\bibinfo {author} {\bibfnamefont {T.}~\bibnamefont
  {Abdelsalhin}}, \bibinfo {author} {\bibfnamefont {L.}~\bibnamefont
  {Gualtieri}}, \ and\ \bibinfo {author} {\bibfnamefont {P.}~\bibnamefont
  {Pani}},\ }\href {\doibase 10.1103/PhysRevD.98.104046} {\bibfield  {journal}
  {\bibinfo  {journal} {Phys. Rev. D}\ }\textbf {\bibinfo {volume} {98}},\
  \bibinfo {pages} {104046} (\bibinfo {year} {2018})},\ \Eprint
  {http://arxiv.org/abs/1805.01487} {arXiv:1805.01487 [gr-qc]} \BibitemShut
  {NoStop}%
\bibitem [{\citenamefont {Landry}(2018)}]{Landry:2018a}%
  \BibitemOpen
  \bibfield  {author} {\bibinfo {author} {\bibfnamefont {P.}~\bibnamefont
  {Landry}},\ }\href@noop {} {\  (\bibinfo {year} {2018})},\ \Eprint
  {http://arxiv.org/abs/1805.01882} {arXiv:1805.01882 [gr-qc]} \BibitemShut
  {NoStop}%
\bibitem [{\citenamefont {Blanchet}\ \emph {et~al.}(2005)\citenamefont
  {Blanchet}, \citenamefont {Faye}, \citenamefont {Iyer},\ and\ \citenamefont
  {Joguet}}]{BlanchetEtAl2005a}%
  \BibitemOpen
  \bibfield  {author} {\bibinfo {author} {\bibfnamefont {L.}~\bibnamefont
  {Blanchet}}, \bibinfo {author} {\bibfnamefont {G.}~\bibnamefont {Faye}},
  \bibinfo {author} {\bibfnamefont {B.~R.}\ \bibnamefont {Iyer}}, \ and\
  \bibinfo {author} {\bibfnamefont {B.}~\bibnamefont {Joguet}},\ }\href
  {\doibase 10.1103/PhysRevD.71.129902} {\bibfield  {journal} {\bibinfo
  {journal} {Phys.\ Rev.\ D}\ }\textbf {\bibinfo {volume} {71}},\ \bibinfo
  {eid} {129902} (\bibinfo {year} {2005})}\BibitemShut {NoStop}%
\bibitem [{\citenamefont {Yagi}(2017)}]{Yagi:2017erratum}%
  \BibitemOpen
  \bibfield  {author} {\bibinfo {author} {\bibfnamefont {K.}~\bibnamefont
  {Yagi}},\ }\href {\doibase 10.1103/PhysRevD.96.129904} {\bibfield  {journal}
  {\bibinfo  {journal} {Phys. Rev. D}\ }\textbf {\bibinfo {volume} {96}},\
  \bibinfo {pages} {129904} (\bibinfo {year} {2017})}\BibitemShut {NoStop}%
\bibitem [{\citenamefont {Yagi}(2018)}]{Yagi:2018erratum}%
  \BibitemOpen
  \bibfield  {author} {\bibinfo {author} {\bibfnamefont {K.}~\bibnamefont
  {Yagi}},\ }\href {\doibase 10.1103/PhysRevD.97.129901} {\bibfield  {journal}
  {\bibinfo  {journal} {Phys. Rev. D}\ }\textbf {\bibinfo {volume} {97}},\
  \bibinfo {pages} {129901} (\bibinfo {year} {2018})}\BibitemShut {NoStop}%
\end{thebibliography}%

\end{document}